\documentclass[iop, apj, numberedappendix]{emulateapj}

\usepackage{xspace}
\usepackage{amsmath}
\usepackage{framed} 
\usepackage{txfonts}
\usepackage{epstopdf} 
\usepackage{color}
\usepackage{rotating}
\usepackage{natbib}
\usepackage{ulem}
\usepackage{perpage}

\usepackage{hyperref}

\hypersetup{
  colorlinks   = true, 
  urlcolor     = black, 
  linkcolor    = black, 
  citecolor   = black 
}

\MakePerPage{footnote}

\journalinfo{The Astrophysical Journal, {\rm 773:119 (23pp), 2013 August 20}}
\submitted{Received 2012 March 5; accepted 2013 June 21; published 2013 August 1}

\def\salt{SALT2\xspace}
\def\saltone{SALT\xspace}

\def\phoenix{Phoenix\xspace}
\def\sedona{Sedona\xspace}
\def\snana{SNANA\xspace}

\def\sna{SNIa\xspace}
\def\sne{SNeIa\xspace}

\def\mb{m_{\rm B}}
\def\Mb{M_{\rm B}}
\def\delM{\Delta_{\rm M}}
\def\dm15{$\Delta m_{15}$}

\def\fav{\avg{f}}
\def\favav{\avg{\avg{f}}}
\def\fxi{f_{\xi^2}}
\def\fxiav{\avg{f_{\xi^2}}}
\def\fxiavav{\avg{\avg{f_{\xi^2}}}}
\def\fp{f_{\rm p}}
\def\fpav{\avg{f_{\rm p}}}
\def\fpavav{\avg{\avg{f_{\rm p}}}}

\def\chinu{\chi_{\nu}^2}

\newcommand{\msun}{\>{\rm M_{\odot}}}

\newcommand{\kmsmpc}{\>{\rm km}\,{\rm s}^{-1}\,{\rm Mpc}^{-1}}
\newcommand{\kms}{\>{\rm km}\,{\rm s}^{-1}}

\def\gcm3{\mathrm{g} / \mathrm{cm}^3}

\def\gtsima{$\; \buildrel > \over \sim \;$}
\def\ltsima{$\; \buildrel < \over \sim \;$}
\def\prosima{$\; \buildrel \propto \over \sim \;$}
\def\gsim{\lower.7ex\hbox{\gtsima}}
\def\lsim{\lower.7ex\hbox{\ltsima}}
\def\simgt{\lower.7ex\hbox{\gtsima}}
\def\simlt{\lower.7ex\hbox{\ltsima}}
\def\simpr{\lower.7ex\hbox{\prosima}}

\newcommand{\avg}[1]{\langle #1 \rangle}

\newcommand{\degree}{\ensuremath{^\circ}}

\shorttitle{Light Curve Comparisons}
\shortauthors{Diemer et al.}

\begin{document}


\def\figdir{.}
\def\figext{pdf}


\title{Comparing the light curves of simulated Type Ia Supernovae with observations using data-driven models}

\author{Benedikt Diemer\altaffilmark{1, 2, 3}, Richard Kessler\altaffilmark{2, 3, 1}, Carlo Graziani\altaffilmark{1, 2}, George C. Jordan IV\altaffilmark{1, 2}, Donald Q. Lamb\altaffilmark{1, 2, 4}, Min Long\altaffilmark{1, 2}, and Daniel R. van Rossum\altaffilmark{1, 2}}

\affil{
$^1$ Flash Center for Computational Science, The University of Chicago, Chicago, IL 60637, USA; bdiemer@oddjob.uchicago.edu \\
$^2$ Department of Astronomy \& Astrophysics, The University of Chicago, Chicago, IL 60637 USA \\
$^3$ Kavli Institute for Cosmological Physics, The University of Chicago, Chicago, IL 60637 USA \\
$^4$ Enrico Fermi Institute, The University of Chicago, Chicago, IL 60637 USA
}


\begin{abstract}
We propose a robust, quantitative method to compare the synthetic light curves of a Type Ia supernova (\sna) explosion model with a large set of observed \sne, and derive a figure of merit for the explosion model's agreement with observations. The synthetic light curves are fit with the data-driven model \salt which returns values for stretch, color, and magnitude at peak brightness, as well as a goodness-of-fit parameter. Each fit is performed multiple times with different choices of filter bands and epoch range in order to quantify the systematic uncertainty on the fitted parameters. We use a parametric population model for the distribution of observed \sna parameters from large surveys, and extend it to represent red, dim, and bright outliers found in a low-redshift \sna data set. We discuss the potential uncertainties of this population model and find it to be reliable given the current uncertainties on cosmological parameters. Using our population model, we assign each set of fitted parameters a likelihood of being observed in nature, and a figure of merit based on this likelihood. We define a second figure of merit based on the quality of the light curve fit, and combine the two measures into an overall figure of merit for each explosion model. We compute figures of merit for a variety of one-, two-, and three-dimensional explosion models and show that our evaluation method allows meaningful inferences across a wide range of light curve quality and fitted parameters. 
\end{abstract}



\section{Introduction} 
\label{sec:intro}

Due to their value as cosmological distance indicators \citep{riess_98, perlmutter_99}, Type Ia supernovae (hereafter \sne) have received intense observational attention, with large-scale SN surveys exponentially increasing the number of observed \sna events \citep[see, e.g.,][for a review]{goobar_11_review}. This trend will continue over the next decade with even larger data sets from the Dark Energy Survey \citep[DES,][]{bernstein_12_des}, Pan-STARRS \citep{tonry_12_panstarrs} and the Large Synoptic Survey Telescope \citep{ivezic_08_lsst}. These surveys also undertake strong efforts to overcome calibration issues, which currently represent the largest source of uncertainty in using \sne to measure of the dark energy equation of state parameter $w$ \citep{conley_11_snls_3year}. As calibration uncertainties are reduced, astrophysical uncertainties become a larger fraction of the total error budget, and a theoretical understanding of \sne may help to reduce those uncertainties. For example, after a decade of observational efforts to reduce the intrinsic Hubble scatter, only modest progress has been made \citep[see, e.g.,][for a thorough analysis]{silverman_12_bsnip3}. Thus, we will need better theoretical models of \sne to understand the nature of intrinsic scatter, and other systematics such as the correlation with host galaxy environment \citep{kelly_10_hostgal, lampeitl_10_hostgal, sullivan_11_snls3}. A theoretical framework might lead to a physically motivated parameterization of these effects with measurable parameters. 

Unfortunately, current theoretical models are not yet able to investigate subtle systematic effects like these. While the general consensus is that \sne arise from the thermonuclear explosion of a white dwarf in a binary system \citep{whelan_73}, it is still unclear whether the companion star is a main sequence star or red giant (single-degenerate scenario), or another white dwarf \citep[double-degenerate scenario,][]{iben_84_dd}. Recent observations disfavor the red giant scenario \citep{li_11_2011fe_progenitor, brown_11_2011fe_progenitor}, but the question is far from settled \citep[see, e.g.,][for a review]{maoz_12_progenitor_review}. Even within the single-degenerate paradigm there is still no consensus as to the exact explosion mechanism \citep[see, e.g.,][for a review]{Hillebrandt_00_review}. The deflagration-to-detonation model \cite[DDT, or delayed detonation, DD, e.g.][]{khokhlov_91, niemeyer_95_flame, niemeyer_99_ddt, kasen_09_models, krueger_12_ddt, seitenzahl_13_ddt} has emerged as the most popular model, but the exact mechanism by which a detonation is initiated is still unknown  \citep{roepke_07_detcondition, woosley_07_detcondition, woosley_09_detcondition}. Furthermore, the field has only recently moved on to three-dimensional (3D) simulations, and results suggest that, due to the effects of turbulence, lower-dimensional results are not reliable \citep{roepke_07_first_3d_ddt}. Another surprising result of multi-dimensional simulations was the discovery of an alternative explosion mechanism, the gravitationally confined detonation \citep[GCD;][]{plewa_04_gcd, townsley_07_gcd_flames, jordan_08_gcd, meakin_09_gcd_detonation, jordan_12_gcd_detonation}. Besides the DDT and GCD, a variety of other models has been investigated, such as the double-degenerate scenario \citep{dan_11_dd, dan_12_dd, pakmor_12_dd}, double-detonation and sub-Chandrasekhar models \citep{livne_95_models, shen_10_doubledet, kromer_10_doubledet}, and failed detonation models \citep{jordan_12_failed, kromer_13_failed}.

With such a multitude of possible explosion models at hand, we need to use observed \sne to discern between promising and invalid models. The main difficulty in performing such validation is that \sna light curves and spectra are quite diverse, meaning that there is no master light curve or spectral template to compare against explosion models \citep{hoeflich_96_comparisons1, nugent_02_template, benetti_05_diversity, branch_06_spec_groups, hayden_10_sdss_lcs, blondin_12_diversity}. Generally, two methods of comparing theoretical models and observations have been employed: visual comparisons of light curves and spectra, and comparisons of characteristic magnitudes, colors and decline rates. Visual comparisons entail overplotting spectra or light curves of the explosion model in question and one or a few observed events. The obvious advantage is simplicity, but (1) one or a few observed events are not representative of \sne as a whole, (2) the results of visual comparisons are subjective, and (3) the human eye easily misses details, especially when plotting on a logarithmic scale. An alternative method is to reduce the high-dimensional space of light curves to a few characteristic quantities such as peak magnitude, \dm15 \citep{phillips_93} and $B-V$ color. For example, \cite{khohklov_93_lcs} and \cite{livne_95_models} used the peak magnitudes and rise times of their explosion model light curves to compare to observed data, and find these quantities to discern between favored and disfavored models. Over time, modelers extended such comparisons to quantities describing the decline rate, such as $\Delta m_{V,20}$ or \dm15 \citep[see, e.g.,][]{hoeflich_96_comparisons1, hoeflich_96_comparisons2} and used multi-color light curves as well as spectra \citep[see, e.g.,][]{kasen_07_failed, blondin_11_comparison, blondin_13_comp1d}. \citet{hoeflich_10_secondary} devised a method to directly infer physical parameters such as the metallicity and central density of the progenitor by comparing their impact on the light curves with observations. Despite all these efforts, robust methods to evaluate models remain elusive. \cite{roepke_12_2011fe} attempted to discern between a double-degenerate model from \cite{pakmor_12_dd} and a DDT model using the recently observed SN 2011fe \citep{nugent_11_2011fe}. Despite the very different explosion mechanisms, they concluded it was difficult to verify or falsify either model.

The main weakness of many of the aforementioned comparison methods is that they employ only a small subset of the available data, either particular \sne which may or may not be representative of the overall sample, or quantities such as peak magnitudes and \dm15 which rely on one or a few measurements in a particular filter band. Besides the somewhat arbitrary definition of those quantities, much attention is often given to the width-luminosity relation \citep[Phillips relation,][]{phillips_93, phillips_99, kasen_07_wlr}, though the color--magnitude relation is equally important for standardizing peak magnitudes, and hence for cosmology \citep{riess_96_colorlaw, tripp_98_colorlaw, wang_03_colormag}. Nevertheless, the comparison methods described above were appropriate when only a few \sne had been observed with good accuracy. With modern data sets in hand, however, we are now in a position to compare theoretical models with a well-understood set of observed \sne. The observational community uses data-driven models such as \salt\ \citep{guy_07_salt2}, SiFTO \citep{conley_08_sifto}, mlcs2k2 \citep{jha_07_mlcs2k2} and SNooPy \citep{burns_11_snoopy} to fit observed light curves and extract a few meaningful parameters which summarize the properties of the light curves in question. The results of those fits, such as stretch, color and magnitude, are used to derive distance moduli for a Hubble diagram. We thus expect that these fitting techniques should also work well for comparing theoretical light curves to observations. 

\citet{blondin_11_comparison} applied data-driven models to the synthetic light curves of \cite{kasen_09_models}, but concluded that neither mlcs2k2, \salt nor SNooPy fit the light curves to sufficient accuracy, and thus focus on rise and decline times, color evolution and spectral comparisons. In this paper, we propose a somewhat different approach. We define a quantitative measure of light curve quality for all explosion models, including those that are not well fit by data-driven models. Though we appreciate the value of spectral comparisons, we focus on broadband quantities. By virtue of averaging over the range of wavelengths in a filter band, the light curves of explosion models are less sensitive to the details of the radiative transport treatment than the spectral features. Secondly, the properties of observed light curves have already been captured in data-driven models which allow us to infer characteristic quantities for a given explosion model. Spectral comparison codes such as SNID \citep{blondin_11_snid} quantify how well a spectrum matches a set of templates, but do not result in any summarizing quantities such as color and stretch.  

Our goal is to design a comparison method which satisfies three main criteria, namely (1) avoids arbitrarily picking a sub-set of the data such as particular \sne, (2) uses quantities which reflect the entire light curve rather than a few particular data points such as the epoch of peak brightness, and (3) results in a well-defined figure of merit, allowing us to rank explosion models by how well they reproduce observations. Our strategy is to fit the explosion model light curves with \salt and compare the fitted stretch, color and peak magnitude to the measured population of normal \sne, ignoring peculiar events.\footnote{Peculiar \sne include events such as 2002cx \citep{woodvasey_02_2002cx, li_03_2002cx}, 2005gj \citep{aldering_06_2005gj, prieto_07_2005gj} and 2005hk \citep{chornock_06_2005hk, phillips_07_2005hk}. See also \citet{foley_13_iax} and references therein.} This procedure returns two separate indicators of light curve quality, namely how well the light curves in question can be fit with \salt, and how likely the fitted stretch, color and peak magnitude are to be observed in nature. These two indicators are combined into an overall figure of merit. We emphasize that this figure of merit should be seen as a relative measure of comparison between a set of explosion models rather than an absolute measure of quality, chiefly because any changes in the \salt model or its uncertainties (e.g. through re-training), or any changes in the exact fitting procedure, will lead to different results. 

The details of our method are described in Section \ref{sec:method}. In Section \ref{sec:results}, we give the results of applying our fitting procedure to a range of explosion models. We discuss potential shortcomings of our method in Section \ref{sec:disc}, and summarize our results in Section \ref{sec:conclusion}.


\section{Method} 
\label{sec:method}

\begin{figure}
\centering
\includegraphics[trim = 16mm 74mm 50mm 36mm, clip, scale=0.5]{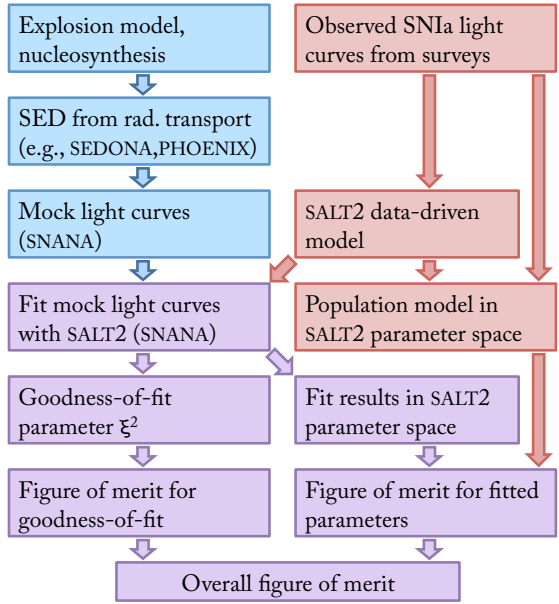}
\caption{Flow chart representation of our procedure for comparing explosion models to observed \sne. Blue boxes represent purely modeling-related steps, i.e., stages in the process where no comparison with observations has taken place yet. Conversely, red boxes mark purely observational stages. Purple boxes represent stages where information from both simulations and observations has been combined.}
\label{fig:flowchart}
\end{figure}

The goal of our analysis is to derive a figure of merit indicating how likely it is that a given explosion model represents observed \sne. Figure \ref{fig:flowchart} gives an overview of the different stages of this procedure. In Section \ref{sec:method:sims} we discuss the hydrodynamic simulations and radiative transfer codes used in this analysis, as well as the conversion from spectral energy distributions (SEDs) to light curves. In Section \ref{sec:method:salt} we introduce data-driven models, and summarize the most important features of the \salt model. In Section \ref{sec:method:fitting} we describe our method for fitting explosion model light curves with the \salt model, and discuss how we derive a goodness-of-fit indicator as well as fitted parameters. In Section \ref{sec:method:popmodel} we describe the population model of \sne in stretch--color--magnitude space. In Section \ref{sec:method:fom} we define figures of merit for goodness-of-fit and fitted parameters, and combine them into an overall figure of merit.

\subsection{Explosion Models and Radiative Transfer}
\label{sec:method:sims}

We are not aiming to introduce new explosion models, or to discuss the details of hydrodynamic simulations, but to use light curves of various explosion models to test our evaluation method. For this purpose, we pick four classes of models which we expect to provide interesting test cases for our method. Namely, we use the well-studied W7 model, the delayed-detonation models of \citet[][hereafter KRW09]{kasen_09_models}, a recent suite of pure deflagration models (M. Long et al. 2013, in preparation), as well as a suite of phenomenological, pre-expanded models. These models represent very different classes of explosions, and were computed in one dimension (1D, W7), two dimensions (2D, KRW09, pre-expanded) and 3D (pure deflagration; light curves for the pure deflagration models were computed in 1D). We briefly discuss these models below. 

The W7 model \citep{nomoto_84_w7, thielemann_86_w7} is a 1D parameterized explosion model. The white dwarf does not detonate, but produces a bright explosion due to a fast deflagration. Despite its simplicity, many radiation transport studies have found W7 to reproduce observed light curves and spectra surprisingly well \citep{jeffery_92, hoeflich_95, nugent_1997, baron_06, kasen_06_sedona, kromer_09_artis, vanrossum_12_phoenix}.

The KRW09 models are discussed in detail in the original paper, as well as in \citet{blondin_11_comparison}. The suite consists of 43 2D delayed-detonation (otherwise known as DDT) explosion models. All models are based on the same progenitor model, but have different distributions of the initial flame bubbles, as well as different detonation criteria. The isotropic (iso) models, as well as one asymmetric model (asym\_1), were given an isotropic distribution of between 20 and 150 flame bubbles. For the other asymmetric models, between 15 and 105 flame bubbles were distributed within a cone of a certain solid angle. The second distinguishing feature is the detonation criterion (dc), where each detonation criterion corresponds to a certain range of densities, as well as a minimum Karlovitz number which has often been used as a criterion for detonation \citep[see, e.g.,][]{niemeyer_97_detonation}. More details on the model setups are given in the supplement to KRW09. The spectral and light curve comparisons in \citet{blondin_11_comparison} showed that some of the KRW09 models agree well with observations, while others are incompatible with observed spectra.

Pure deflagration models rely on the initial deflagration to consume and unbind the white dwarf. Thus, they tend to burn less material to $^{56}$Ni than detonating models, and produce events which are dimmer than standard \sne. We use the light curves of a recently completed suite of six pure deflagration models (M. Long et al. 2013, in preparation). These high-resolution, 3D simulations were computed using the FLASH code \citep{fryxell_00_flash} and follow the deflagration of a $1.365 \msun$ white dwarf. The white dwarf was ignited with a certain number of ignition bubbles, distributed in a sphere centered at the origin. The models are labeled by the number of ignition bubbles (between 63 and 3500) as well as the radius of the sphere (128-384 km), e.g., ``PureDef\_0063\_128''. The explosions produce between 0.13 and 0.29 $\msun$ of $^{56}$Ni.

Finally, we use three 2D, phenomenological, off-centered detonation models. These models were created by expanding an initially stable, Chandrasekhar-mass, white dwarf with a radial velocity. The total expansion energy was equivalent to 30\%, 60\% and 80\% of the binding energy, respectively. We allow the white dwarf to expand, reach its maximum expansion and contract. We trigger a detonation during the contracting phase of the WD by artificially creating a high-temperature region offset from the WD's center. This scenario is designed to investigate strong asphericity, and thus strong variations of light curves with viewing angle. The simulations were computed with the FLASH code. In the rest of the paper, we refer to these models as ``pre-expanded'' models.

SEDs for each explosion model were computed using two different radiative transfer codes, \sedona (KRW09 and pre-expanded models) and \phoenix (W7 and pure deflagration models).  

\sedona \citep{kasen_06_sedona} is a Monte-Carlo code which sends a large number of photon packets through the ejecta, and eventually bins them into an SED. The version of \sedona used can handle 2D, and thus returns spectra as a function of viewing angle. Due to the nature of the Monte-Carlo method, the code returns Poisson uncertainties on the SEDs due to the finite number of photon packages used. However, when integrating the SEDs to obtain light curves, the number of photon packets in each broadband filter is generally so large that the Poisson error is negligible. For all models with \sedona radiative transport, 30 viewing angles were computed. The cosine of the viewing angles were evenly spaced, so that each viewing angle has equal chance of being observed. The viewing angles quoted refer to the mean cosine of the segment in solid angle, ranging from 15\degree\ (the south pole) to 165\degree\ (the north pole).

The second radiation transport code used was \phoenix \citep{hauschildt_99_phoenix, hauschildt_04_phoenix, vanrossum_12_phoenix}. \phoenix numerically solves the special relativistic radiative transfer equation using the efficient and highly accurate short characteristic and operator splitting methods. It samples millions of atomic lines individually, and solves for the two-way interaction between the radiation field and matter in a self-consistent manner (i.e., avoiding the use of the local thermodynamic equilibrium approximation, and thus free scattering parameters). \phoenix solves for the time evolution using the radiation energy balance method. The code does not use the Sobolev approximation, diffusion approximations or opacity binning. \phoenix operates in 1D, assuming spherical symmetry.

The SEDs from \sedona start 2.5 days after explosion, with a cadence of one day, and in wavelength bins of 10 \AA. The \phoenix SEDs for W7 start 4 days after explosion, with a cadence of two days; the SEDs for the pure deflagration models start 1 day after explosion, with a cadence of one day. All Phoenix SEDs are binned in about 1000 irregularly spaced wavelength bins with an average width of 25 \AA. As discussed in Section \ref{sec:method:fitting}, the Poisson uncertainty on the fluxes plays almost no role in the light curve fits and was set to a small number.

Finally, we generate mock light curve observations of the explosion models by multiplying their SEDs with filter transmission functions representing either the Bessell filter system \citep[$UBVRI$;][]{bessell_90} or the SDSS filter system \citep[$ugriz$;][]{fukugita_96_sloanfilters}. This operation is performed using the publicly available Supernova Analysis software \citep[\snana;][]{kessler_09_snana}. The mock observations are generated with a cadence of one day, and at a very low redshift of $z=0.002$ so that K-corrections are unnecessary.

\subsection{The \salt Data-driven Model}
\label{sec:method:salt}

With light curves for a given explosion model in hand, the simplest comparison with observations is to pick one or a few well-observed \sne, and overplot their light curves. However, \sna light curves are heterogeneous, and picking one or a few is inevitably a biased representation of light curves as a whole. Fortunately, the observational community has developed interpolation models, called ``data-driven models'', in order to standardize the brightness of each \sna. They generally constitute complex functions for the spectral evolution of \sne with a few free parameters which are determined from a least-squares fit to light curves. All other degrees of freedom (i.e., the exact shape of the epoch-wavelength surface) are fixed by training the model on large sets of observed events. Thus, data-driven models summarize information from many \sne, and allow a much more robust estimation of light curve characteristics than comparing to a few hand-picked \sne. A drawback of using data-driven models is that they are trained on normal \sne, meaning they do not capture peculiar \sna events.

For the purposes of this analysis, we choose the Spectral Adaptive Light curve Template method \citep[\salt;][hereafter G10]{guy_05_salt, guy_07_salt2, guy_10_snls_3year}. \salt fits broadband photometric light curves using a stretch-dependent SED for each epoch, and a color law. The model is agnostic as to the physical mechanisms which cause decline rate and color variations in \sne. A plethora of other light curve models exists in the literature, almost all of which could be used instead of \salt. For example, SiFTO \citep{conley_08_sifto} is similar to the \salt model, but uses the SED of \citet{hsiao_07}. The Multicolor Light Curve Shapes method \citep[mlcs2k2;][]{riess_96_mlcs, jha_07_mlcs2k2} is built on the underlying assumption that all color variations are due to dust reddening similar to that in our own galaxy. SNooPy \citep{burns_11_snoopy} was especially designed for infrared (IR) light curve fitting, and could be used to extend our analysis into the infrared (given reliable IR light curves from explosion models). 

We choose the \salt model as it is publicly available and has been trained on the most recent SNLS3 sample (G10). Since we use the model extensively in this analysis, we review its main features here. \salt models the flux as a function of wavelength, $\lambda$, and time, $t$, such that
\begin{equation}
\label{equ:salt2_flux}
F(t, \lambda) = x_0 \left[F_0(t, \lambda) + x_1 F_1(t, \lambda) \right] \times e^{c \times CL(\lambda)} \,.
\end{equation}
The spectral surfaces $F_0$ and $F_1$, as well as the color law $CL(\lambda)$, are functions with thousands of parameters, all of which are fixed by ``training'' the model on observed data.  The remaining parameters, $x_0$, $x_1$ and $c$, are derived from a least-squares fit to an individual \sna event, where light curves in all filter bands are fit simultaneously (using the nominal \salt model errors which include a covariance term). Since we refer to these quantities frequently, we shall briefly review their physical meaning. Though the fitted parameter $x_0$ is a flux normalization rather than a magnitude, we will generally refer to the corresponding magnitude
\begin{equation}
\mb = -2.5 \log_{10}(x_0) - 10.095.
\end{equation}
As evident from Equation (\ref{equ:salt2_flux}), $x_1$ is the coefficient of the second flux surface $F_1$, rather than a simple stretch quantity, $s$, in the sense of the original \saltone model ($t \rightarrow t/s$). However, the main component of $F_1$ does correspond to the difference of light curves with different stretch values, so that there are nearly linear relations between $x_1$ and $s$ \citep{guy_07_salt2}, as well as between $x_1$ and $\Delta m_{15}$. Thus, we will refer to $x_1$ as stretch. Lastly, $c$ is the coefficient of the color law of Equation (\ref{equ:salt2_flux}), and is virtually equal to $B-V$ color \citep{kessler_13_var}, 
\begin{equation}
\label{eq:BVc}
B-V = 1.016 c + 0.0008 x_1 + 0.0232.
\end{equation}
Due to these close relationships with observable quantities, we shall speak of the \salt parameter space as a stretch--color--magnitude space. In the rest of this paper we refer to the stretch, color and magnitude derived from a light curve fit interchangeably as fitted parameters, fit results, or simply parameters.

Though \salt's spectral training covers wavelengths from 2000 \AA\ to 9200 \AA, the central wavelengths of filter bands are not recommended to lie beyond 2800 \AA\ and 7000 \AA. We adhere to this standard by using $UBVR$ and $ugr$. The central wavelength for the next set of filters, $I$-band and $i$-band, lie beyond 7000 \AA. Extending filters past the recommended wavelength range would introduce uncertainties of a few percent which would be acceptable for the purpose of fitting explosion models, since they typically deviate from the \salt light curves by a larger amount anyway. However, computing reliable radiative transfer results in the IR has traditionally been challenging because the results are highly sensitive to the ejecta temperature \citep[see, e.g.,][]{kasen_06_ir}. Thus, we refrain from including $I$-band in our comparisons. Our method, however, can easily be extended to include the IR. 

So far, we have discussed the necessary background for performing light curve fits and deriving $x_1$, $c$ and $\mb$. In Section \ref{sec:method:popmodel} we will also derive a population model for those parameters, and thus be concerned with the {\it correlations} between the populations in each parameter. \salt models the correlation between stretch and peak brightness with a parameter $\alpha$, and the correlation between color and peak brightness with a parameter $\beta$. In addition, a magnitude offset $M_0$ allows us to translate to absolute magnitudes \citep{guy_07_salt2}. The three additional parameters are {\it not} derived from the light curve fit, but in a second, independent fit which, given a certain set of observed \sne, minimizes the scatter in the Hubble diagram. For each \sna in the set, the distance modulus is estimated as
\begin{equation}
\label{equ:salt2_mu}
\mu_{\rm est} = \mb - M_0 + \alpha x_1 - \beta c
\end{equation}
where $x_1$, $c$ and $\mb$ are derived from the light curve fit, and $M_0$, $\alpha$ and $\beta$ are determined by the Hubble diagram fit \citep[see, e.g.,][]{marriner_11_salt2mu}. Thus, the results for $M_0$, $\alpha$ and $\beta$ depend on the sample of \sne used for the Hubble diagram fit, and their values may vary between surveys (see Section \ref{sec:disc:popmodel}). It is important to note that using \salt for our light curve fits does {\it not} force us to adopt the model of Equation (\ref{equ:salt2_mu}) when describing the correlations between stretch, color and magnitude. It is, however, by far the most intuitive and well described way to use the results of \salt light curve fits to describe those correlations, and we thus adopt it. 

For explosion models, we can simplify Equation (\ref{equ:salt2_mu}) since we know the true distance modulus $\mu$ at which the mock observations were generated. We translate the observed magnitude into an absolute magnitude, $\Mb = \mb - \mu$, giving an expression for the absolute magnitude we expect a \sna to have based on its stretch and color,
\begin{equation}
\label{equ:salt2_MBfit2}
\Mb = M_0 - \alpha x_1 + \beta c\,.
\end{equation}
We note that $\Mb$ represents the best-fit peak magnitude based on light curves in all filter bands, and is not the magnitude of the $B$-band at the epoch of peak brightness. Furthermore, the relation between $\Mb$ and $\mb$, and thus Equation (\ref{equ:salt2_MBfit2}), are only valid at $z=0$. Since we generate the mock light curves of our explosion models at a negligible redshift, this caveat does not pose a problem. We describe how we derive $\Mb$ for higher redshift data in Appendix \ref{sec:app:data:restframe}.

\subsection{Fitting Explosion Model Light Curves}
\label{sec:method:fitting}

\begin{figure*}
\centering
\includegraphics[trim = 0mm 0mm 0mm 0mm, clip, scale=0.45]{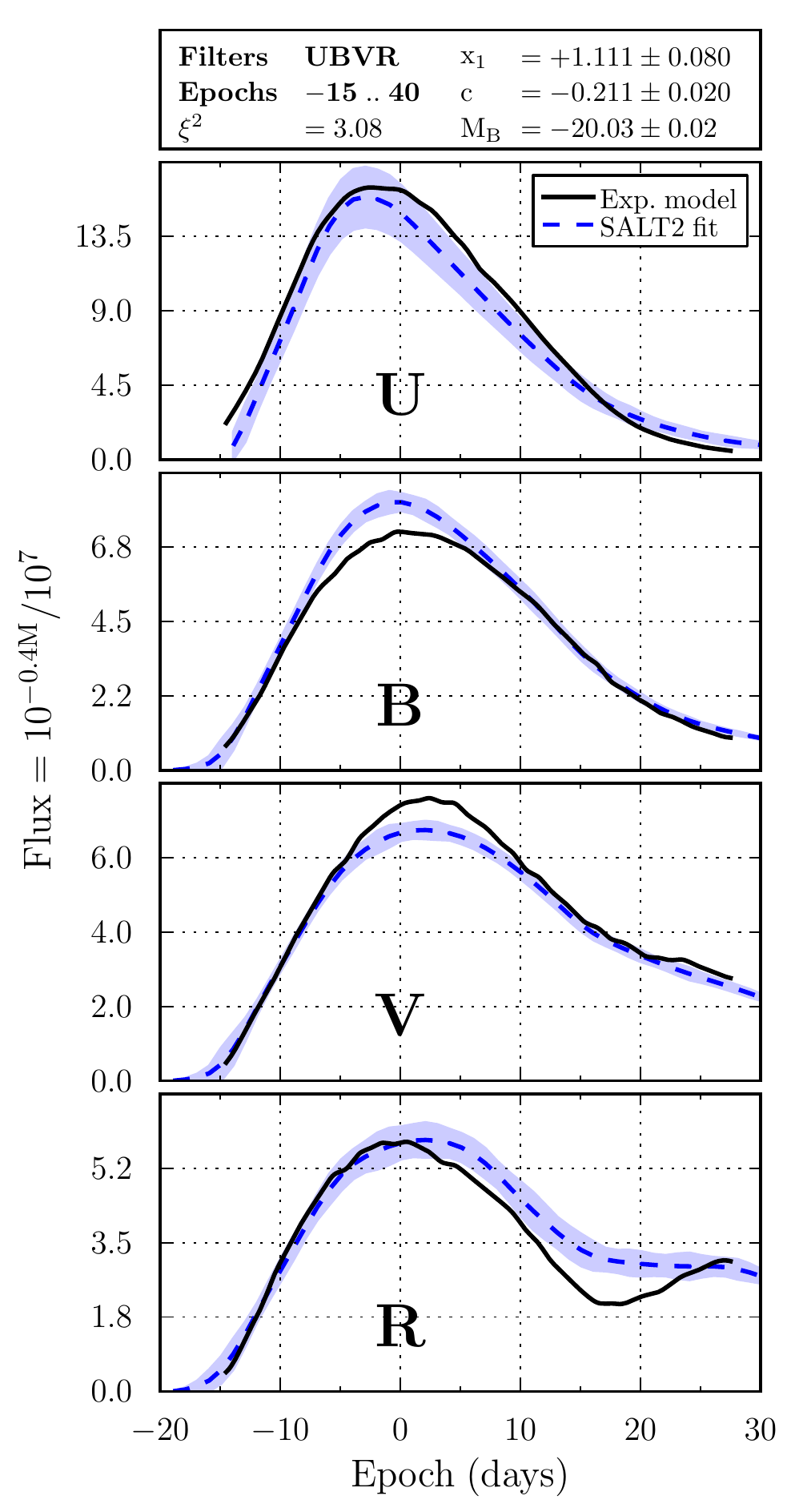}
\includegraphics[trim = 10mm -39mm 0mm 0mm, clip, scale=0.45]{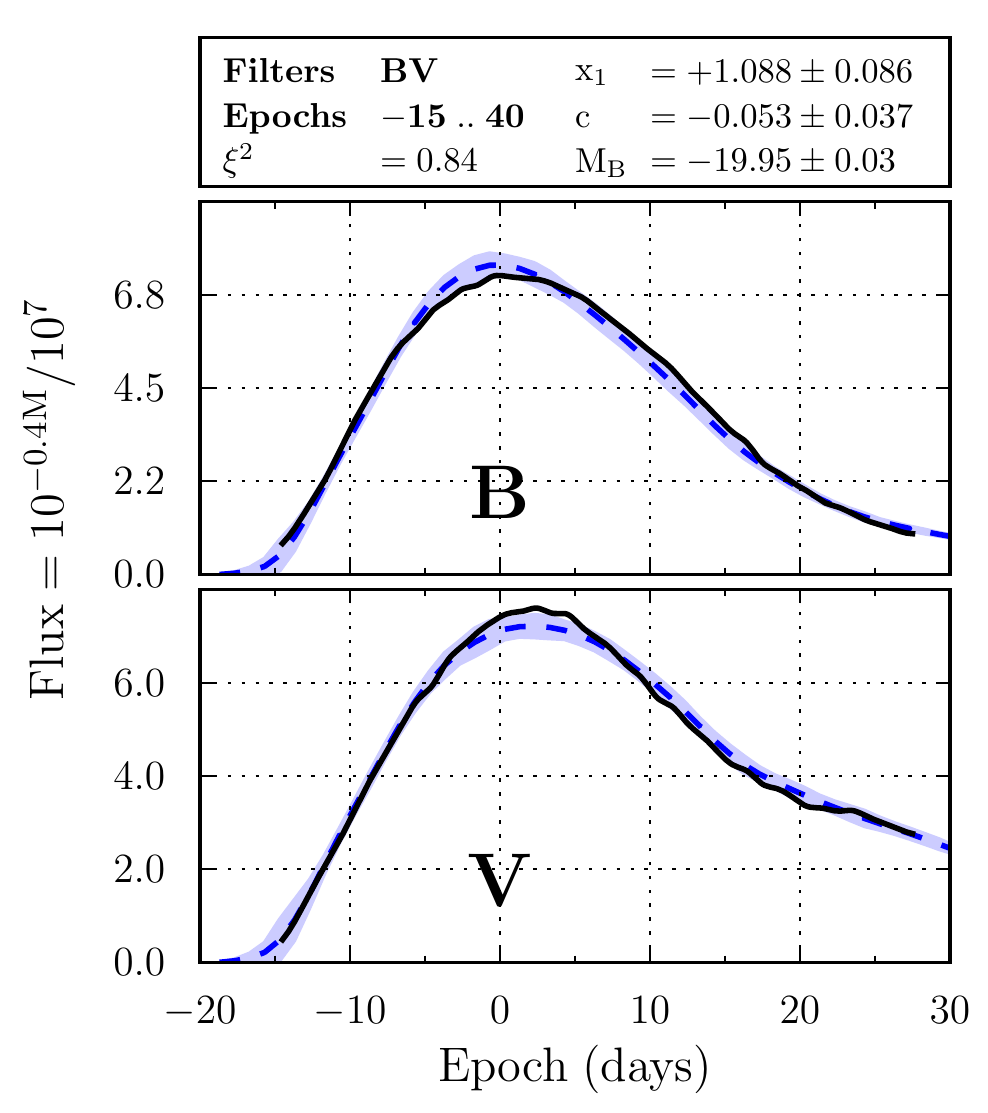}
\includegraphics[trim = 10mm 0mm 0mm 0mm, clip, scale=0.45]{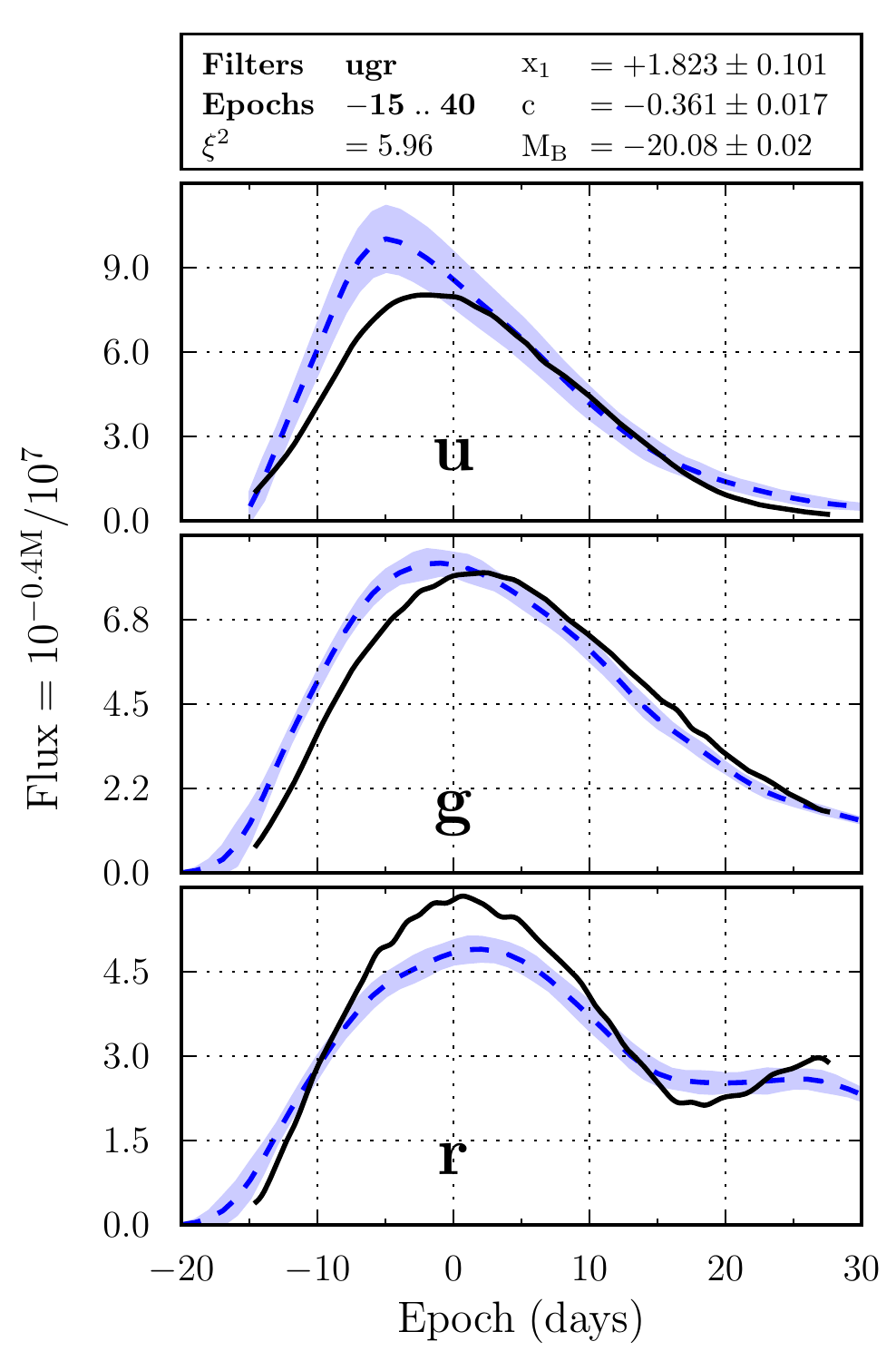}
\includegraphics[trim = 10mm 0mm 0mm 0mm, clip, scale=0.45]{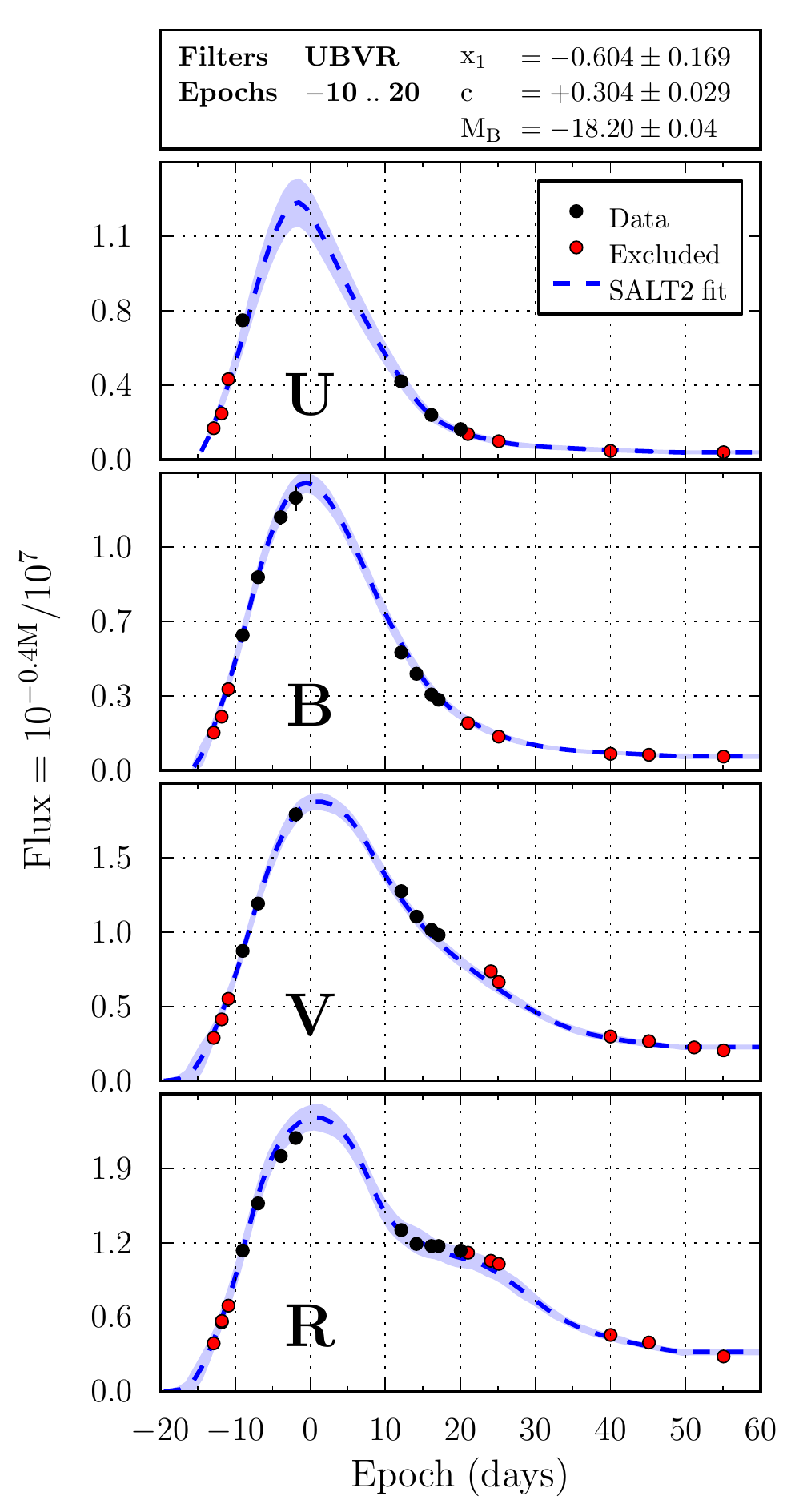}
\caption{The influence of filter band choice on \salt fit results. Each of the left three columns represents a fit to the same explosion model, but using different filter bands. The choices of filters and epoch range, as well as the fit results, are indicated in the top panels. The black curves show the 68\degree viewing angle of the KRW09\_asym\_6\_dc\_2 model. All fits include epochs between $-15$ and $+40$ days, which includes all epochs of this viewing angle. The dashed blue curves show the \salt fits, with shaded blue regions indicating the 1$\sigma$ confidence regions of the \salt model. The three columns correspond to three different choices of filters ($UBVR$, $BV$ and $ugr$), demonstrating that this choice can influence the fit results, particularly when the fit is not very good. In the $UBVR$ fit, the \salt model overestimates the $B$-band peak magnitude, and underestimates the $V$-band peak. In the $BV$ fit, the peak magnitudes of the $B$ and $V$ bands can be fit simultaneously, resulting in a significantly redder color ($c = -0.05$ instead of $-0.21$). The $ugr$ fit, however, results in a bluer color than $UBVR$ ($c = -0.36$). Right column: $UBVR$ fits to the observed SN 2002bo (CfA3 data, see Appendix \ref{sec:app:data:cfa3}). Even though epochs before $-10$ and after $+20$ days are excluded from the fit, the fit is compatible with the data at all epochs.}
\label{fig:fits:filters}
\end{figure*}

\begin{figure*}
\centering
\includegraphics[trim = 0mm 0mm 0mm 0mm, clip, scale=0.45]{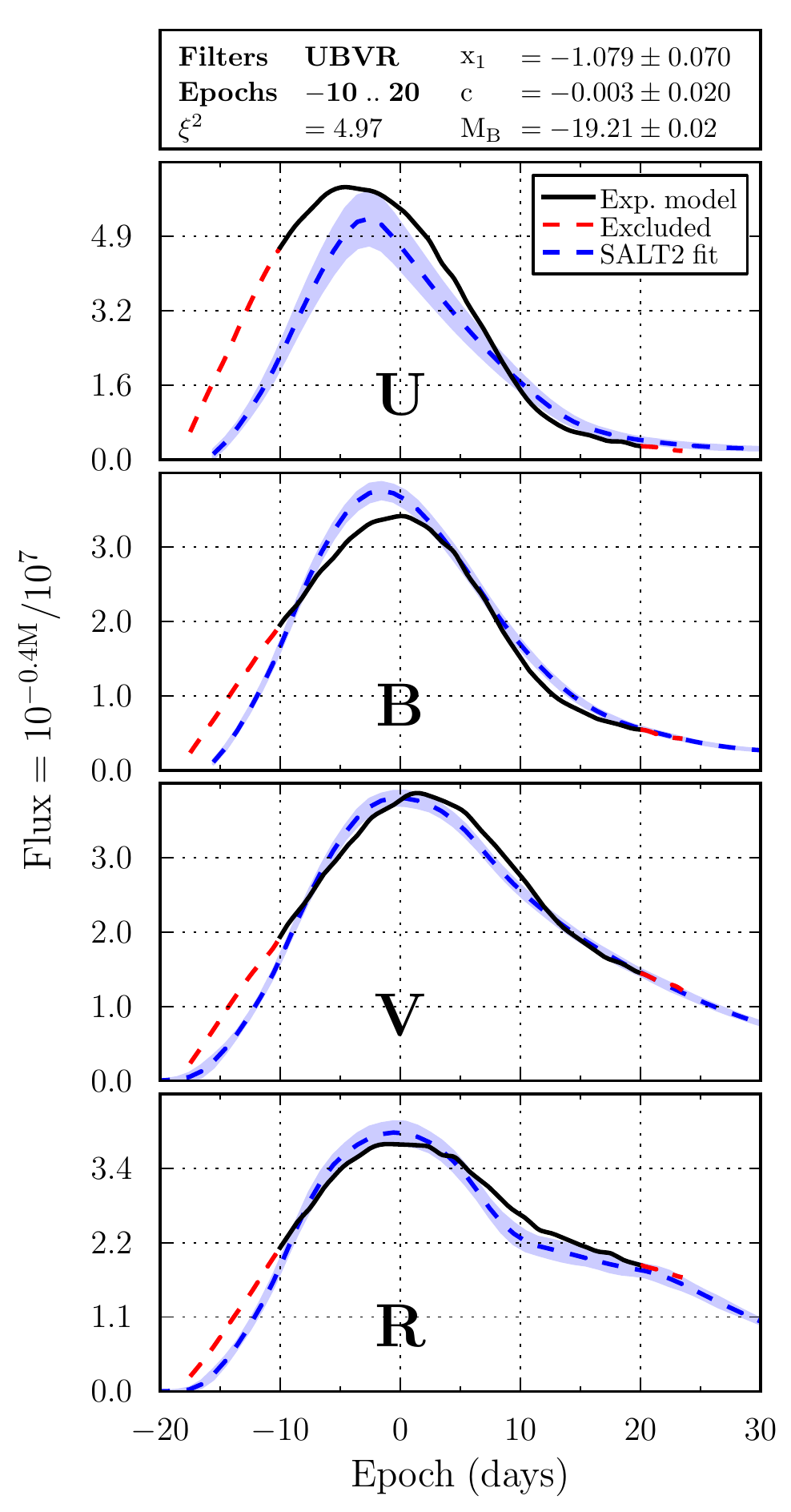}
\includegraphics[trim = 10mm 0mm 0mm 0mm, clip, scale=0.45]{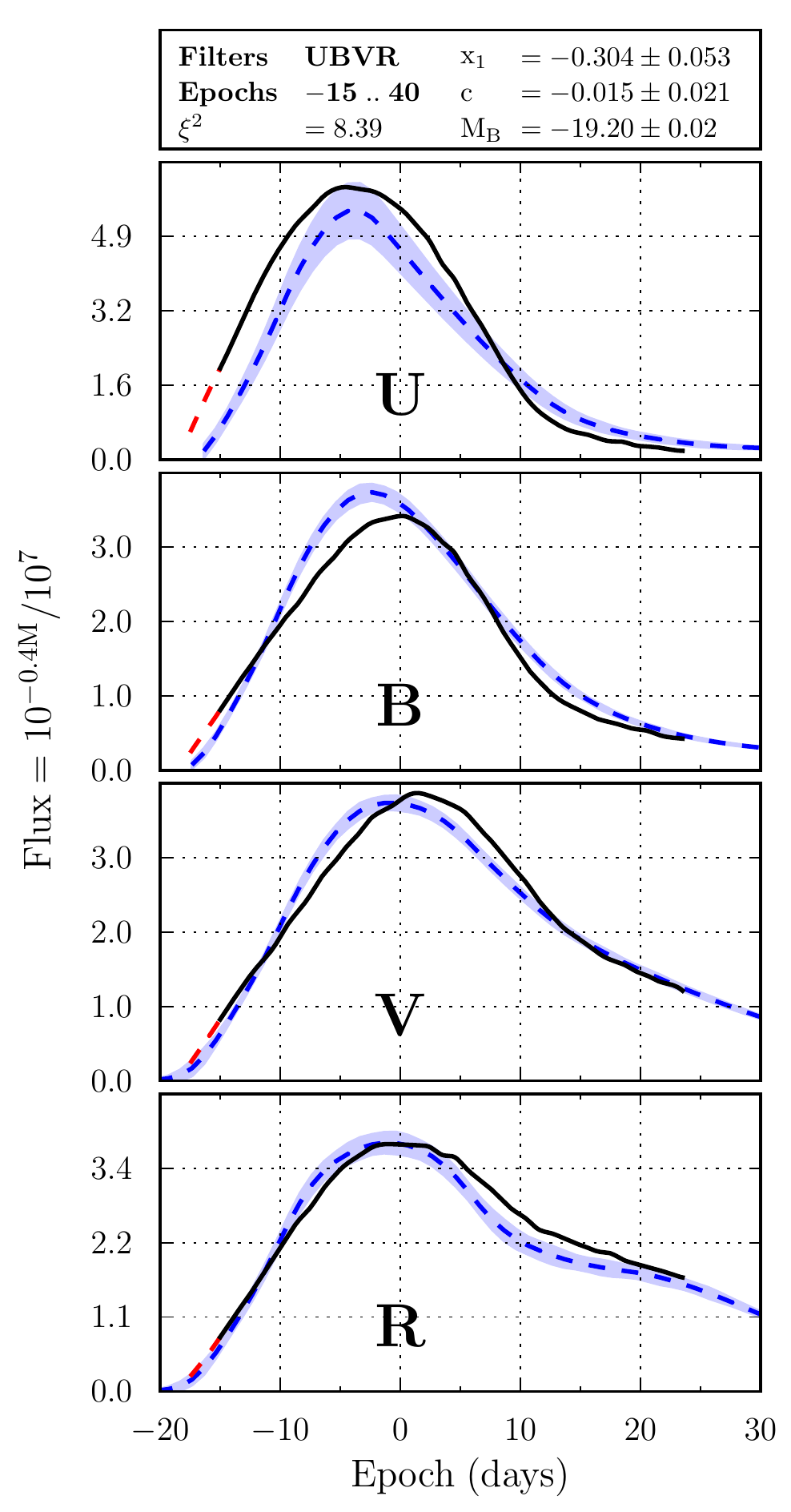}
\includegraphics[trim = 10mm 0mm 0mm 0mm, clip, scale=0.45]{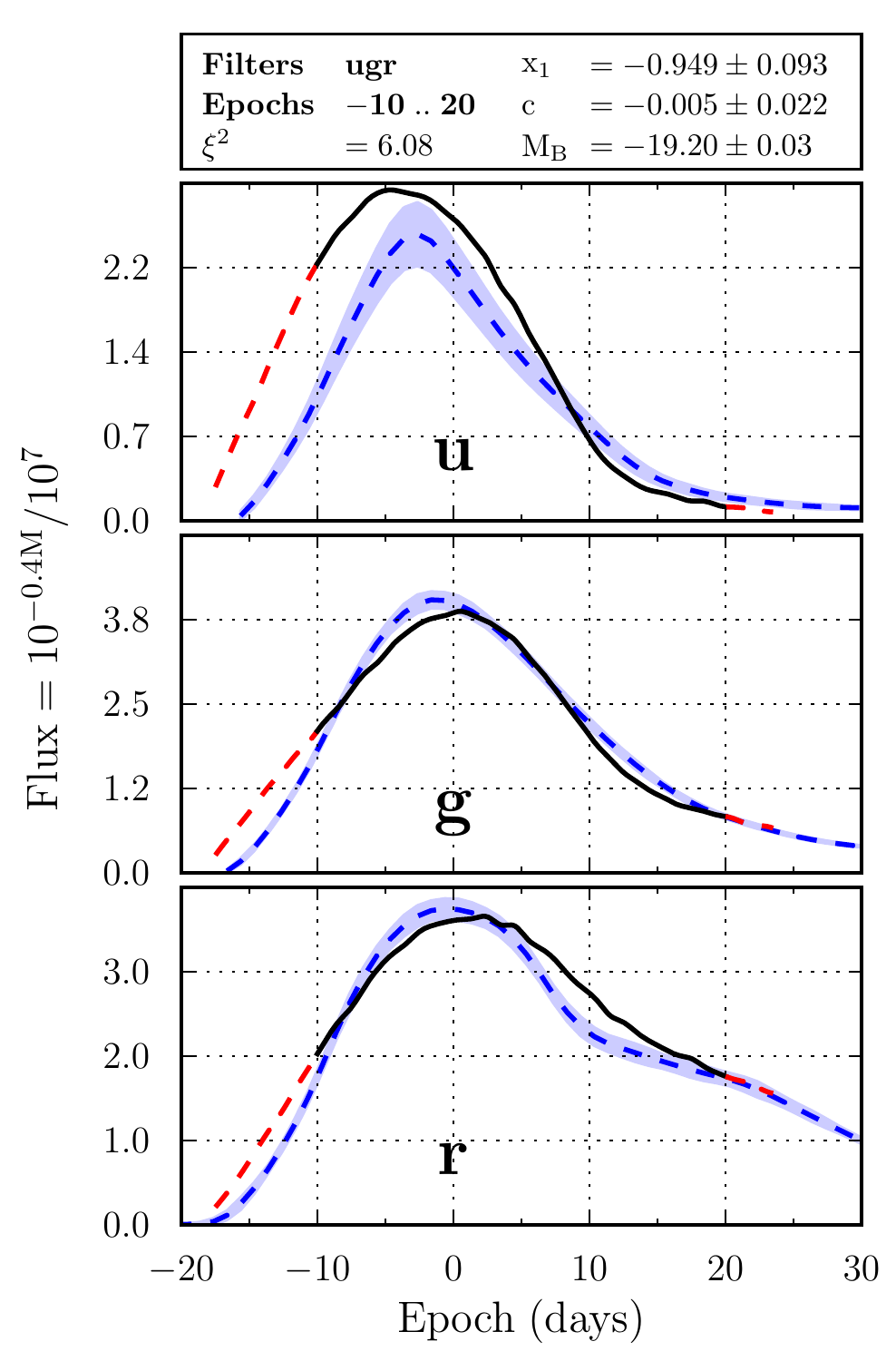}
\includegraphics[trim = 10mm 0mm 0mm 0mm, clip, scale=0.45]{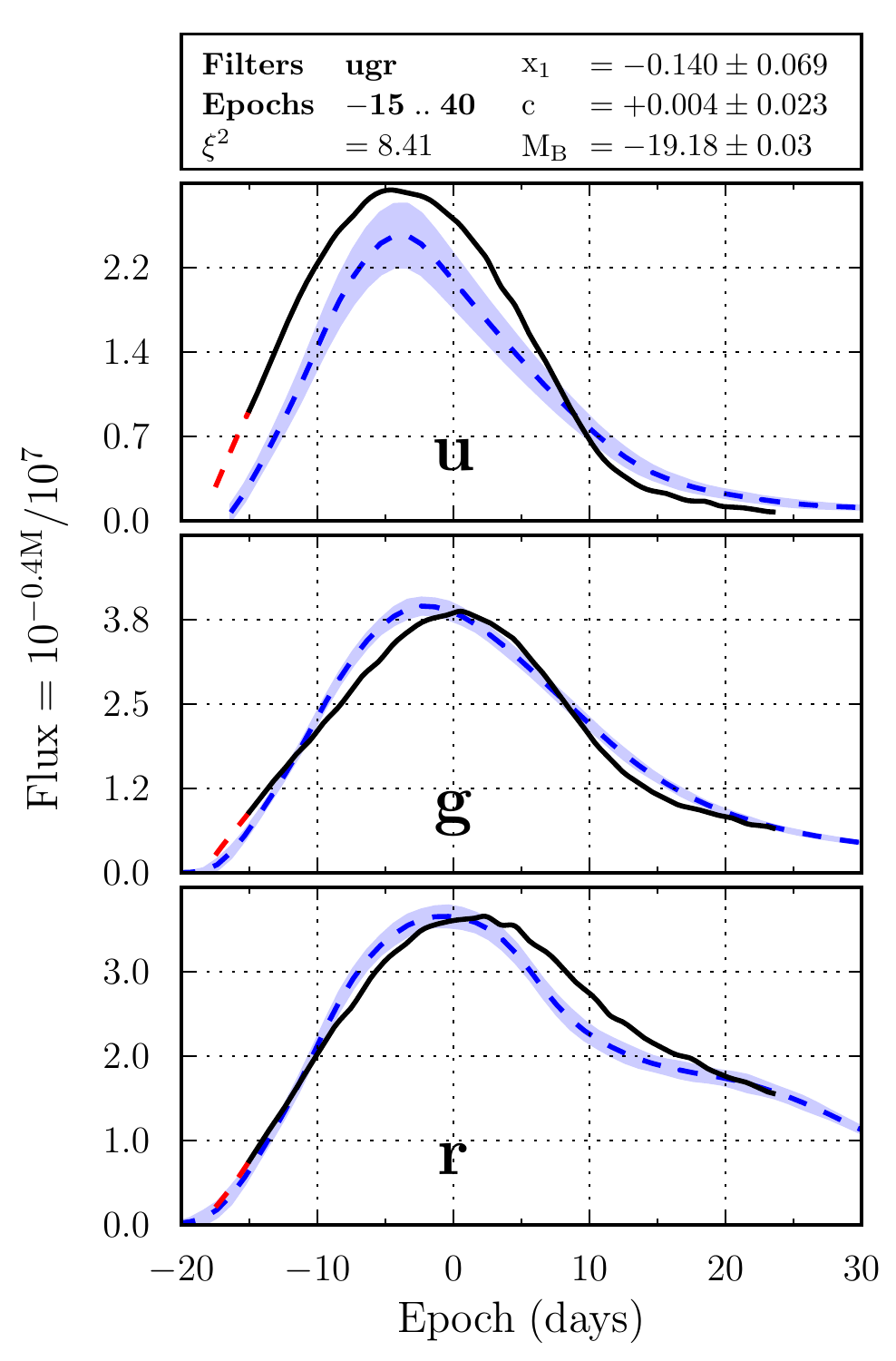}
\caption{Same as Figure \ref{fig:fits:filters}, but for different fitted epoch ranges. The light curves show the 15\degree viewing angle of the KRW09\_iso\_6\_dc\_4 model. In the left column, only epochs from $-10$ to $+20$ days from $B$-band peak brightness were included in the fit, as indicated by the dashed red, excluded regions. In the second column, early and late epochs were included as well ($-15$ to $+40$ days), significantly degrading the fit as the \salt model cannot accommodate the shallow rising slope of the light curves. While the color and magnitude results of the two fits are almost identical, the stretch varies dramatically from $\approx -1$ to $\approx -0.3$. The same is true when the $ugr$ filters are used instead of $UBVR$ (right two columns), with the stretch varying from $\approx -0.95$ to $-0.14$. Note that the main reason for the large discrepancies is that the light curve fit is relatively poor in the first place, at least when the early epochs are included.}
\label{fig:fits:epochrange}
\end{figure*}

The goal of fitting simulated light curves to a data-driven model is to estimate the likelihood that the underlying explosion model could represent an observed \sna. Accepting the paradigm that \sna light curves are described by an analytical function with a few free parameters ($x_1$, $c$ and $\mb$, in the case of \salt), we split this likelihood into two separate components. First, the quality of the fit itself indicates whether the synthetic light curves represent a \sna. Secondly, the resulting best-fit values for $x_1$, $c$ and $\mb$ correspond to a likelihood of observing a \sna with those values. The second part will be discussed in Section \ref{sec:method:popmodel}, while this section deals with how the \salt model is used to fit explosion model light curves rather than observed \sne.

In the ordinary case of fitting observed data, a light curve fit refers to a least-squares fit of a set of given \sna observations to Equation (\ref{equ:salt2_flux}), redshifted into the observer frame and multiplied by the filter bands used in the observations. We use the \snana light curve fitting program to apply the same procedure to our mock observations of explosion models. However, there are some differences between explosion models and observed data which deserve further discussion before we describe the more technical aspects of the fitting procedure.

When we fit explosion model light curves, \salt takes on the role of ``the model'' while the simulated light curves serve as ``the data''. Of course, the simulated light curves do not carry the measurement uncertainties typically associated with \sna observations. In order to follow the model-data approach faithfully, we could use \snana to simulate the survey conditions of, say, SDSS, and generate mock observations at a variety of redshifts, introduce reddening and atmospheric conditions, and propagate them through a virtual telescope, camera and processing pipeline. Such mock observations would then carry flux uncertainties similar to those of the  survey which was simulated. There are, however, two reasons not to follow this procedure. First, we would inevitably introduce selection biases, and discard information about the explosion model light curves. Secondly, the simulated statistical uncertainties would carry no information about the systematic uncertainties of explosion models, which are much larger than the typical statistical uncertainties in the data.

The systematic uncertainties can be split into two categories, parametric and non-parametric errors. Parametric sources of error include the progenitor's mass and C/O abundance ratio, and the number, locations and sizes of the ignition bubbles. Non-parametric systematic errors are caused by missing or uncertain physics, for example in the flame model, the detonation model, the energy deposition of gamma rays, the ray tracing algorithm, or the number and properties of spectral lines in the radiation transfer calculation. More mundane sources of systematic errors include inadequate spatial and temporal resolution in the explosion or radiation transfer phase, an inadequate number or distribution of Lagrangian tracer particles used to determine the abundance of elements, or an inadequate number of rays, spectral bins, or angular bins in the radiation transfer phase. Unless the simulated survey conditions are extremely poor, a combination of all those sources of systematic uncertainty far outweighs the statistical errors.

As systematic errors typically dominate over statistical uncertainties, and because we wish to avoid discarding light curve information by simulating survey conditions, we choose to ignore statistical uncertainties entirely. Instead, we place the simulated \sne at a low redshift and set their statistical uncertainties to a small number, ensuring that the uncertainties on the \salt model dominate the error term in the $\chi^2$ computation. This strategy has two significant consequences. First, the $\chi^2$ per degree of freedom ($\chinu$) of the light curve fits is generally much greater than for observed data. Secondly, we cannot interpret the results of the \salt fits in a statistically meaningful fashion. For example, while $-\chi^2/2$ from the fit to a set of observed \sna light curves is ordinarily interpretable as the log-likelihood of the event having been a \sna, we do not assign such interpretation to an explosion model.

While the lack of rigorous error treatment may be a sobering realization, we are not trying to compute the likelihood of a certain model representing a \sna, but rather we are solving an optimization problem. We are trying to identify which explosion models, or which classes of explosion models, come closest to reproducing the light curves of real \sne. Our methodological approach can yield information about the values of the model parameters that are favored and disfavored given the current sample of observed \sne. It can also provide clues as to non-parametric sources of systematic error. For these purposes, we do not need to derive an absolute likelihood. Thus, we shall avoid quoting quantities which are usually assigned statistical meaning, such as $\chinu$, and instead focus on indicators of light curve quality which can be computed without invoking uncertainties on the light curves themselves.

Having decided to forego a statistical interpretation of the light curve fits, we face another issue which causes systematic uncertainty in the fit results. We need to make certain choices regarding the light curve fits, two of which turn out to influence the results significantly: the choice of filter bands, and the epoch range included in the fit. Figure \ref{fig:fits:filters} demonstrates the impact of the first choice. Adding or omitting filter bands can change the fit results, as shown by the first two columns in Figure \ref{fig:fits:filters}. While fitting the entire $UBVR$ set analyzes a larger wavelength range, fits to a subset may result in more accurate fits to those bands. For example, fits to $BV$ only (center column of Figure \ref{fig:fits:filters}) might lead to a better estimate of $B-V$ color or $B$-band peak brightness. Furthermore, there is no clear reason to prefer, for example, the Bessell filter system ($UBVR$) over the SDSS system ($ugr$), but they result in slightly different fitted parameters (third column of Figure \ref{fig:fits:filters}). Especially the color parameter, $c$, is affected by the choice of filters, since it correlates closely with $B-V$ color (Equation (\ref{eq:BVc})). 

Secondly, the fit results can depend on the choice of fitted epoch range, particularly when the light curve fit is poor, as demonstrated in Figure \ref{fig:fits:epochrange}. A simple argument could be made that one should always include as many epochs as possible, if the radiative transfer code provides them. In practice, however, that is not the best choice. First, the early and late epochs can contribute a disproportionate amount of $\chi^2$ per epoch. This emphasis appears when an explosion model fails to match the rising or falling slope, since the flux uncertainties on the \salt model are small at early and late times (though the magnitude errors are quite large). While such disagreement highlights a real problem with the explosion model, it can drive the fit away from the best fit around peak. In principle, an explosion model should of course be able to reproduce observed light curves at all epochs, but almost no explosion model to date has reached this level of agreement with data. Secondly, one might be interested in the performance of a particular explosion model around peak, not at early or late times. Third, when comparing explosion models we do not wish to penalize a model for going out further in time than others. Such a $\chi^2$ penalty would inevitably occur if the fitted epoch range was varied between models.

The biasing of fit results due to our choices of filter bands and epoch range is expected, since those choices simply correspond to weighting certain regions of the epoch-wavelength surfaces stronger than others. For explosion models which perfectly fit the \salt light curves, however, we expect no such bias. We tested this assumption by generating ideal light curves from the epoch-wavelength surfaces of the \salt model, and then fitting them with \salt. As expected, fits to these ideal light curves reproduced the input magnitude, stretch and color to better than $0.1\%$ error, regardless of which epoch range or filter bands were fit. As a second check, we tested the impact of epoch range on the fitted stretch parameter when fitting observed data. Since data are measured in one particular filter system (e.g., $UBVRI$), we cannot test the impact of using other filter systems (e.g., $ugriz$). We chose two small sets of events from the CfA3 survey (Appendix \ref{sec:app:data:cfa3}) which have data points before $-10$ and after $+20$ days. The first set was chosen to contain only very good fits with $\chinu \approx 1$, while the second set consisted of \sne with relatively poor fits, $\chinu > 1$, such as the sub-luminous 2005hk. All events except 2005hk were part of the \salt training set (G10). When fitted with different epoch ranges and subsets of filters, the fitted parameters for the set with good fits show very little variation (typically about $0.1 \sim 0.2$ in the stretch parameter $x_1$, corresponding to a $0.01 \sim 0.02$ correction in peak magnitude). For example, the light curve fit to 2002bo (shown in the right column of Figure \ref{fig:fits:filters}) is compatible with the data at all epochs, even though epochs before $-10$ and after $+20$ days were excluded from the fit. The differences between fits of the poorly fit set were about twice as large as for the well-fit set, as expected. However, even the poorly fit set shows modest variations with epoch range compared to many explosion models (Section \ref{sec:results}). This test demonstrates once again that the results of high-quality light curve fits are less sensitive to the fitted epoch range than the results of poor fits. Since observed data generally fit much better than explosion models, their fitted parameters change relatively little when changing the fitted epoch range.

In conclusion, the choice of filter bands and epoch range systematically bias the results of light curve fits to explosion models which are not well fit by \salt. The choice of filter bands has a particularly strong impact on color, and the epoch range on stretch. The peak brightness is generally less sensitive to the details of the fitting procedure. Since there is no clear choice, we fit models with a wide range of reasonable filter and epoch range choices. Here, we fit each light curve with three sets of filters ($UBVR$, $BV$ and $ugr$) as well as four epoch ranges ($-10$ to $+20$, $-10$ to $+40$, $-15$ to $+20$ and $-15$ to $+40$), resulting in a total of 12 fits. The three sets of filter bands allow us to fit the maximum wavelength range ($UBVR$), get the best fit to the well-constrained $B$ and $V$ bands ($BV$), and investigate an alternative filter system ($ugr$). The set of epoch ranges was chosen to include and exclude the regions of $-15$ to $-10$ days, and $+20$ to $+40$ days, since many explosion models struggle to match the \salt model at those epochs. While the \salt model is defined to $-20$ days, the data coverage becomes sparse at such early times. Furthermore, many explosion model light curves do not include epochs before $-15$ days.

The dispersion in the fitted parameters $x_1$, $c$ and $\mb$ from the 12 fits is interpreted as a systematic uncertainty on those parameters. As outlined above, this dispersion is generally larger for explosion models which are poorly fit in the first place, meaning that the results are better constrained for explosion models which agree best with data. This correlation is reassuring, since we care somewhat less about how poorly a model fits the data, but {\it do} care in cases where the model fits the data well.

Finally, we wish to define a goodness-of-fit parameter for a light curve fit. As discussed above, $\chinu$ cannot be interpreted in the usual statistical context, since, in the absence of uncertainties on the explosion model, the formal $\chinu$ gives a small likelihood for any of the light curve fits in this analysis. Thus, we define a $\chinu$-like goodness-of-fit parameter,
\begin{equation}
\label{eq:xi2}
\xi^2 \equiv \frac{1}{N} \sum_{i=1}^N \left( \frac{F_{\rm i,model}-F_{\rm i,SALT}}{\sigma_{\rm i,SALT}} \right)^2
\end{equation}
where $N$ is the number of epochs, $F_i$ the flux at epoch $i$, and $\sigma_{\rm i,SALT}$ the uncertainty on the \salt model at that epoch. The main differences between $\xi^2$ and $\chinu$ are that $\xi^2$ is agnostic of uncertainties on the explosion model light curves, and that it ignores the covariances between the fitted parameters. Thus, we once again emphasize that no statistical interpretation should be derived from the value of $\xi^2$.

\subsection{\sna Population in Color--Stretch--Magnitude Space}
\label{sec:method:popmodel}

\begin{deluxetable*}{llll}
\tablecaption{Parameters for the Population Model of SNe in $x_1-c-\Mb$ Parameter Space
\label{table:pop_model}}
\tablewidth{0pt}
\tablehead{
\colhead{Parameter} &
\colhead{Value} &
\colhead{Description} &
\colhead{Reference} 
}
\startdata
$\alpha$ & 0.13 & Slope of $\Mb$--$x_1$ relation (derived from Hubble diagram fit) & \citet{conley_11_snls_3year, sullivan_11_snls3} \\
$\beta$ & 3.2 & Slope of $\Mb$--$c$ relation (derived from Hubble diagram fit) & \citet{conley_11_snls_3year, sullivan_11_snls3} \\
$M_0$ & $-19.095$ & Uncorrected $B$-band peak magnitude (derived from Hubble diagram fit) & \citet{guy_10_snls_3year} \\
$H_0$ & $70 \kmsmpc$ & Hubble constant (degenerate with $M_0$) & \citet{guy_10_snls_3year} \\
\hline
$\sigma_M$ & 0.13 & Intrinsic scatter in $\Mb$ (including uncertainty in $H_0$) & \citet{kessler_13_var} \\
$\mu_{\rm x1}$ & 0.5 & Mean of distribution in $x_1$ & \citet{kessler_13_var} \\
$\sigma_{\rm x1-}$ & 1.4 & Standard deviation for $x_1 > \mu_{\rm x1}$ & \citet{kessler_13_var} \\
$\sigma_{\rm x1+}$ & 0.7 & Standard deviation for $x_1 < \mu_{\rm x1}$ & \citet{kessler_13_var} \\
$\mu_c$ & 0 & Mean of distribution in $c$  & \citet{kessler_13_var} \\
$\sigma_{\rm c-}$ & 0.08  & Standard deviation for $c > \mu_c$ & \citet{kessler_13_var} \\
$\sigma_{\rm c+}$ & 0.13 & Standard deviation for $c < \mu_c$  & \citet{kessler_13_var}  \\
\hline
$w_{\rm c,2}$ & 0.04 & Weight of secondary distribution in $c$ & This paper \\
$\mu_{\rm c,2}$ & 0.3 & Secondary distribution: mean in $c$  & This paper \\
$\sigma_{\rm c-,2}$ & 0.2  & Secondary distribution: standard deviation for $c > \mu_{c,2}$ & This paper \\
$\sigma_{\rm c+,2}$ & 0.2 & Secondary distribution: standard deviation for $c < \mu_{c,2}$  & This paper  \\
$w_{\rm M,2}$ & 0.04 & Weight of secondary distribution in $\Mb$ & This paper \\
$\mu_{\rm M,2}$ & $-19.095$ & Secondary distribution: mean in $\Mb$ & This paper  \\
$\sigma_{\rm M,2}$ & 0.35 & Secondary distribution: standard deviation for $\Mb$  & This paper  \\
\enddata
\tablecomments{The quoted value for $M_0$ given in G10 refers to the SiFTO light curve fitter rather than \salt. Because SiFTO has different corrections for color and stretch, the uncorrected magnitude $M_0$ is slightly different, but can be converted to the \salt value using linear transformations (J. Guy 2012, private communication). The corresponding value of $H_0$ used in their analysis is $70 \kmsmpc$. See Section \ref{sec:disc:popmodel} for a detailed discussion.}
\end{deluxetable*}

In the previous step, we quantified the agreement between explosion model light curves and the \salt model, indicating whether they represent a Type Ia-like explosion. However, there are three free parameters in the \salt fit, $x_1$, $c$ and $\mb$. Even if the light curves in question are well fit with the \salt model, there is no guarantee that the resulting parameters correspond to values observed in nature. Namely, the explosion model could suffer from physically unreasonable color and stretch properties, or not obey the magnitude--stretch and magnitude--color relations. Furthermore, we wish to evaluate the likelihood that a certain set of fit parameters could be observed in nature. Hence, we need a model for the population of \sne in the $\mb$--$x_1$--$c$ parameter space. In the context of \salt, $x_1$ and $c$ are independent parameters, whereas the peak magnitude is correlated with $x_1$ and $c$. Thus, generating a population model consists of two somewhat independent steps: quantifying the observed distributions of stretch and color, and quantifying the relations between stretch, color and magnitude.

The observed distributions of $x_1$ and $c$ have been quantified for various surveys, but surveys are subject to significant observational biases. Here we are interested in the {\it underlying} distribution of $x_1$ and $c$, corresponding to the results of an ideal, unbiased survey. These unbiased distributions were modeled by \citet[][hereafter K13]{kessler_13_var}, who determined the distributions by comparing Monte-Carlo simulations with survey data. The Monte-Carlo simulations, run using \snana, assumed a certain parent population in stretch and color, and forward-modeled various observational biases. These effects included an improved signal-to-noise model and the corresponding cuts, Malmquist bias, search efficiency, as well as the efficiency of the spectroscopic selection. The most likely underlying population was determined by matching the results of the simulation with the observed distributions of 251 \sne from the SDSS (Appendix \ref{sec:app:data:sdss}) and 191 \sne from the Supernova Legacy Survey \citep[SNLS;][G10]{astier_06_snls}. The underlying populations were modeled as asymmetric, continuous Gaussians such that for a parameter $\theta = (x_1, c)$,
\begin{equation}
\frac{dP(\theta)}{d\theta}  = \sqrt{\frac{2}{\pi}} \frac{1}{(\sigma_{\theta+} + \sigma_{\theta-})} \left\{
  \begin{array}{l l l}
    \mathrm{exp}\left(\frac{(\theta - \mu_{\theta})^2}{2 {\sigma_{\theta+}}^2} \right) & \forall & \theta > \mu_{\theta} \\
    \mathrm{exp}\left(\frac{(\theta - \mu_{\theta})^2}{2 {\sigma_{\theta-}}^2} \right) & \forall & \theta < \mu_{\theta} \,.\\
  \end{array} \right.
\end{equation}
Table \ref{table:pop_model} shows the values for $\mu_{\theta}$ and the various standard deviations used, as determined by K13. 

Having quantified the distributions of stretch and color, we now move on to their correlations with magnitude. In Section \ref{sec:method:salt} we described how the \salt model deals with the relations between $x_1$, $c$, $\mb$, and $\Mb$ using the parameters $M_0$, $\alpha$ and $\beta$ which are derived from a Hubble diagram fit. Since explosion models are not guaranteed to follow the relations between stretch, color and magnitude observed in nature, we do not use a set of explosion models to derive $M_0$, $\alpha$ and $\beta$. Instead, we pick sensible values derived from large \sna surveys (Table \ref{table:pop_model}). We discuss these choices in detail in Section \ref{sec:disc:popmodel}.

Equation (\ref{equ:salt2_mu}) suggests that the fitted $\mb$ (and thus $\Mb$) is perfectly correlated with $x_1$ and $c$, and assumes no residual scatter in the Hubble diagram which is unrealistic. Thus, we add a scatter term, $\delM$ to the absolute magnitude, giving the final relation
\begin{equation}
\label{equ:salt2_MB}
\Mb = M_0 - \alpha x_1 + \beta c + \delM
\end{equation}
where $\delM$ is assumed to be drawn from a symmetric Gaussian distribution with mean zero and standard deviation $\sigma_M$. The standard deviation includes both the intrinsic scatter as well as the uncertainty of the Hubble constant, which is a sub-dominant but relevant contribution. We are now in a position to write down the probability density of \sne in the 3D space of stretch, color and absolute magnitude. The probability distribution is the product of three independent Gaussians in $x_1$, $c$ and $\delM$, and can easily be transformed from the magnitude scatter $\delM$ to absolute magnitude,
\begin{equation}
\label{equ:popmodel_basic}
dP(x_1, c, \Mb) \varpropto e^{\frac{1}{2} \chi^2} d x_1\, dc\, d\Mb
\end{equation}
with
\begin{eqnarray}
\chi^2 &=& \frac{(x_1 - \mu_{\rm x1})^2}{\sigma_{\rm x_1}^2} + \frac{(c-\mu_{\rm c})^2}{\sigma_{\rm c}^2} + \frac{\delM^2}{\sigma_{\rm M}^2} \nonumber \\
&=& \frac{(x_1 - \mu_{\rm x1})^2}{\sigma_{\rm x_1}^2} + \frac{(c-\mu_{\rm c})^2}{\sigma_{\rm c}^2} + \frac{(\Mb + \alpha x_1 - \beta c - M_0)^2}{\sigma_{\rm M}^2} \nonumber
\end{eqnarray}
where $\sigma_{x_1}$ and $\sigma_{c}$ are understood to take on different values above and below the respective means of their distributions. We note that all $\sigma$ denote the widths of the population distributions rather than fitted errors. A more formal and general derivation of this result is given in Appendix \ref{sec:app:popmodel}. 

Figure \ref{fig:pop_model}(a) shows confidence contours for the likelihood of Equation (\ref{equ:popmodel_basic}). Each panel shows the contours in two variables, marginalized over the third variable. Note that in the cases of the $x_1-\Mb$ and $c-\Mb$ planes, the marginalization over asymmetric Gaussian distributions leads to a 2D probability which itself is not Gaussian any more. Expressions for the probability density shown in Figure \ref{fig:pop_model} are given in Appendix \ref{sec:app:popmodel}. 

To check that our population model differs from biased survey data in ways which we expect and can explain, we overplot data from the SDSS survey in Figure \ref{fig:pop_model}(a). We review the SDSS survey, the cuts applied to the sample, and our procedure for estimating the rest-frame peak magnitude in Appendix \ref{sec:app:data}. The differences between the overplotted SDSS data and the population model clearly demonstrate why simply comparing the fit results from explosion models to those of survey data would be inappropriate. The SDSS data is strongly biased toward blue \sne (low $c$) and slow decliners (high $x_1$), which are synonymous with bright events (Malmquist bias). All outliers are compatible with the 3$\sigma$ confidence contours, at least within their $1\sigma$ error bars. 

However, there is one potential issue with the population model derived from SDSS and SNLS data. From the survey data used to determine the $x_1$ and $c$ distributions (see Figures 1 and 2 in K13), one might suspect that the dim ends (red, fast decliners) of these distributions are poorly constrained. In order to investigate these regions of parameter space, we turn to a low-redshift survey, the CfA3 sample of nearby \sne (reviewed in Appendix \ref{sec:app:data:cfa3}). The CfA3 sample was not considered in the original population model since its survey biases could not be modeled as accurately as those of SDSS and SNLS, but it contains a higher fraction of red, dim events. Thus, the CfA3 survey should be ideal for the purpose of finding outliers in the \sna population. In addition, we use the well-observed set of \sne used in \citet{jha_07_mlcs2k2}, and refer to the combined sample as the nearby sample. Figure \ref{fig:pop_model}(b) shows the nearby sample plotted over the contours of our population model. The comparison demonstrates that the nearby sample contains some objects which lie significantly outside the $x_1-c$ range expected from the population model of K13 (with a sample of this size, there should be zero or one objects outside the 99.7\% contour). In particular, there appears to exist a very red population not represented by the population model. Secondly, the $x_1-\Mb$ plot suggests that there are objects with both brighter and dimmer magnitudes than expected. The cuts described in Appendix \ref{sec:app:data:cuts} ensure that the extreme fit results are not artifacts of poor light curve fits. However, we cannot exclude the possibility that at least a fraction of the red, dim \sne experienced significant host galaxy reddening. Nevertheless, as long as there is a possibility that the red colors and dim magnitudes are intrinsic, the population model should not exclude explosion models with similar characteristics.

\begin{figure*}[p]
\centering
\includegraphics[trim = 0mm 16mm 3mm 3mm, clip, scale=0.7]{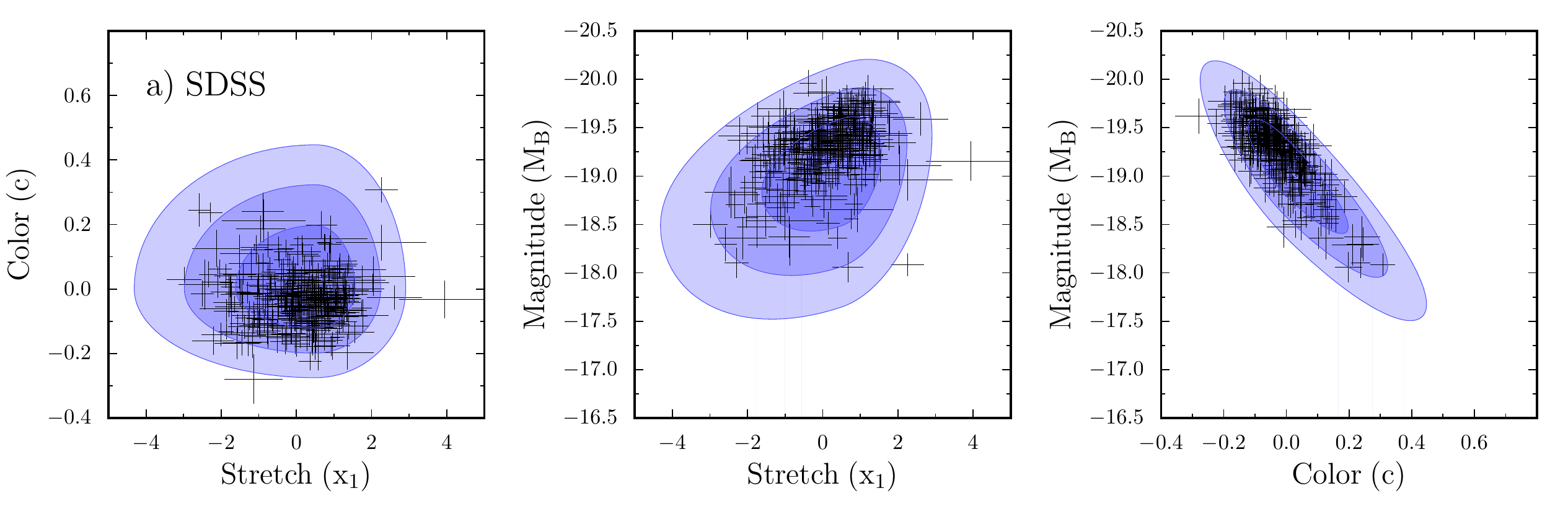}
\includegraphics[trim = 0mm 16mm 3mm 3mm, clip, scale=0.7]{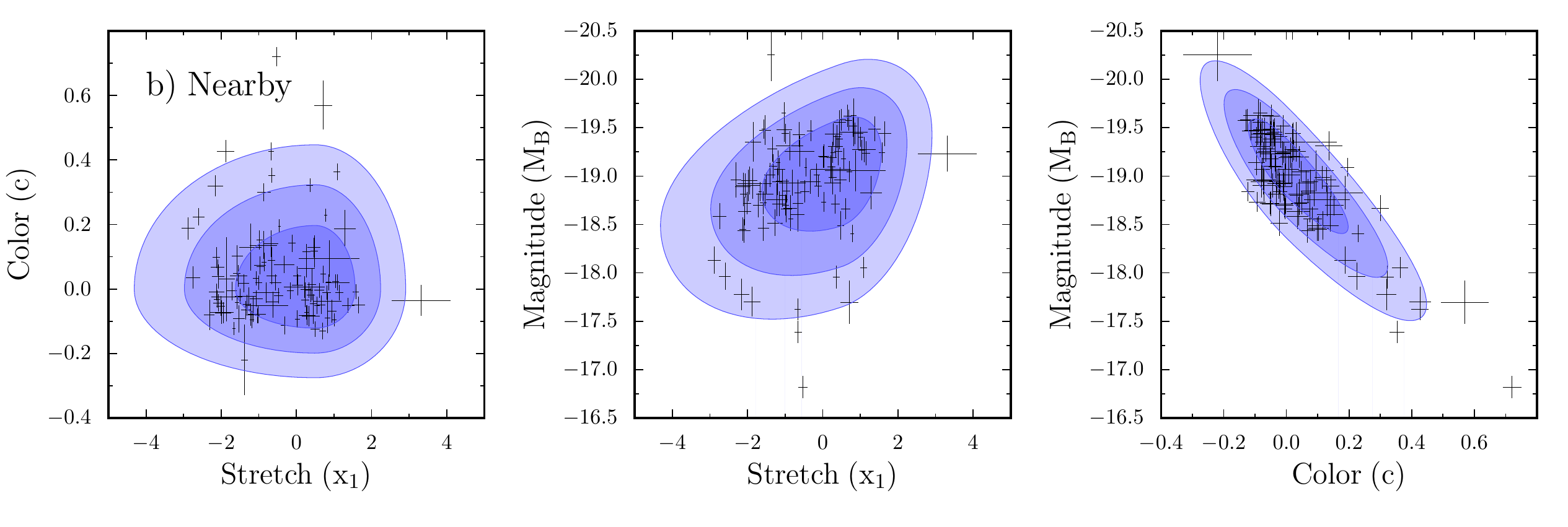}
\includegraphics[trim = 0mm 4mm 3mm 3mm, clip, scale=0.7]{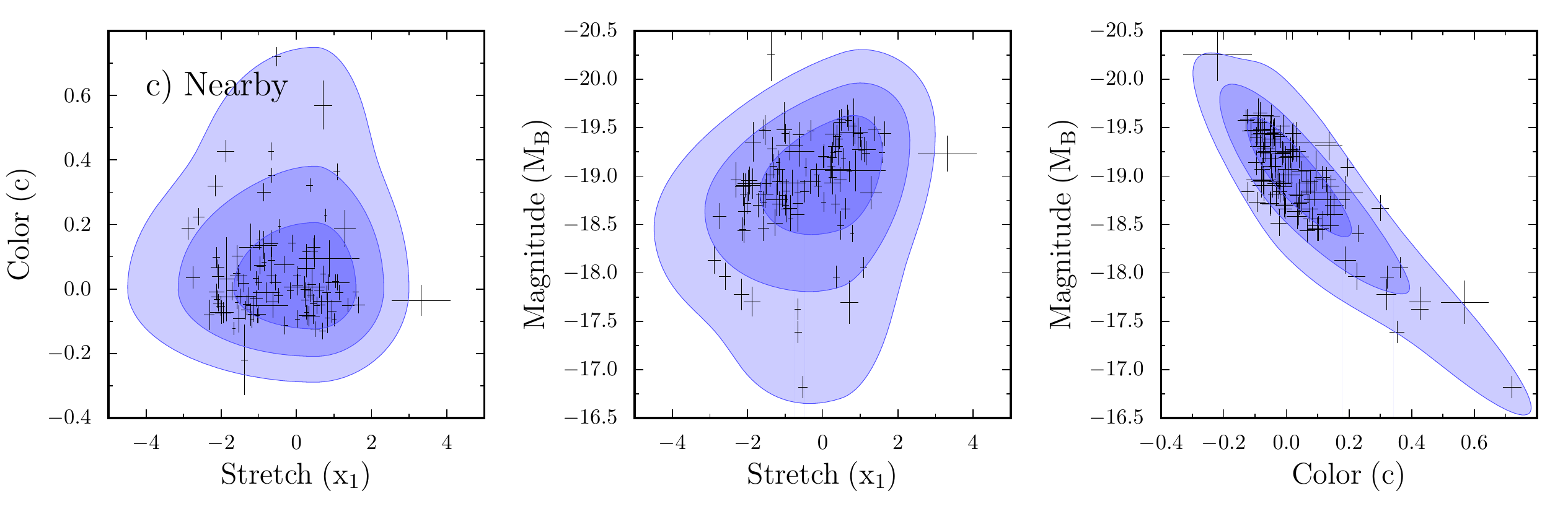}
\caption{Constructing a population model. Each row shows confidence contours (68.3\%, 95.4\% and 99.7\%) in $x_1-c$, $x_1-\Mb$ and $c-\Mb$ space, marginalized over the third variable which is not plotted. Row (a): the population model of \citet{kessler_13_var}, with survey data from SDSS. The comparison demonstrates the necessity of a population model, since the survey data are highly biased toward blue (low $c$), slow-declining (high $x_1$) events. Row (b): the nearby sample contains outliers which should not exist according to the population model. Row (c): the extended population model does a better job accounting for the nearby outliers (see Section \ref{sec:method:popmodel} for details).}
\label{fig:pop_model}
\end{figure*}

\begin{figure*}[p]
\centering
\includegraphics[trim = 0mm 4mm 3mm 3mm, clip, scale=0.7]{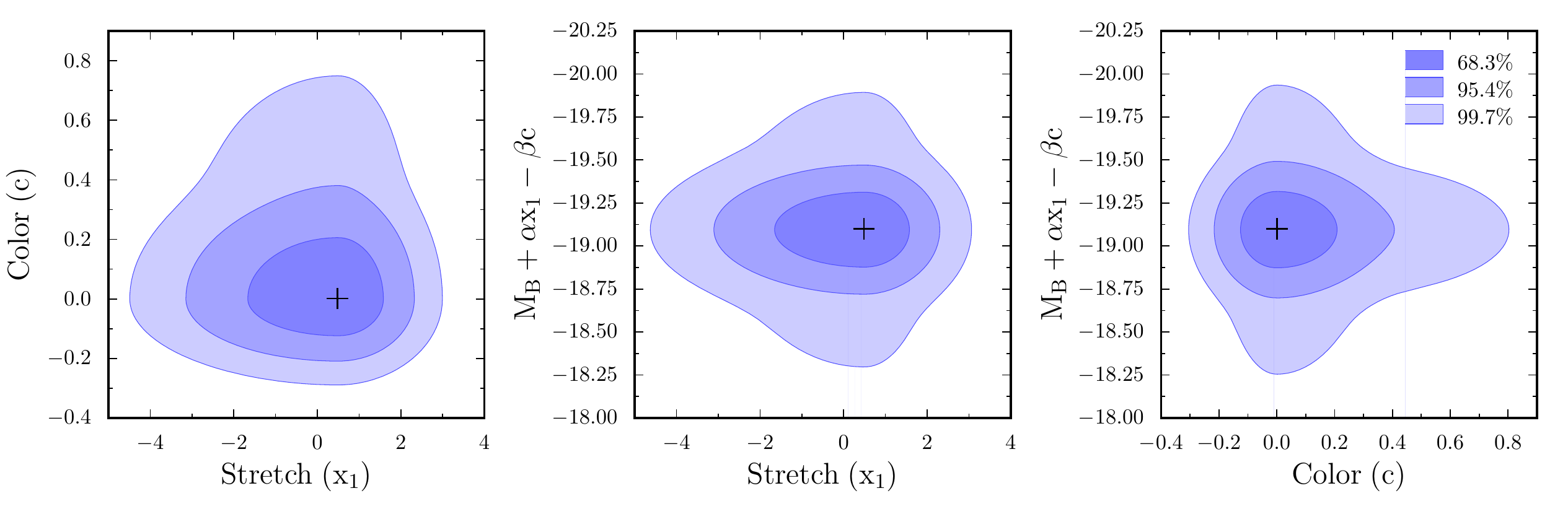}
\caption{The extended population model as in Figure \ref{fig:pop_model}(c), but plotting the corrected magnitude instead of $\Mb$. In corrected magnitude space, the basic population model consists of three independent Gaussians, and it is easier to visually evaluate how likely an event is according to the model. The black crosses mark the loci of maximum probability density.}
\label{fig:pop_model2}
\end{figure*}

To account for the tails in the \sna population, we extend the population model to include a second Gaussian, making up 4\% of the overall population, in both $c$ and $\Mb$. The means and standard deviations for the secondary populations are listed in Table \ref{table:pop_model}. The second Gaussian in color is centered at $c=0.3$, with $\sigma=0.2$, adding a total 1.3\% likelihood for a \sna to be observed with a value of $c$ greater than the previous $3 \sigma$ limit of $c=0.39$. Since the second Gaussian is centered at red colors, it adds virtually no probability where $c < 0$. The second Gaussian in $\Mb$ is centered at the original value of $M_0$, and thus simply allows for an outlier population at low and high magnitudes, adding a total of 0.5\% likelihood each above and below the previous $3 \sigma$ limits. Contrary to color and magnitude, Figure \ref{fig:pop_model}(b) shows no evidence for an extreme population in $x_1$. Figure \ref{fig:pop_model}(c) demonstrates that the extended population model does a better job of accounting for the extreme objects of the nearby sample. Figure \ref{fig:pop_model2} shows the same extended population model, for the corrected magnitude, $\Mb + \alpha x_1 - \beta c$, instead of $\Mb$. The advantage of this variable transformation is that the Gaussians are independent in all three variables, and it is thus easier to see which parts of the 3D parameter space are allowed. The plots of $\Mb$ can be somewhat misleading, since they show projected confidence contours, not the 3D likelihood. A more detailed mathematical description of the population model is given in Appendix \ref{sec:app:popmodel}.

Furthermore, we note that the contours in the $x_1-\Mb$ plane of Figure \ref{fig:pop_model}(c) indicate a less tight correlation between stretch and peak magnitude than the Phillips relation between \dm15 and $\Mb$ (see, e.g., Figure 1 of \citet{hillebrandt_13_review} or Figure 7 of \citet{blondin_11_comparison}). There are a few reasons for the weaker correlation. First, tight Phillips relation fits are often derived from {\it color-corrected} peak magnitudes \citep[e.g.,][]{folatelli_10_csp}, while Figure \ref{fig:pop_model}(c) shows uncorrected magnitudes. Removing the component of the scatter in magnitude which is correlated with color reduces the scatter in the stretch--magnitude relation significantly. Secondly, our extended population model accounts for dim objects which have been shown to lie off the main Phillips relation \citep{foley_13_iax}. Finally, the stretch parameter $x_1$ is derived from an entire set of light curves, rather than two particular epochs in $B$-band, and there is no reason to assume that its correlation with peak magnitude would be as tight as that of \dm15.

While the extended population model seems to do well at describing the nearby sample, we emphasize that the purpose of the extension is {\it not} to make accurate statements about the actual, underlying population of \sne, since it was not derived from data in a rigorous fashion. Given the scarcity of current data, such an investigation will be a future project (see Section \ref{sec:disc:popmodel}). Here, we simply make sure to include explosion models with extreme fitted parameters if there are {\it any} reasonably well observed events with similar parameters. Adding secondary populations with a few percent of the overall population changes the likelihood closer to the peak of the distribution very little. In the context of comparing explosion models to the population model, the extended model only differs from the basic model when the fit results are somewhat extreme. The validity of our population model is further discussed in Section \ref{sec:disc:popmodel}.

\subsection{Figure of Merit}
\label{sec:method:fom}

In the last two sections we described procedures for fitting explosion models with \salt, and derived a goodness-of-fit indicator $\xi^2$ as well as a likelihood of the fit results corresponding to values observed in nature. We now wish to merge these two components into one figure of merit, $f$, which represents a weighted sum of its two constituents. Of course, there are countless ways to define $f$, meaning any procedure is somewhat arbitrary and can be changed for any particular purpose. Here, we aim to assign equal weight to both the goodness-of-fit and the likelihood of fit parameters. 

For a given set of fitted parameters, we define $p$ to be the fraction of the population with a {\it lower} likelihood of being observed in nature than the set of fit parameters in question. This definition means that $p=1$ corresponds to $x_1$, $c$ and $\Mb$ at the point of highest likelihood, and, for example, $p=0.01$ indicates than only 1\% of observed \sne have parameter values more extreme than those at hand. It makes sense to put a lower limit on the figure of merit for $p$ because the population model is not defined at very low likelihoods, and the computation can return 0 due to round-off error. Furthermore, we re-normalize the figure of merit to take on values between 0 (worst) and 1 (best) to simplify its interpretation. We thus define
\begin{equation}
\fp \equiv \left\{
  \begin{array}{l l l}
     \frac{\log_{10}(p_{\rm min}) - \log_{10}(p)}{\log_{10}(p_{\rm min})}  & \forall & p \geq p_{\rm min} \\
     0 & \forall & p < p_{\rm min}  \,.\\
  \end{array} \right.
\end{equation}
Similarly, we need to define a $\xi^2_{\rm max}$ above which we consider a light curve fit to have failed. Incidentally, some light curves fit so poorly that the fit fails within \snana, and we penalize such failed fits by assigning them $\xi^2 = \xi^2_{\rm max}$. In either case, failed fits are not included when computing $\fp$, since they contain no information about the best-fit parameters. Furthermore, we assign $\xi^2_{\rm max}$ to certain {\it epochs} (not entire fits) if both the flux and uncertainty of the best-fit \salt model go to zero. In this rare case, $\xi^2$ for the epoch is not well-defined (see Equation (\ref{eq:xi2})).

While $p$ is automatically limited to $p \leq 1$, there is no clear lower limit for $\xi^2$. A $\xi^2$ below 1 corresponds to a set of light curves which, on average, lie within the statistical uncertainty of the \salt model. Such an excellent match is achieved by a few individual fits to the explosion models considered here, but no model achieves $\xi^2<1$ when averaged over the 12 sets of fits with different filters and epoch ranges. Nevertheless, we wish to ensure that $\fxi$ has both lower and upper bounds, and thus define
\begin{equation}
\fxi \equiv \left\{
  \begin{array}{l l l}
     1 & \forall & \xi^2 < 1 \\
     \frac{\log_{10}(\xi^2_{\rm max})-\log_{10}(\xi^2)}{\log_{10}(\xi^2_{\rm max})}  & \forall & 1 \leq \xi^2 \leq \xi^2_{\rm max} \\
     0 & \forall & \xi^2 > \xi^2_{\rm max}  \,.\\
  \end{array} \right.
\end{equation}
All the information about the quality of the light curve fits and the fitted parameters is contained in $\fxi$ and $\fp$. In order to rank explosion models by their quality, we wish to combine these two figures of merit into an overall figure of merit, $f$, which we define to be a weighted sum of $\fp$ and $\fxi$, with weights $w_{\rm p}$ and $w_{\xi^2}$. As both $\fp$ and $\fxi$ are logarithms of $p$ and $\xi^2$, summing them is equivalent to multiplying $p$ and $\xi^2$ with the weights as exponents. The weights can be set arbitrarily to emphasize the quality of either light curve fits or fit results. Here, we wish to weight them equally, meaning $\fp$ and $\fxi$ should contribute roughly similar values to the overall figure of merit. For the models considered in this analysis, we find that this is roughly satisfied for the simple average of $\fp$ and $\fxi$, $w_p=w_{\xi^2}=1$, leaving
\begin{equation}
f \equiv \frac{w_{\rm p} \fp + w_{\xi^2} \fxi}{w_{\rm p} + w_{\xi^2}} = \frac{\fp + \fxi}{2}\,.
\end{equation}
We are left with the task of choosing sensible values of $p_{\rm min}$ and $\xi^2_{\rm max}$ such that we do not unnecessarily exclude regions of parameter space, but cut off regions where the fits and population model are not informative. For $\fp$, we conservatively choose a limit of $p_{\rm min}=10^{-4}$ which corresponds to an explosion model lying almost $4\sigma$ off our population model. At such low levels of likelihood, our model is not constrained, and the explosion model has been ruled out anyway. For $\fxi$, a limit of $\xi^2_{\rm max} = 25$ seems sensible, corresponding to the light curves deviating $5\sigma$ from the \salt model on average.

We emphasize once again that $f$ should not be assigned any statistical meaning, since $\fxi$ has no statistical meaning; $f$ merely serves as a convenient measure of light curve fit quality. Since each set of light curves is fit multiple times with different filter bands and epoch ranges, we average the figures of merit of the individual fits, and denote the averaged quantities $\fxiav$, $\fpav$ and $\fav$. When error bars on these quantities are given, they refer to the root mean squared (rms) deviation from the mean figure of merit. Note that this procedure does not imply averaging over viewing angles, which for now we treat as individual \sna events. Where the figures of merit are also averaged over viewing angle, we denote them $\fxiavav$, $\fpavav$ and $\favav$.

When discussing the fitted parameters for particular explosion models (or a particular viewing angle of a model), it would be cumbersome to refer to each of the 12 sets of light curve fits individually. Instead, we wish to compute the most likely $x_1$, $c$ and $\Mb$ for the explosion model. Thus, we define averages over all fits, $\avg{x_1}$, $\avg{c}$ and $\avg{\Mb}$. We note that these quantities are not used to compute figures of merit, but are used for plotting the fit results on the contour plots of the population model. Since fits with a better goodness-of-fit parameter $\xi^2$ are bound to give more accurate estimates of the fitted parameters, we define the averaged parameters as a weighted average, 
\begin{equation}
\avg{x_1}=\sum_i \frac{x_{1,i}}{\xi^2_i} \left/ \sum_i \frac{1}{\xi^2_i} \right.
\end{equation}
where $i$ runs over all 12 fits except those which failed (meaning they have $\xi^2>\xi^2_{\rm max}$, or the \snana fit failed to converge in the first place). $\avg{c}$ and $\avg{\Mb}$ are defined similarly. Furthermore, we compute the rms deviation from the mean parameters separately for values above and below the mean, respectively. The rms deviations are {\it not} weighted by $\xi^2$, and serve as the error bars shown in Section \ref{sec:results}. We emphasize that these error bars do not represent 68\% intervals of the likelihood of obtaining a certain parameter, or any other quantity which can be interpreted statistically, but merely give an estimate of the dispersion of a fitted parameter between fits. When the error bars are large, the large rms deviation is usually due to one or a few failed fits which return unreasonable results. Since the mean parameters, $\avg{x_1}$, $\avg{c}$ and $\avg{\Mb}$, are weighted by $\xi^2$, they are much more reliable in those cases than the error bars indicate. Furthermore, we note that we do not compute fit results averaged over viewing angle (such as $\avg{\avg{x_1}}$), but instead plot separate fit results for each viewing angle, giving an impression of the variations of fit results with viewing angle.


\section{Results} 
\label{sec:results}

\begin{deluxetable*}{llllllllllllllll}
\tablecaption{Figures of Merit for All Analyzed Explosion Models
\label{table:results}}
\tablewidth{0pt}
\tablehead{
\colhead{Explosion Model} &
\colhead{$f^{\,\rm a}$} &
\colhead{$\sigma_{\theta}^{\,\rm b}$} &
\colhead{$\theta_{\rm min}$} &
\colhead{$\theta_{\rm max}$} &
\colhead{$\sigma_{\rm fr}^{\,\rm c}$} &
\colhead{$\fxi$} &
\colhead{$\sigma_{\theta}$} &
\colhead{$\theta_{\rm min}$} &
\colhead{$\theta_{\rm max}$} &
\colhead{$\sigma_{\rm fr}$} &
\colhead{$\fp$} &
\colhead{$\sigma_{\theta}$} &
\colhead{$\theta_{\rm min}$} &
\colhead{$\theta_{\rm max}$} &
\colhead{$\sigma_{\rm fr}$}
}
\startdata
KRW09\_iso\_5\_dc\_1 & $\mathbf{0.909}$ & $0.009$ & $0.885$ & $0.924$ & $0.078$ & $\mathbf{0.93}$ & $0.02$ & $0.88$ & $0.95$ & $0.07$ & $\mathbf{0.89}$ & $0.01$ & $0.87$ & $0.90$ & $0.04$ \\
KRW09\_asym\_1\_dc\_3 & $\mathbf{0.891}$ & $0.031$ & $0.821$ & $0.925$ & $0.100$ & $\mathbf{0.86}$ & $0.06$ & $0.73$ & $0.94$ & $0.09$ & $\mathbf{0.92}$ & $0.03$ & $0.83$ & $0.96$ & $0.04$ \\
KRW09\_iso\_8\_dc\_1 & $\mathbf{0.890}$ & $0.029$ & $0.808$ & $0.919$ & $0.119$ & $\mathbf{0.89}$ & $0.04$ & $0.80$ & $0.93$ & $0.11$ & $\mathbf{0.89}$ & $0.02$ & $0.81$ & $0.93$ & $0.04$ \\
KRW09\_asym\_1\_dc\_2 & $\mathbf{0.884}$ & $0.029$ & $0.797$ & $0.914$ & $0.106$ & $\mathbf{0.85}$ & $0.04$ & $0.74$ & $0.90$ & $0.10$ & $\mathbf{0.92}$ & $0.04$ & $0.82$ & $0.97$ & $0.03$ \\
KRW09\_iso\_3\_dc\_1 & $\mathbf{0.873}$ & $0.014$ & $0.833$ & $0.889$ & $0.105$ & $\mathbf{0.89}$ & $0.03$ & $0.82$ & $0.92$ & $0.10$ & $\mathbf{0.86}$ & $0.00$ & $0.85$ & $0.86$ & $0.04$ \\
KRW09\_iso\_4\_dc\_1 & $\mathbf{0.869}$ & $0.029$ & $0.816$ & $0.909$ & $0.103$ & $\mathbf{0.88}$ & $0.03$ & $0.81$ & $0.92$ & $0.09$ & $\mathbf{0.86}$ & $0.03$ & $0.79$ & $0.90$ & $0.04$ \\
KRW09\_iso\_3\_dc\_2 & $\mathbf{0.839}$ & $0.022$ & $0.813$ & $0.887$ & $0.087$ & $\mathbf{0.71}$ & $0.04$ & $0.66$ & $0.81$ & $0.09$ & $\mathbf{0.97}$ & $0.00$ & $0.96$ & $0.97$ & $0.02$ \\
KRW09\_asym\_2\_dc\_2 & $\mathbf{0.833}$ & $0.023$ & $0.790$ & $0.873$ & $0.080$ & $\mathbf{0.72}$ & $0.05$ & $0.64$ & $0.80$ & $0.07$ & $\mathbf{0.95}$ & $0.01$ & $0.93$ & $0.97$ & $0.03$ \\
KRW09\_iso\_8\_dc\_2 & $\mathbf{0.831}$ & $0.040$ & $0.701$ & $0.864$ & $0.087$ & $\mathbf{0.68}$ & $0.08$ & $0.44$ & $0.74$ & $0.09$ & $\mathbf{0.99}$ & $0.01$ & $0.96$ & $0.99$ & $0.01$ \\
KRW09\_iso\_5\_dc\_2 & $\mathbf{0.828}$ & $0.015$ & $0.791$ & $0.846$ & $0.064$ & $\mathbf{0.67}$ & $0.03$ & $0.59$ & $0.70$ & $0.06$ & $\mathbf{0.99}$ & $0.00$ & $0.98$ & $0.99$ & $0.01$ \\
KRW09\_iso\_6\_dc\_1 & $\mathbf{0.820}$ & $0.034$ & $0.690$ & $0.846$ & $0.127$ & $\mathbf{0.87}$ & $0.06$ & $0.61$ & $0.91$ & $0.10$ & $\mathbf{0.77}$ & $0.01$ & $0.74$ & $0.79$ & $0.07$ \\
KRW09\_iso\_2\_dc\_3 & $\mathbf{0.819}$ & $0.055$ & $0.695$ & $0.873$ & $0.130$ & $\mathbf{0.83}$ & $0.06$ & $0.69$ & $0.89$ & $0.10$ & $\mathbf{0.81}$ & $0.05$ & $0.70$ & $0.86$ & $0.08$ \\
KRW09\_iso\_4\_dc\_2 & $\mathbf{0.817}$ & $0.024$ & $0.784$ & $0.862$ & $0.087$ & $\mathbf{0.68}$ & $0.04$ & $0.62$ & $0.75$ & $0.08$ & $\mathbf{0.96}$ & $0.03$ & $0.87$ & $0.99$ & $0.02$ \\
KRW09\_iso\_6\_dc\_2 & $\mathbf{0.816}$ & $0.055$ & $0.683$ & $0.874$ & $0.117$ & $\mathbf{0.76}$ & $0.08$ & $0.57$ & $0.85$ & $0.09$ & $\mathbf{0.87}$ & $0.03$ & $0.79$ & $0.91$ & $0.07$ \\
KRW09\_iso\_2\_dc\_2 & $\mathbf{0.799}$ & $0.036$ & $0.709$ & $0.835$ & $0.137$ & $\mathbf{0.85}$ & $0.04$ & $0.76$ & $0.89$ & $0.11$ & $\mathbf{0.75}$ & $0.04$ & $0.65$ & $0.78$ & $0.08$ \\
KRW09\_asym\_5\_dc\_3 & $\mathbf{0.786}$ & $0.026$ & $0.735$ & $0.824$ & $0.140$ & $\mathbf{0.82}$ & $0.04$ & $0.74$ & $0.88$ & $0.10$ & $\mathbf{0.75}$ & $0.01$ & $0.73$ & $0.77$ & $0.10$ \\
KRW09\_iso\_7\_dc\_2 & $\mathbf{0.786}$ & $0.019$ & $0.756$ & $0.824$ & $0.063$ & $\mathbf{0.58}$ & $0.04$ & $0.52$ & $0.65$ & $0.06$ & $\mathbf{0.99}$ & $0.01$ & $0.96$ & $1.00$ & $0.01$ \\
KRW09\_asym\_4\_dc\_3 & $\mathbf{0.785}$ & $0.017$ & $0.736$ & $0.808$ & $0.127$ & $\mathbf{0.87}$ & $0.03$ & $0.76$ & $0.91$ & $0.10$ & $\mathbf{0.70}$ & $0.04$ & $0.63$ & $0.78$ & $0.08$ \\
KRW09\_asym\_5\_dc\_2 & $\mathbf{0.785}$ & $0.018$ & $0.743$ & $0.810$ & $0.138$ & $\mathbf{0.85}$ & $0.03$ & $0.79$ & $0.89$ & $0.11$ & $\mathbf{0.72}$ & $0.01$ & $0.68$ & $0.74$ & $0.09$ \\
KRW09\_iso\_1\_dc\_5 & $\mathbf{0.785}$ & $0.016$ & $0.751$ & $0.809$ & $0.166$ & $\mathbf{0.79}$ & $0.03$ & $0.73$ & $0.83$ & $0.12$ & $\mathbf{0.78}$ & $0.01$ & $0.75$ & $0.79$ & $0.12$ \\
KRW09\_iso\_1\_dc\_3 & $\mathbf{0.784}$ & $0.010$ & $0.762$ & $0.799$ & $0.150$ & $\mathbf{0.82}$ & $0.02$ & $0.78$ & $0.85$ & $0.11$ & $\mathbf{0.75}$ & $0.01$ & $0.73$ & $0.76$ & $0.10$ \\
KRW09\_iso\_1\_dc\_4 & $\mathbf{0.784}$ & $0.010$ & $0.762$ & $0.799$ & $0.150$ & $\mathbf{0.82}$ & $0.02$ & $0.78$ & $0.85$ & $0.11$ & $\mathbf{0.75}$ & $0.01$ & $0.73$ & $0.76$ & $0.10$ \\
KRW09\_iso\_1\_dc\_2 & $\mathbf{0.783}$ & $0.007$ & $0.772$ & $0.795$ & $0.147$ & $\mathbf{0.84}$ & $0.02$ & $0.81$ & $0.87$ & $0.12$ & $\mathbf{0.72}$ & $0.01$ & $0.70$ & $0.74$ & $0.09$ \\
KRW09\_asym\_4\_dc\_2 & $\mathbf{0.773}$ & $0.021$ & $0.688$ & $0.794$ & $0.138$ & $\mathbf{0.86}$ & $0.05$ & $0.73$ & $0.91$ & $0.11$ & $\mathbf{0.68}$ & $0.04$ & $0.63$ & $0.76$ & $0.08$ \\
KRW09\_iso\_6\_dc\_3 & $\mathbf{0.770}$ & $0.047$ & $0.667$ & $0.824$ & $0.104$ & $\mathbf{0.62}$ & $0.06$ & $0.47$ & $0.69$ & $0.08$ & $\mathbf{0.92}$ & $0.03$ & $0.87$ & $0.96$ & $0.06$ \\
KRW09\_iso\_2\_dc\_5 & $\mathbf{0.764}$ & $0.045$ & $0.664$ & $0.813$ & $0.102$ & $\mathbf{0.63}$ & $0.05$ & $0.50$ & $0.69$ & $0.08$ & $\mathbf{0.90}$ & $0.04$ & $0.77$ & $0.94$ & $0.06$ \\
KRW09\_asym\_3\_dc\_3 & $\mathbf{0.764}$ & $0.022$ & $0.716$ & $0.789$ & $0.132$ & $\mathbf{0.83}$ & $0.03$ & $0.72$ & $0.86$ & $0.12$ & $\mathbf{0.69}$ & $0.04$ & $0.61$ & $0.75$ & $0.06$ \\
KRW09\_asym\_2\_dc\_3 & $\mathbf{0.757}$ & $0.021$ & $0.722$ & $0.806$ & $0.064$ & $\mathbf{0.53}$ & $0.04$ & $0.45$ & $0.62$ & $0.06$ & $\mathbf{0.99}$ & $0.00$ & $0.98$ & $1.00$ & $0.01$ \\
KRW09\_asym\_3\_dc\_2 & $\mathbf{0.750}$ & $0.018$ & $0.703$ & $0.769$ & $0.139$ & $\mathbf{0.82}$ & $0.02$ & $0.76$ & $0.85$ & $0.12$ & $\mathbf{0.68}$ & $0.04$ & $0.60$ & $0.73$ & $0.06$ \\
KRW09\_iso\_3\_dc\_3 & $\mathbf{0.747}$ & $0.028$ & $0.715$ & $0.803$ & $0.052$ & $\mathbf{0.50}$ & $0.05$ & $0.45$ & $0.61$ & $0.05$ & $\mathbf{0.99}$ & $0.00$ & $0.98$ & $1.00$ & $0.01$ \\
KRW09\_iso\_8\_dc\_3 & $\mathbf{0.743}$ & $0.029$ & $0.670$ & $0.770$ & $0.083$ & $\mathbf{0.52}$ & $0.06$ & $0.37$ & $0.58$ & $0.08$ & $\mathbf{0.96}$ & $0.01$ & $0.92$ & $0.98$ & $0.02$ \\
KRW09\_asym\_7\_dc\_3 & $\mathbf{0.729}$ & $0.027$ & $0.632$ & $0.759$ & $0.160$ & $\mathbf{0.81}$ & $0.04$ & $0.65$ & $0.84$ & $0.14$ & $\mathbf{0.65}$ & $0.02$ & $0.59$ & $0.68$ & $0.08$ \\
KRW09\_iso\_6\_dc\_4 & $\mathbf{0.725}$ & $0.034$ & $0.651$ & $0.764$ & $0.098$ & $\mathbf{0.50}$ & $0.05$ & $0.36$ & $0.55$ & $0.09$ & $\mathbf{0.95}$ & $0.02$ & $0.90$ & $0.98$ & $0.04$ \\
KRW09\_iso\_4\_dc\_3 & $\mathbf{0.718}$ & $0.021$ & $0.687$ & $0.756$ & $0.077$ & $\mathbf{0.47}$ & $0.05$ & $0.41$ & $0.58$ & $0.07$ & $\mathbf{0.97}$ & $0.02$ & $0.87$ & $0.98$ & $0.02$ \\
KRW09\_asym\_7\_dc\_2 & $\mathbf{0.715}$ & $0.040$ & $0.544$ & $0.745$ & $0.175$ & $\mathbf{0.80}$ & $0.06$ & $0.57$ & $0.84$ & $0.16$ & $\mathbf{0.63}$ & $0.03$ & $0.52$ & $0.66$ & $0.08$ \\
KRW09\_iso\_7\_dc\_3 & $\mathbf{0.710}$ & $0.028$ & $0.670$ & $0.775$ & $0.073$ & $\mathbf{0.47}$ & $0.04$ & $0.40$ & $0.57$ & $0.06$ & $\mathbf{0.95}$ & $0.02$ & $0.93$ & $0.98$ & $0.03$ \\
KRW09\_iso\_6\_dc\_5 & $\mathbf{0.710}$ & $0.031$ & $0.646$ & $0.746$ & $0.091$ & $\mathbf{0.45}$ & $0.05$ & $0.32$ & $0.51$ & $0.09$ & $\mathbf{0.97}$ & $0.02$ & $0.93$ & $0.99$ & $0.03$ \\
W7 & $\mathbf{0.700}$ & & & & $0.146$ & $\mathbf{0.60}$ & & & & $0.09$ & $\mathbf{0.80}$ & & & & $0.12$ \\
KRW09\_asym\_8\_dc\_3 & $\mathbf{0.692}$ & $0.021$ & $0.643$ & $0.717$ & $0.172$ & $\mathbf{0.79}$ & $0.02$ & $0.74$ & $0.83$ & $0.15$ & $\mathbf{0.59}$ & $0.04$ & $0.52$ & $0.65$ & $0.09$ \\
KRW09\_iso\_4\_dc\_4 & $\mathbf{0.690}$ & $0.031$ & $0.654$ & $0.747$ & $0.090$ & $\mathbf{0.44}$ & $0.07$ & $0.38$ & $0.63$ & $0.08$ & $\mathbf{0.94}$ & $0.03$ & $0.84$ & $0.96$ & $0.04$ \\
KRW09\_asym\_6\_dc\_3 & $\mathbf{0.686}$ & $0.028$ & $0.621$ & $0.715$ & $0.182$ & $\mathbf{0.77}$ & $0.05$ & $0.61$ & $0.81$ & $0.16$ & $\mathbf{0.60}$ & $0.03$ & $0.54$ & $0.63$ & $0.08$ \\
PreExp60 & $\mathbf{0.685}$ & $0.055$ & $0.604$ & $0.821$ & $0.403$ & $\mathbf{0.63}$ & $0.10$ & $0.41$ & $0.82$ & $0.32$ & $\mathbf{0.74}$ & $0.11$ & $0.57$ & $0.91$ & $0.23$ \\
KRW09\_iso\_5\_dc\_3 & $\mathbf{0.676}$ & $0.026$ & $0.622$ & $0.705$ & $0.064$ & $\mathbf{0.41}$ & $0.04$ & $0.34$ & $0.45$ & $0.05$ & $\mathbf{0.94}$ & $0.02$ & $0.91$ & $0.96$ & $0.04$ \\
KRW09\_asym\_6\_dc\_2 & $\mathbf{0.676}$ & $0.021$ & $0.620$ & $0.700$ & $0.184$ & $\mathbf{0.77}$ & $0.03$ & $0.69$ & $0.80$ & $0.16$ & $\mathbf{0.58}$ & $0.03$ & $0.49$ & $0.61$ & $0.09$ \\
KRW09\_asym\_8\_dc\_2 & $\mathbf{0.668}$ & $0.016$ & $0.629$ & $0.688$ & $0.187$ & $\mathbf{0.79}$ & $0.03$ & $0.72$ & $0.82$ & $0.15$ & $\mathbf{0.55}$ & $0.04$ & $0.47$ & $0.61$ & $0.11$ \\
KRW09\_iso\_5\_dc\_4 & $\mathbf{0.642}$ & $0.027$ & $0.586$ & $0.668$ & $0.090$ & $\mathbf{0.41}$ & $0.04$ & $0.34$ & $0.46$ & $0.06$ & $\mathbf{0.87}$ & $0.02$ & $0.83$ & $0.89$ & $0.06$ \\
PreExp30 & $\mathbf{0.571}$ & $0.092$ & $0.285$ & $0.655$ & $0.222$ & $\mathbf{0.54}$ & $0.11$ & $0.23$ & $0.65$ & $0.20$ & $\mathbf{0.60}$ & $0.08$ & $0.34$ & $0.66$ & $0.09$ \\
PreExp80 & $\mathbf{0.380}$ & $0.101$ & $0.251$ & $0.569$ & $0.333$ & $\mathbf{0.55}$ & $0.07$ & $0.46$ & $0.73$ & $0.31$ & $\mathbf{0.21}$ & $0.17$ & $0.00$ & $0.53$ & $0.08$ \\
PureDef\_0128\_256 & $\mathbf{0.331}$ & & & & $0.042$ & $\mathbf{0.02}$ & & & & $0.04$ & $\mathbf{0.64}$ & & & & $0.01$ \\
PureDef\_0150\_128 & $\mathbf{0.316}$ & & & & $0.187$ & $\mathbf{0.12}$ & & & & $0.12$ & $\mathbf{0.51}$ & & & & $0.14$ \\
PureDef\_0063\_128 & $\mathbf{0.302}$ & & & & $0.174$ & $\mathbf{0.05}$ & & & & $0.07$ & $\mathbf{0.56}$ & & & & $0.16$ \\
PureDef\_1700\_384 & $\mathbf{0.142}$ & & & & $0.338$ & $\mathbf{0.28}$ & & & & $0.34$ & $\mathbf{0.00}$ & & & & $0.00$ \\
PureDef\_1100\_256 & $\mathbf{0.142}$ & & & & $0.347$ & $\mathbf{0.28}$ & & & & $0.35$ & $\mathbf{0.00}$ & & & & $0.00$ \\
PureDef\_3500\_384 & $\mathbf{0.124}$ & & & & $0.299$ & $\mathbf{0.25}$ & & & & $0.30$ & $\mathbf{0.00}$ & & & & $0.00$ \\
\enddata
\tablecomments{The models in this table are sorted by a descending overall figure of merit, $\favav$. Some of the column headers take on slightly different meanings compared to the text, as explained in the comments below. The original KRW09 paper lists 44 models, but accidentally no radiative transfer results were computed for iso\_1\_dc\_4 \citep{blondin_11_comparison}. See Section \ref{sec:results} for a discussion of these results. \\
$^{\rm a}$ The bold columns $f$, $\fxi$ and $\fp$ refer to the viewing angle averaged figures of merit, $\favav$, $\fxiavav$ and $\fpavav$, where multiple viewing angles were present. Otherwise, they refer to the figures of merit averaged over fits, $\fav$, $\fxiav$ and $\fpav$. \\
$^{\rm b}$ The three columns to the right of each figure of merit show the standard deviation in viewing angle, $\sigma_{\theta}$, as well as the minimum and maximum figure of merit in any one viewing angle, denoted $\theta_{\rm min}$ and $\theta_{\rm max}$. For models with only one viewing angle, these columns are blank. \\
$^{\rm c}$ The fourth column to the right of each figure of merit shows the standard deviation over the 12 fits performed, $\sigma_{\rm fr}$. For models with multiple viewing angles, $\sigma_{\rm fr}$ is the average of the standard deviations in each viewing angle. Due to failed fit runs, the overall figure of merit can only be computed for the average of all fits, i.e. $\fav$, and the standard deviation of $\fav$ is the quadratic sum of the standard deviations on $\fpav$ and $\fxiav$. Generally, the standard deviation of the figures of merit over the 12 fits, $\sigma_{\rm fr}$, is large which is expected, particularly for models which exhibit poor light curve fits (see Section \ref{sec:method:fitting}).}
\end{deluxetable*}

\begin{figure}
\centering
\includegraphics[trim = 6mm 8mm 5mm 9mm, clip, scale=0.8]{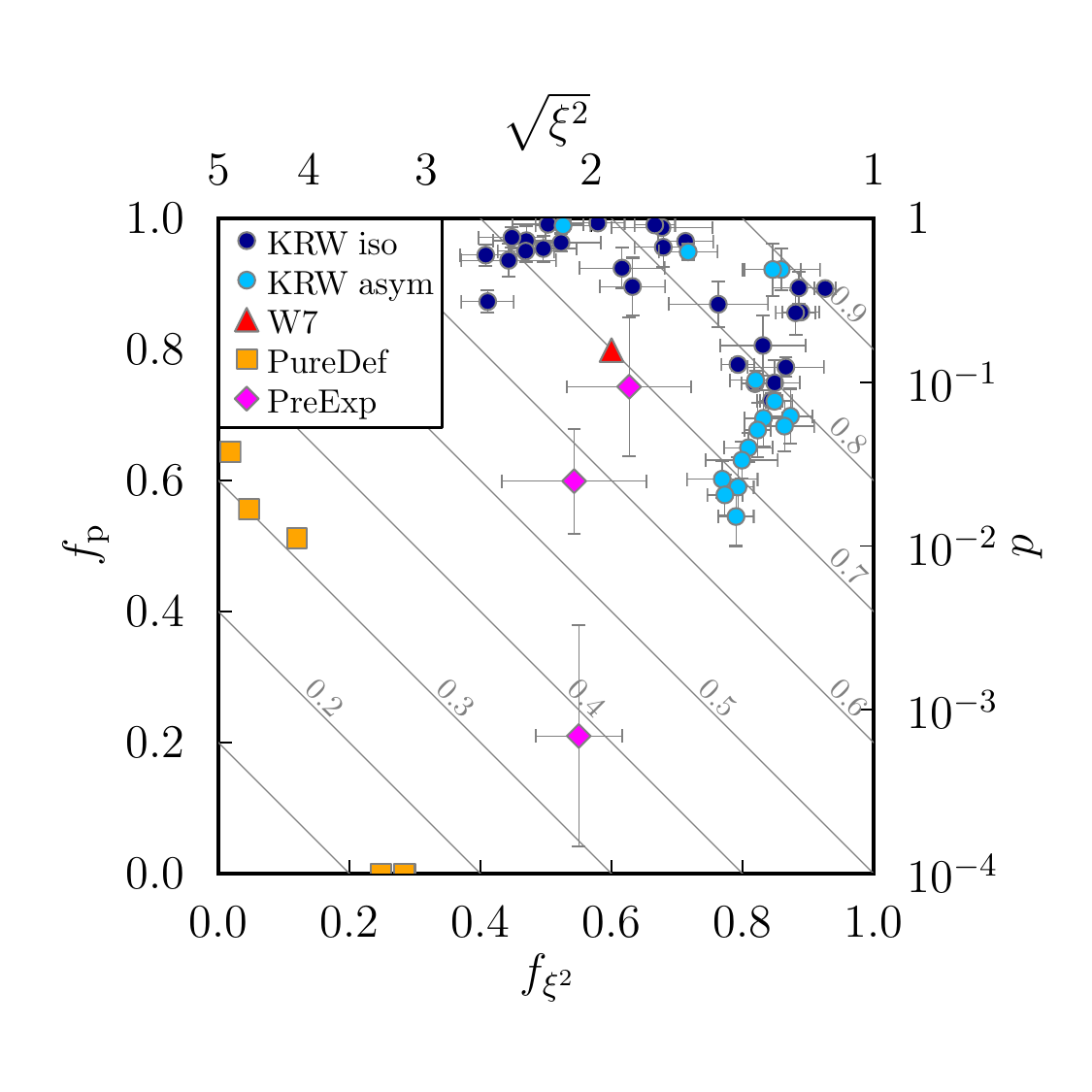}
\caption{The figures of merit from Table \ref{table:results}, with the quality of the light curve fit, $\fxiavav$, increasing to the right, and the likelihood to observe a \sna with the fitted parameters in nature, $\fpavav$, increasing toward the top. The diagonal gray lines and labels show lines of constant overall figure of merit. The error bars show the dispersion of the figures of merit between the viewing angles of a given model. The top axis indicates how many standard deviations the model light curves deviate from the \salt model on average as a function of $\fxi$, while the right axis shows the $p$-values corresponding to $\fp$.}
\label{fig:foms}
\end{figure}

With the machinery developed in the previous section, we are now in a position to compute figures of merit in a fully automated way. Table \ref{table:results} lists the figures of merit for goodness-of-fit, fitted parameters, and the overall figure of merit for all explosion models introduced in Section \ref{sec:method:sims}. Figure \ref{fig:foms} shows the distribution of $\fxiavav$ and $\fpavav$ graphically. The overall figures of merit span a range of $0.124 \leq \favav \leq 0.909$, most of the allowed range of $0 < \favav <1$. The figure of merit for fitted parameters, $\fpavav$, spans almost the entire possible range, $0 \leq \fpavav \leq 0.99$, meaning that the explosion models lie both at the very center of the population and entirely outside it. The figure of merit for light curve fit, $\fxiavav$, exhibits only a slightly smaller range with $0.02 \leq \fxiavav \leq 0.93$. In the following sections we discuss the results for each family of explosion models.

\subsection{The KRW09 Delayed-detonation Models}
\label{sec:results:krw09}

\begin{figure*}
\centering
\includegraphics[trim = 0mm 0mm 0mm 0mm, clip, scale=0.45]{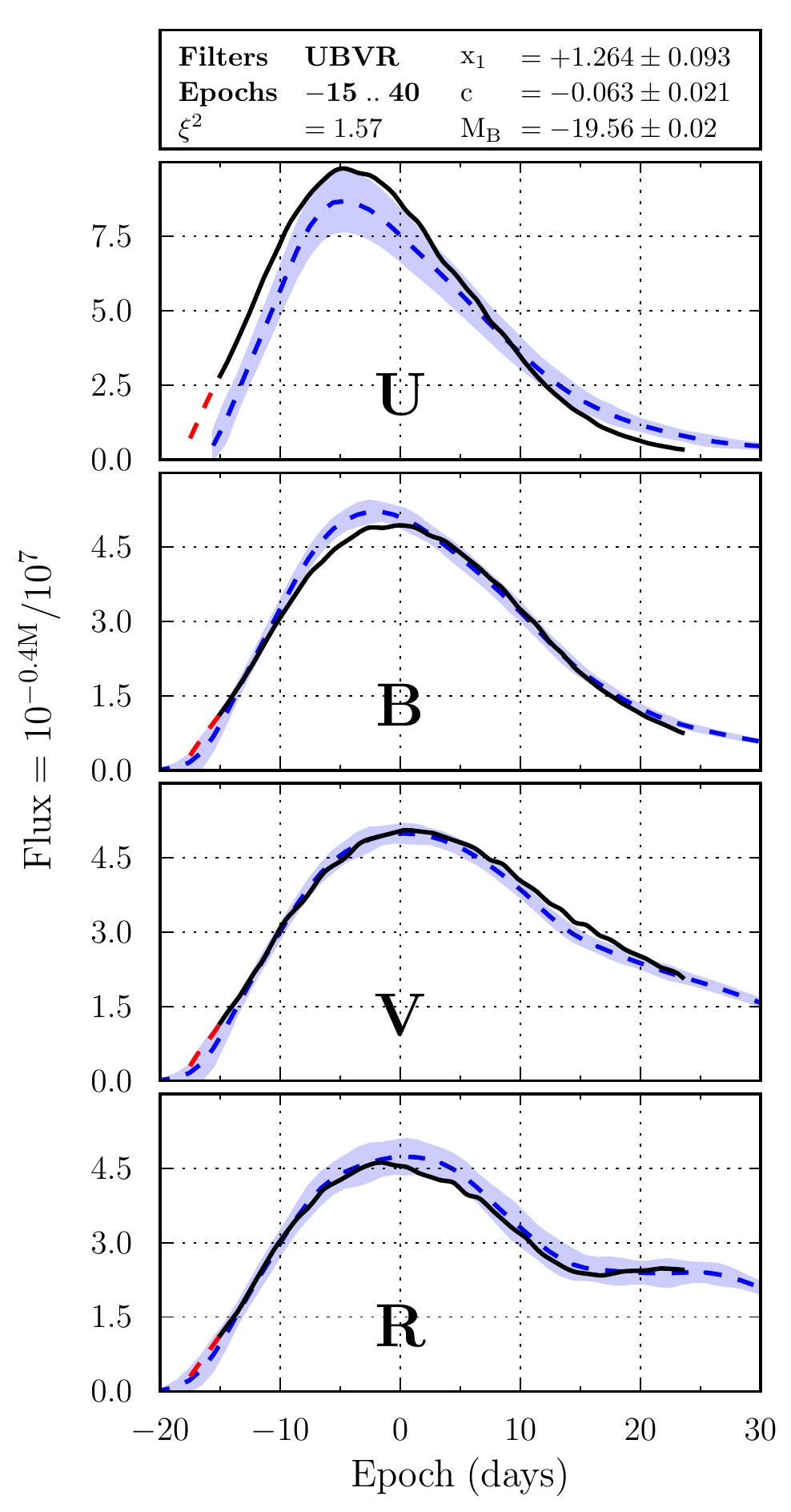}
\includegraphics[trim = 10mm 0mm 0mm 0mm, clip, scale=0.45]{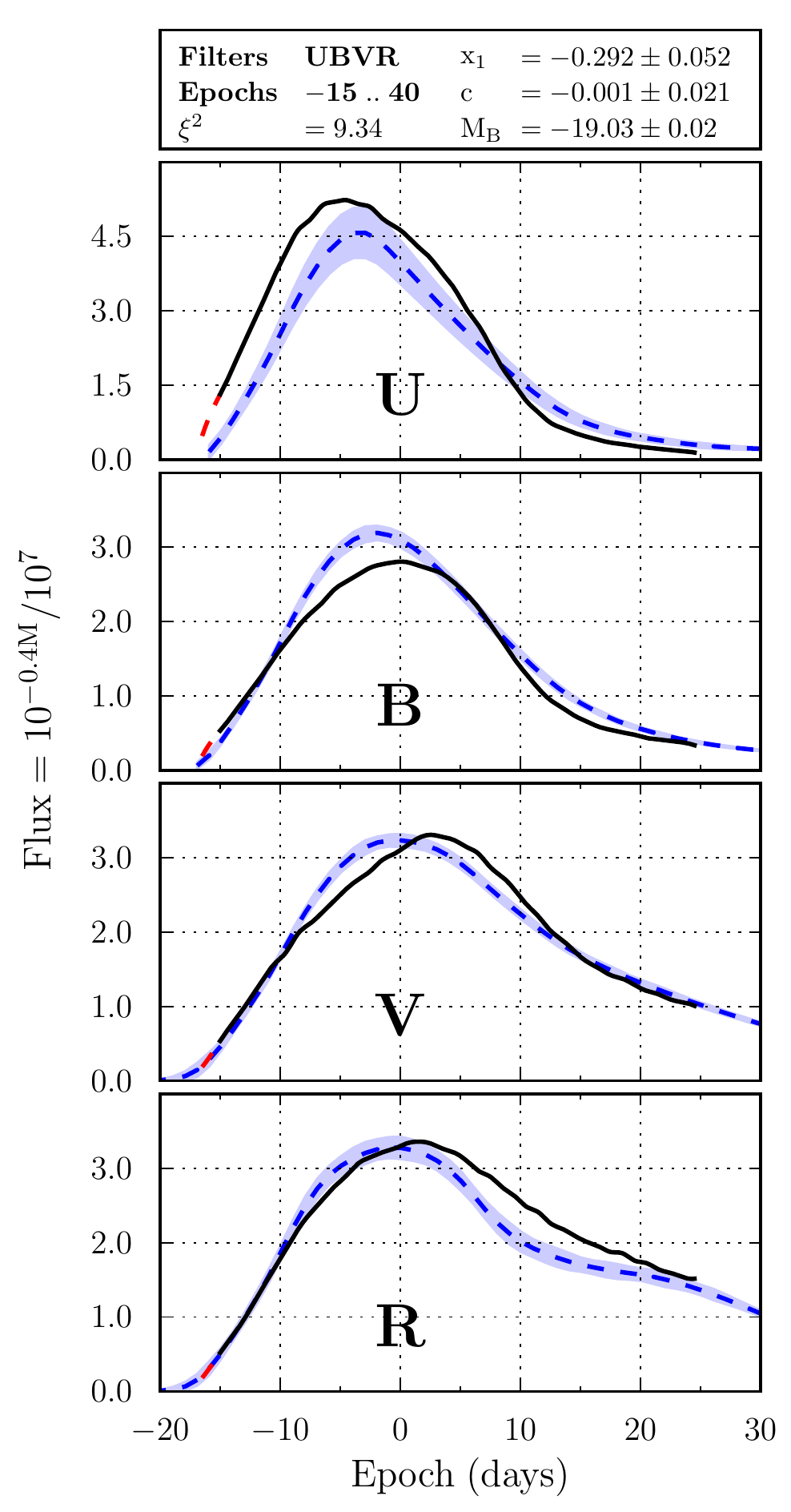}
\includegraphics[trim = 10mm 0mm 0mm 0mm, clip, scale=0.45]{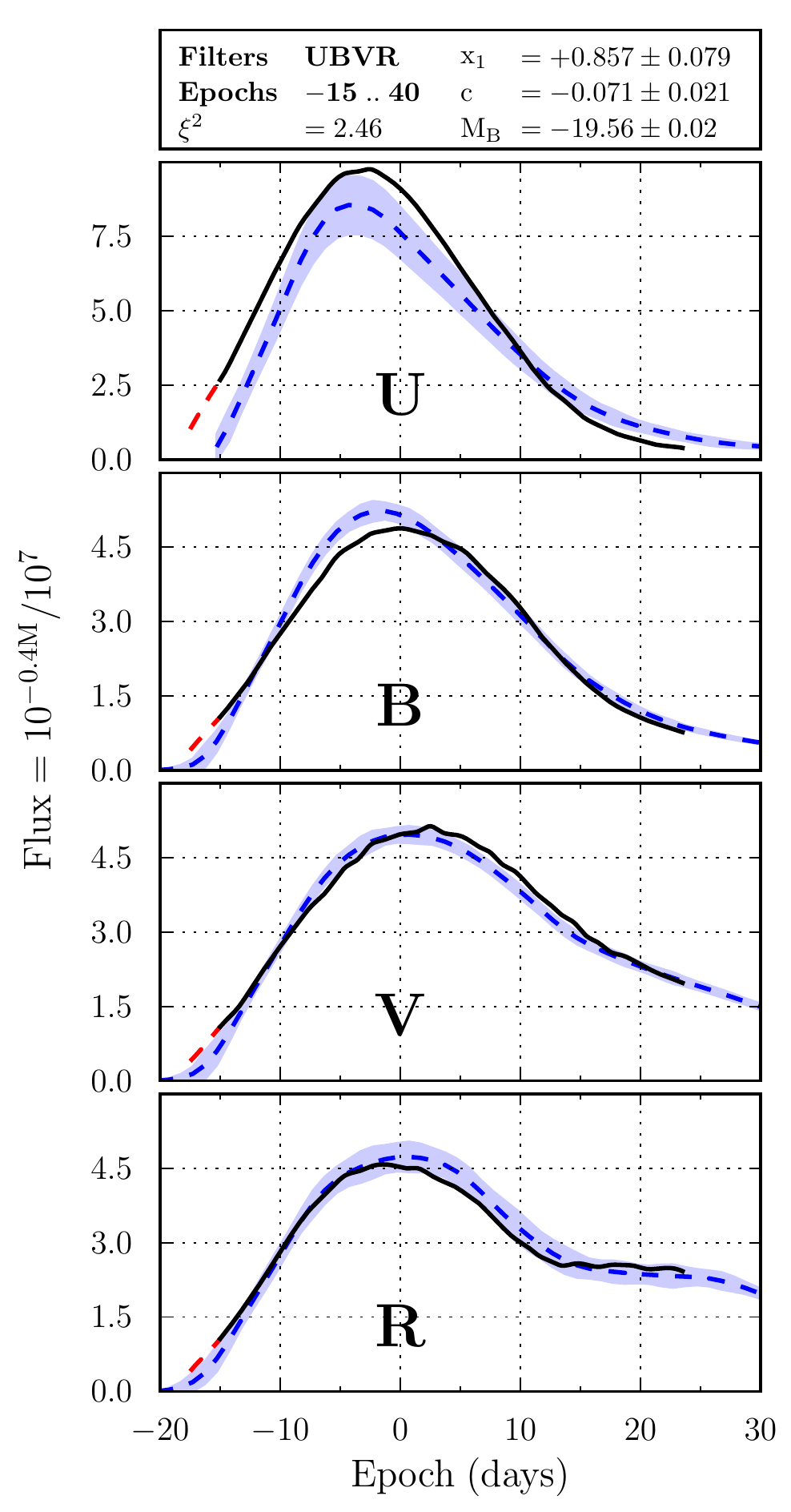}
\includegraphics[trim = 10mm 0mm 0mm 0mm, clip, scale=0.45]{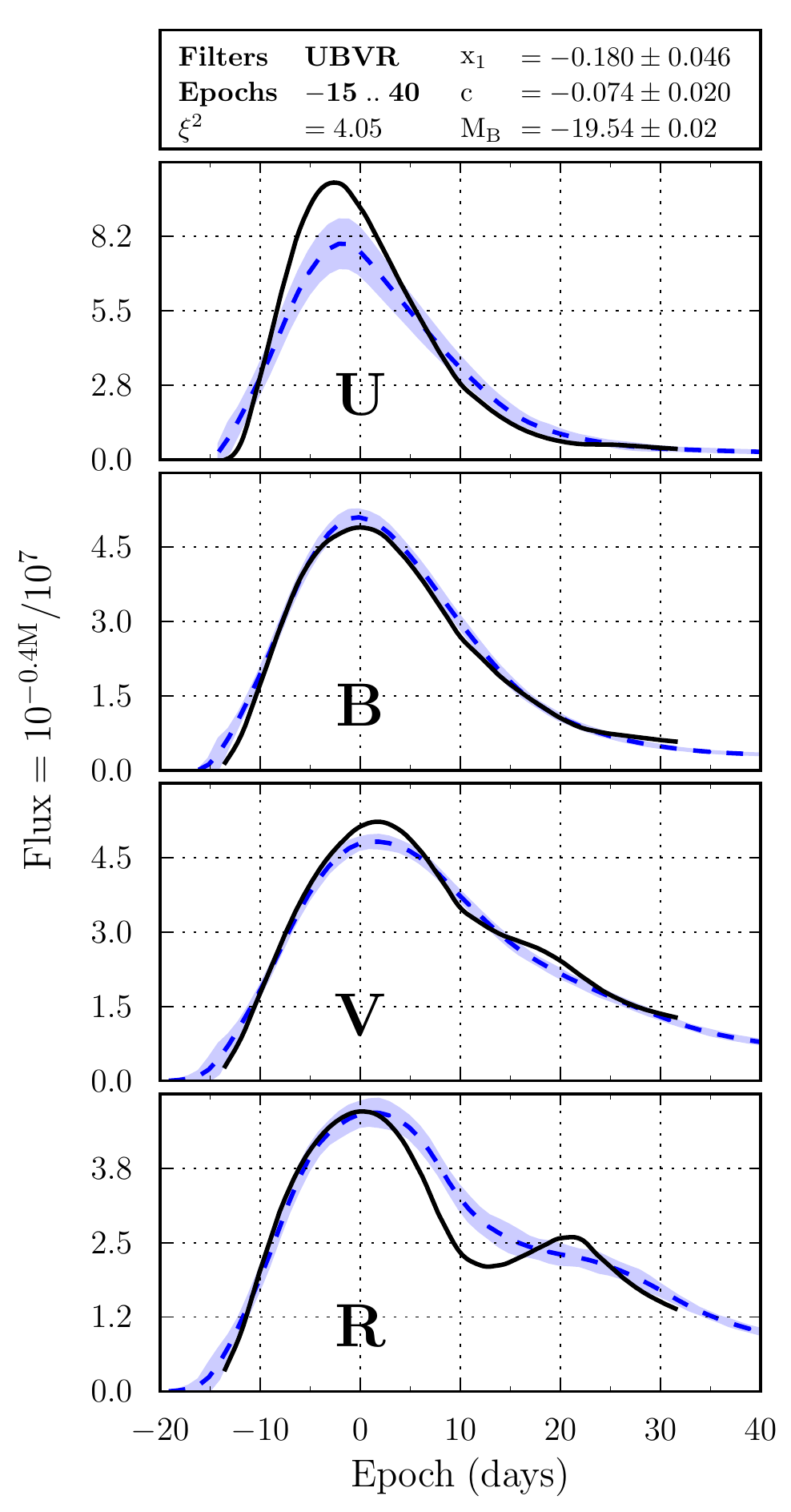}
\caption{$UBVR$ light curve fits to three KRW09 models (left three columns) and W7 (right column), including all epochs from $-15$ to $+40$ days. The lines have the same meaning as in Figure \ref{fig:fits:filters}. Left column: the 51\degree viewing angle of the iso\_5\_dc\_1 model, the viewing angle with the highest $\fxiav=0.95$. Second column: the 165\degree viewing angle of the iso\_6\_dc\_5 model, the viewing angle with the lowest $\fxiav=0.32$. Third column: the 88\degree\ viewing angle of the iso\_6\_dc\_2 model which we will analyze in detail in Section \ref{sec:disc:directcomp}. Right column: the W7 model is well fit with \salt across the entire epoch range, particularly in the $B$ and $V$ bands ($\favav = 0.700$).}
\label{fig:fits_krw}
\end{figure*}

\begin{figure*}
\centering
\includegraphics[trim = 0mm 16mm 3mm 3mm, clip, scale=0.7]{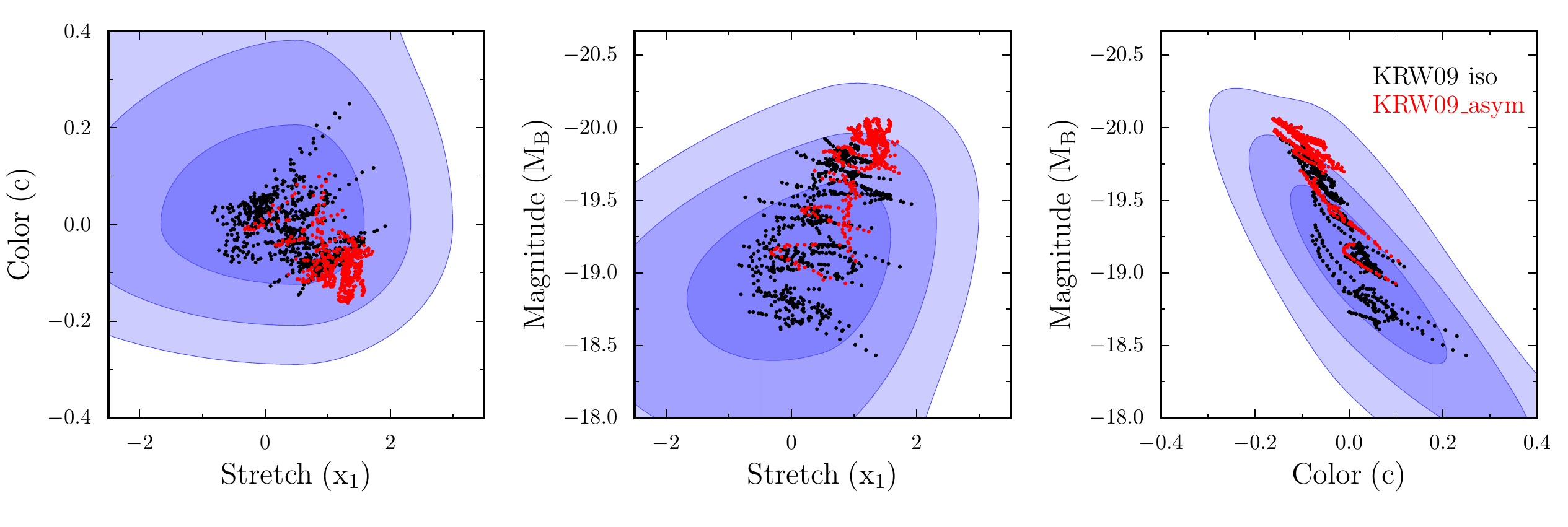}
\includegraphics[trim = 0mm 4mm 3mm 3mm, clip, scale=0.7]{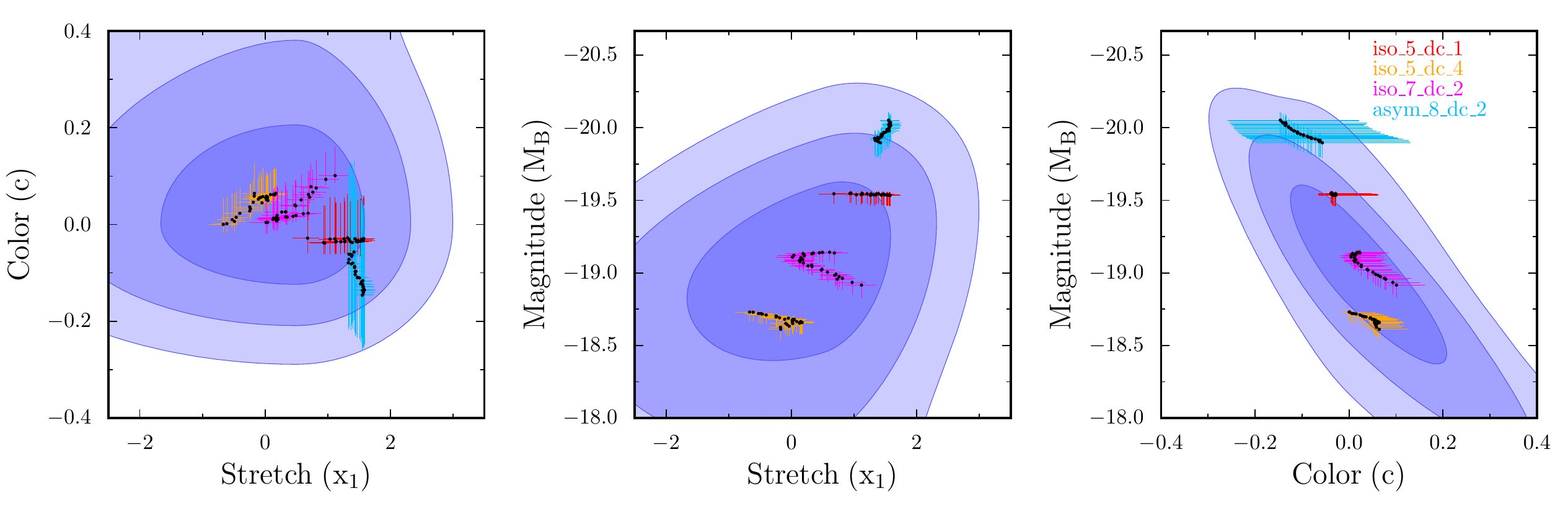}
\caption{Fit results for the KRW09 models. Top row: Fit results for all viewing angles of all KRW09 models, plotted without error bars for clarity. The KRW09 models occupy a favorable region of parameter space, with a few models outside the 2$\sigma$ contours in the stretch--magnitude and color--magnitude planes. On average, the isotropic models (black) are slightly favored over the asymmetric models (red). Bottom row: Fit results for several individual models including iso\_5\_dc\_1 (red, highest $\favav$), iso\_5\_dc\_4 (orange, lowest $\favav$), iso\_7\_dc\_2 (pink, highest $\fpavav$), and asym\_8\_dc\_2 (cyan, lowest $\fpavav$). Even the farthest outlying viewing angle of asym\_8\_dc\_2 is not excluded by the population model (see the light curve fit for this viewing angle in Figure \ref{fig:fits_krw}).}
\label{fig:pop_model_krw}
\end{figure*}

Generally, all KRW09 models do relatively well ($0.642 \leq \favav \leq 0.909$), particularly in matching the population model ($0.55 \leq \fxiavav \leq 0.99$). With 43 models in the KRW09 set, a total of 1290 viewing angles and 15,480 fits, it is difficult to give a visual impression of the various light curves and fits. Figure \ref{fig:fits_krw} shows three selected light curve fits for some of the highest and lowest ranked viewing angles. In order to give an overview of the region of \salt parameter space occupied by the KRW09 models, the top row of Figure \ref{fig:pop_model_krw} shows the location of all KRW09 models on the parameter space plots. For clarity, the error bars are omitted. On average, the asymmetric models tend to be slightly brighter and redder than the isotropic models. The isotropic models lie somewhat closer to the center of the population distribution in all three panels. This trend is reflected in the higher $\fpavav$ that the isotropic models receive on average (Figure \ref{fig:foms}). We emphasize that this observation refers only to the isotropic and asymmetric cases investigated by KRW09, and is not a statement about isotropic or asymmetric models in general.

The bottom row of Figure \ref{fig:pop_model_krw} shows the locations of a few selected models, including those with the highest and lowest $\favav$, and highest and lowest $\fpavav$. While the error bars on the color parameter can be quite large, we note that the error bars represent the standard deviation given the set of 12 fits performed, and are usually dominated by one or a few poor fits. Since the average values plotted are {\it weighted} averages, the fitted parameters from the bad fit runs are essentially ignored, and the average value is presumably more reliable than the error bars indicate (see discussion in Section \ref{sec:method:fom}).

One might suspect that the asymmetric models would exhibit systematically different light curves with viewing angle, and thus a larger dependence of their figure of merit with viewing angle. Such a trend, however, is not apparent in the dispersions of $\fav$ with viewing angle ($\sigma_{\theta}$ in Table \ref{table:results}), or in the parameter space plots in Figure \ref{fig:pop_model_krw}. In fact, the two models with the highest dispersion are iso\_2\_dc\_3 and iso\_6\_dc\_2 ($\sigma_{\theta}$ = 0.055), though a low dispersion in the figure of merit does not automatically mean that the fitted parameters show a low dispersion as well.

The asymmetric models show a few interesting trends (light blue points in Figure \ref{fig:foms}). First, the asym\_1 models (dc 2 and 3) both receive excellent figures of merit, and lie in the top right corner of the plot. The asym\_2 models have slightly brighter magnitudes, but receive equally excellent $\fpavav$. Their light curve fit, however, is degraded by a shallower rising slope, moving them to the top center of the plot. All other asym models share roughly the same $\fxiavav \approx 0.8$ because their light curves exhibit very similar shapes. Their magnitudes, however, become brighter, roughly in the order asym\_5, 4, 3, 7, 6 and 8, moving them away from the center of the population model, with the dc\_3 versions performing slightly better than their dc\_2 counterparts. Overall, it appears that those models with more ignition bubbles (asym\_1 and 2) perform better than those with fewer bubbles. More bubbles mean a stronger deflagration, and thus a less powerful detonation and a dimmer peak magnitude, in better agreement with the population model. For the isotropic models, the comparable trends are much less clear (dark blue points in Figure \ref{fig:foms}).

The KRW09 models have been compared with observations before, namely in the original KRW09 paper and in \citet{blondin_11_comparison}. Figure 3 of KRW09 demonstrates that the models broadly follow the Phillips relation, with only a few models outside the 1$\sigma$ contour. Furthermore, the models adhere to the observed color--magnitude relation, with some scatter. We confirm both results, though some models lie outside our $2\sigma$ contours because their peak magnitudes range toward the bright end of the observed distribution. 

\citet{blondin_11_comparison} excluded six of the models on the basis of spectral comparisons, namely the dc\_2 and dc\_3 versions of asym\_3, asym\_6 and asym\_8. The asym\_6 and asym\_8 models are the lowest ranked asymmetric models according to our method ($0.668 \leq \favav \leq 0.692$), while the asym\_3 models range in mid-field ($0.75 \leq \favav \leq 0.764$), meaning that none of these models are excluded in our analysis. Out of a subset of eight models which were analyzed in detail in \citet{blondin_11_comparison}, iso\_3\_dc\_1 and asym\_1\_dc\_3 showed the best and third-best spectral matches with observations. These model are among the highest ranked models in our analysis as well ($\favav = 0.873$ and $0.891$, respectively). On the other hand, the second-best spectral match, iso\_6\_dc\_5, ranges toward the bottom of the table in our analysis ($\favav = 0.710$). Such discrepancies highlight that a good spectral match and good light curve fit are not necessarily correlated. When spectra are integrated over broad-band filters in order to generate magnitudes, discrepant features which might degrade a spectral comparison are washed out in favor of the overall normalization, which is in turn ignored in most spectral comparisons. The resulting differences between spectral and light curve comparisons were highlighted, for example, in Figure 7 of \citet{blondin_11_comparison} which demonstrates that those KRW09 models with good spectral matches are only slightly more likely to match the Phillips relation.

\subsection{The W7 Model}
\label{sec:results:w7}

As expected from the discussion in Section \ref{sec:method:sims}, Table \ref{table:results} shows that W7 agrees with observed light curves quite well. The average figures of merit for the light curve fits are $\fxiav=0.60$, $\fpav=0.80$ and $\fav=0.700$, as shown by the red triangle in Figure \ref{fig:foms}. The right column of Figure \ref{fig:fits_krw} shows $UBVR$ light curve fits to the W7 model, fit over the entire epoch range computed by \phoenix. The goodness-of-fit parameter $\xi^2$ suffers mainly from two features of the light curves, namely the high $U$-band peak brightness and the strong second peak in $R$-band.

The fitted parameters for W7 match observations as well as the light curves, as shown with the black data points in Figure \ref{fig:pop_model_w7_puredef}. W7 also lies well within the range of the population model, though slightly on the bright side for its stretch and color. Since W7 is not a very physical model (see Section \ref{sec:method:sims}), there are no deep insights to be taken away from these results. They simply demonstrate that our method arrives at the accepted conclusion that W7 represents observed light curves surprisingly well.

\subsection{Pure Deflagration Models}
\label{sec:results:puredef}

\begin{figure}
\centering
\includegraphics[trim = 0mm 0mm 3mm 0mm, clip, scale=0.45]{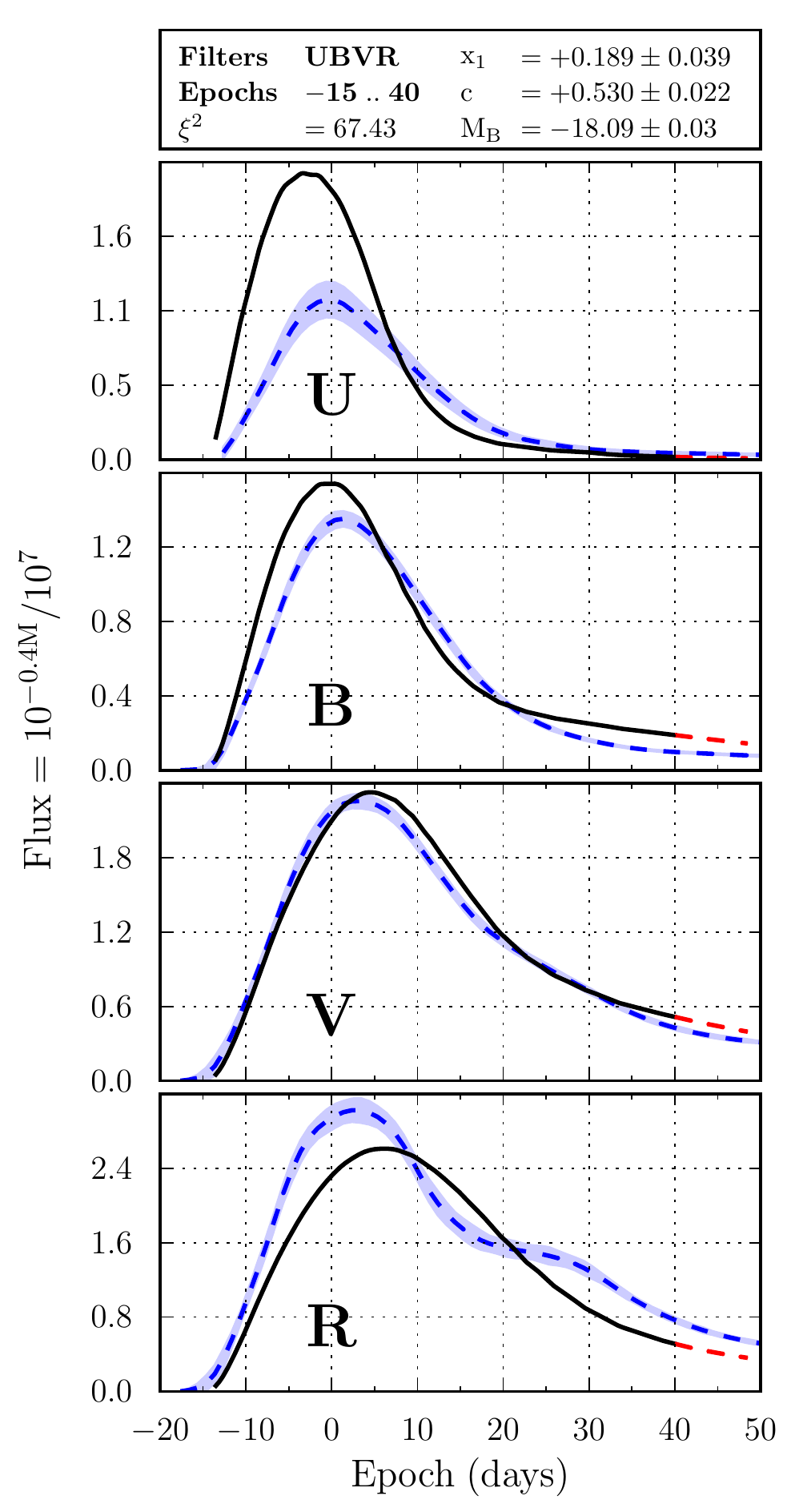}
\includegraphics[trim = 10mm 0mm 3mm 0mm, clip, scale=0.45]{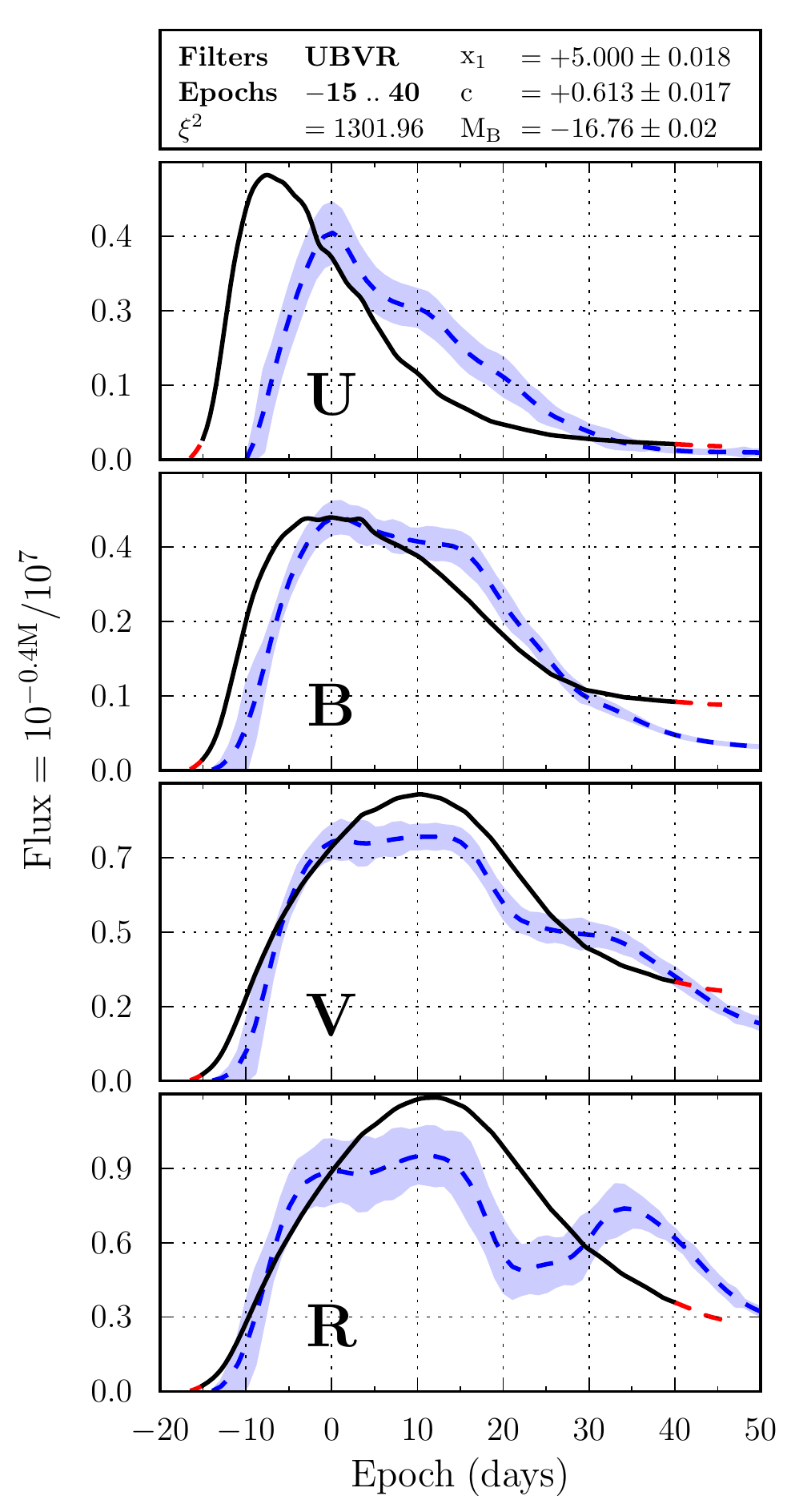}
\caption{$UBVR$ light curve fits to the PureDef\_0128\_256 (left column) and PureDef\_3500\_384 (right column) models. The lines have the same meaning as in Figure \ref{fig:fits:filters}. PureDef\_0128\_256 receives the best overall figure of merit of all PureDef models, PureDef\_3500\_384 the worst. See Section \ref{sec:results:puredef} for a discussion of the large uncertainties on the \salt model in the PureDef\_3500\_384 fit.}
\label{fig:fits_w7_puredef}
\end{figure}

\begin{figure*}
\centering
\includegraphics[trim = 0mm 4mm 3mm 3mm, clip, scale=0.7]{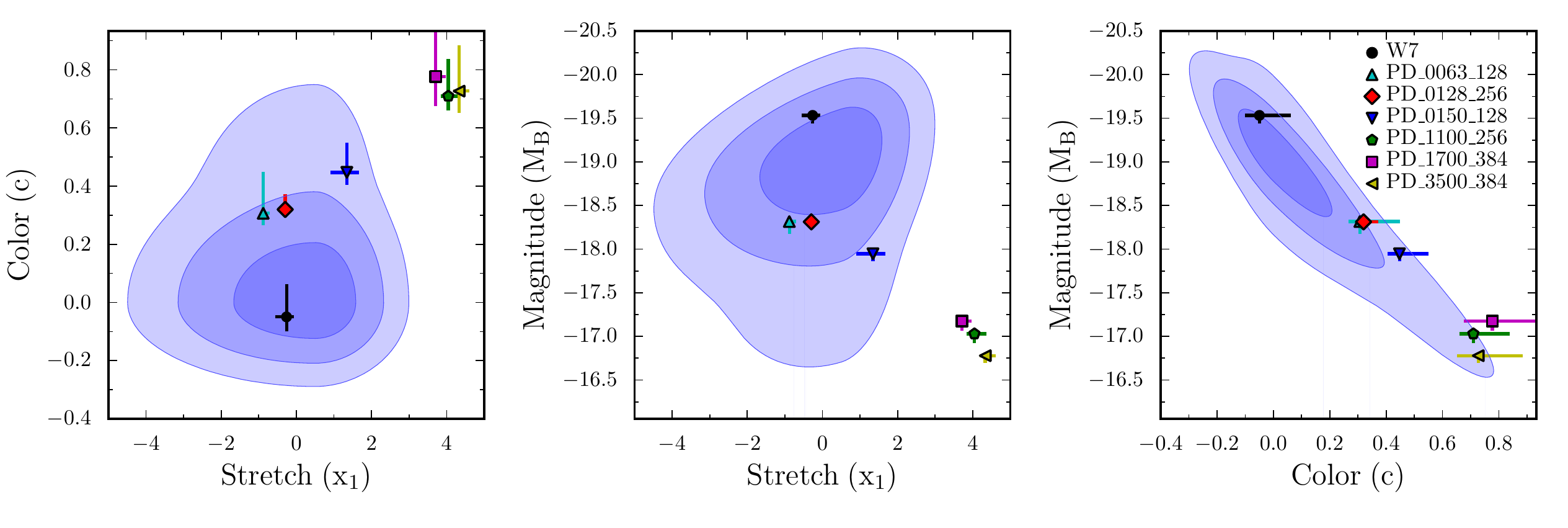}
\caption{Fit results for the W7 and pure deflagration models. The W7 model (black data point) is known to reproduce observed light curves well which is confirmed here, though W7 lies toward the bright end of the stretch--magnitude and color--magnitude relations. The pure deflagration models split into two groups, those with a few ignition bubbles (PureDef\_0128\_256, PureDef\_0150\_128 and PureDef\_0063\_128; red, cyan and blue data points) which are not in the center of the distribution, but certainly not ruled out either, and three models with many ignition bubbles. The latter are clearly ruled out by the population model as the values of both their stretch and color are higher than observed \sne.}
\label{fig:pop_model_w7_puredef}
\end{figure*}

Though underluminous \sne have recently received increased attention \citep{foley_13_iax}, they constitute at best a sub-class of \sne. They are not well represented in observed samples because they are both less frequent and dim, and thus hard to detect. Pure deflagration models are generally known to result in underluminous \sne and may afford an explanation of these events. However, we expect pure deflagration models to receive relatively low figures of merit for two reasons: first, the \salt model was trained on normal \sne and typically does not fit peculiar Ia well; and secondly, only a small fraction of the \sne observed in nature are underluminous, and the population model does not favor them as a result. We return to this issue in Section \ref{sec:conclusion}. Once again, we emphasize that the six pure deflagration models investigated here are not necessarily characteristic of pure deflagration models in general, as they represent a small subset of the possible ignition configurations and other parameters.

Figure \ref{fig:foms} demonstrates that the pure deflagration models, shown with orange squares, separate into two groups: those with few ignition points (63, 128 and 150, $0.302 \leq \fav \leq 0.331$) and those with many (1100, 1700 and 3500, $0.124 \leq \fav \leq 0.142$). The few-bubble models occupy an untypical location in  $\fxiav$--$\fpav$ space, as they combine poor light curve fits with favorable fitted parameters.

The left column of Figure \ref{fig:fits_w7_puredef} shows light curve fits to the few-bubble model PureDef\_0128\_256. The fit is extremely poor in $U$-band, and thus receives a low figure of merit, as do the other few-bubble models ($0.05 \leq \fxiav \leq 0.12$). The fitted parameters shown in Figure \ref{fig:pop_model_w7_puredef}, however, reveal a different picture. The few-bubble models (red, cyan and blue data points) can be accommodated by the population model, with figures of merit of $0.51 \leq \fpav \leq 0.64$. We note that the few-bubble models exhibit very small error bars on stretch and magnitude, despite their mediocre light curve fits. As discussed in Section \ref{sec:method:fitting}, poor fits are generally more sensitive to changes in filter bands and epoch range, and we might thus expect large uncertainties on the PureDef models. In this case, however, the discrepancy with the \salt model is largely driven by the excessive $U$-band flux of the few-bubble models. When only the $B$ and $V$ bands are fit, the fitted color is much redder than that of the full $UBVR$ fit. As a result, the uncertainty in the fitted color parameter is significantly larger than the uncertainty in stretch and magnitude.

In contrast to the few-bubble models, the many-bubble models (purple, yellow and green data points in Figure \ref{fig:pop_model_w7_puredef}) are clearly excluded by the population model ($\fpav = 0$). Their light curve fits, however, receive slightly higher figures of merit ($0.25 \leq \fxiav \leq 0.28$), seemingly in conflict with the poor light curve fit to the many-bubble model PureDef\_3500\_384 (right column of Figure \ref{fig:fits_w7_puredef}). While the particular fit shown receives $\fxi = 0$ due to the large $U$-band discrepancy with the \salt model, the $BV$ bands can be fit reasonably well, and the model receives a goodness-of-fit score of $\fxiav = 0.25$. The reason for this comparatively high score is that the uncertainties on the \salt model light curves are much larger than for the fits to other models. \salt assigns these increased uncertainties whenever the fit is driven to an extreme region of parameter space, in this case large stretch, red color and dim magnitude. Since very few such events have been observed in nature, the \salt flux surfaces are poorly constrained which is reflected in the larger uncertainties. At first sight, the larger uncertainties seem to pose a problem since they cause $\fxiav$ to reward fits with extreme parameters. However, extreme fitted parameters automatically mean a very small likelihood of observing a comparable \sna in nature, and thus a very low $\fpav$. Indeed, $\fpav = 0$ for the PureDef\_3500\_384 model, meaning that the model is deemed incompatible with observations despite its ``unfairly'' reasonable $\fxiav$. Since we are not particularly interested in the exact numerical values of $\fav$ for poorly fitting models, the imbalance in $\fxiav$ does not lead to false conclusions.

Overall, our analysis has revealed an interesting distinction between models with few and many bubbles. At least for this particular set of pure deflagration models, realizations with too many ignition bubbles are ruled out by the population model. Those with few bubbles cannot be accommodated by the \salt model, but since \salt was not trained on peculiar \sne, the PureDef light curves need to be compared to peculiar events directly (M. Long et al. 2013, in preparation).

\subsection{Pre-expanded Models}
\label{sec:results:preexp}

\begin{figure}
\centering
\includegraphics[trim = 85mm 4mm 90mm 3mm, clip, scale=0.56]{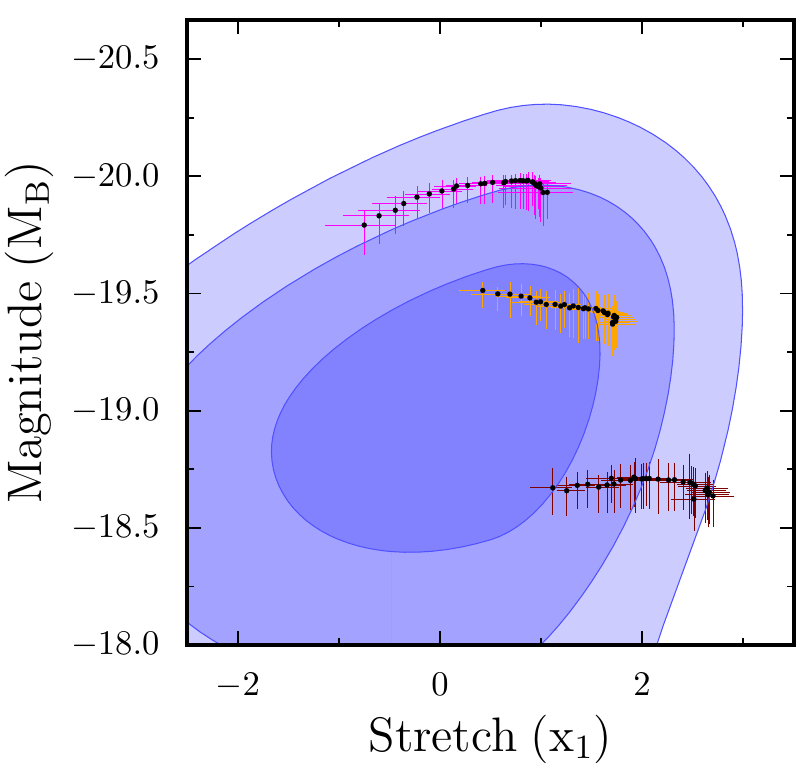}
\includegraphics[trim = 190mm 4mm 3mm 3mm, clip, scale=0.56]{\figdir/PopModel_PreExp.\figext}
\caption{Fit results for the three pre-expanded models. Almost all viewing angles lie within the allowed region of color-stretch space (not shown). The PreExp80 model exhibits the largest scatter of $\fav$ with viewing angle of any model. The amount of pre-expansion (30\%, 60\% and 80\% of the binding energy, respectively) appears to mostly change the peak magnitude of the light curves, but not the decline rate. As a consequence, the three models lie on a trajectory perpendicular to the Phillips relation.}
\label{fig:pop_model_preexp}
\end{figure}

Since the pre-expansion of the white dwarf was varied over a large range (30\% to 80\% of the binding energy), we expect the pre-expanded models to give diverse results. The fitted parameters shown in Figure \ref{fig:pop_model_preexp} confirm this expectation. The PreExp30 and PreExp60 models can be accommodated by the population model, though some viewing angles lie beyond the 2$\sigma$ contours of the stretch--magnitude and color--magnitude planes ($0.6 \leq \fpavav \leq 0.74$). Some viewing angles of the third model, PreExp80, lie outside the 3$\sigma$ contours, causing a poor figure of merit for the model overall ($\fpavav=0.21$). The amount of pre-expansion appears to mostly change the peak magnitude (since it determines how much material is burned to $^{56}$Ni), but not the decline rate. Thus, the models end up on a trajectory perpendicular to the Phillips relation. Since the detonation was initiated significantly off-center, we expect the pre-expanded models to exhibit strong variations with viewing angle. Indeed, the PreExp80 model shows the strongest variations in $f$ between viewing angles of any model ($\sigma_{\theta}=0.1$). 

Overall, the analysis shows that the PreExp60 model matches observed light curves surprisingly well, considering that it is a very simple, parametric, pure detonation model. This agreement is very sensitive to the amount of pre-expansion though, as brighter or dimmer peak magnitudes move the models away from the Phillips relation. Thus, it is unclear whether such simple pre-expanded models can fill in the rest of the parameter space in color and stretch. The results allow us to constrain which combinations of pre-expansion, contraction and detonation criteria result in a reasonable amount of $^{56}$Ni, and thus reasonable magnitudes.


\section{Discussion} 
\label{sec:disc}

We have introduced a new method for evaluating simulated \sna light curves with observations, and applied it to a range of explosion models. In this section we address some remaining questions. In Section \ref{sec:disc:popmodel} we discuss the parameter choices we made in designing the population model, and the reliability of those choices. In Section \ref{sec:disc:directcomp} we investigate an apparent contradiction between our evaluation method and visual light curve comparisons for one particular explosion model, and discuss potential issues with visual comparisons in general.

\subsection{Uncertainties in the Population Model}
\label{sec:disc:popmodel}

In this section, we discuss the various uncertainties which may affect our population model, namely the reliability of the global \salt parameters, the potential effects of host galaxy extinction, as well as trends with redshift and host galaxy type.

\subsubsection{The Global \salt Parameters}
\label{sec:disc:popmodel:param}

\begin{deluxetable*}{llcccccc}
\tablecaption{Values for the Global \salt Parameters from the Literature
\label{table:salt2_globalvars}}
\tablewidth{0pt}
\tablehead{
\colhead{Reference} &
\colhead{Data Set} &
\colhead{Model} &
\colhead{$\alpha$} &
\colhead{$\beta$} &
\colhead{$M_0$} &
\colhead{$H_0$} &
\colhead{$\sigma_{\rm int}$}
}
\startdata
\citet{sullivan_11_snls3}\,\tablenotemark{a} & SNLS & SiFTO & $0.132 - 0.138$ & $3.17 - 3.25$ &  &  & \\
\citet{conley_11_snls_3year}\,\tablenotemark{a} & SNLS, SDSS, low-$z$, $HST$ & SiFTO & $0.132 - 0.138$ & $3.17 - 3.25$ & & & $0.068 - 0.113$ \\
\citet{guy_10_snls_3year}\,\tablenotemark{a} & SNLS & \salt / SiFTO & $0.124 \pm 0.012$ & $3.17 \pm 0.13$ & $-19.095 \pm 0.032$\,\tablenotemark{$\dagger$}  & $70$ & 0.087 \\
\citet{marriner_11_salt2mu} & SDSS (four seasons) & \salt & $0.135_{-0.017}^{+0.033}$ & $3.19_{-0.24}^{+0.14}$ & & & \\
\citet{campbell_13_sdss} & SDSS (all seasons) & \salt & $0.22 \pm 0.02$ & $3.12 \pm 0.12$ & & & \\
\enddata
\tablecomments{The model column refers to the fit model for which values are given in the reference, even though other fit models may also have been used for the analysis. Wherever a range of parameters is given, different fits were performed, and the differences between their results were greater than the statistical error quoted.}
\tablenotetext{a}{The parameters from the recent SNLS3 analyses were re-scaled from SiFTO. Since SiFTO uses slightly different color and stretch corrections, $\alpha$, $\beta$ and the uncorrected magnitude $M_0$ also take on slightly different meanings. Linear re-scaling formulae for $\alpha$ and $\beta$ are given in G10. In their paper, the SiFTO $M_0$ is given as $M_0 = −19.218 \pm 0.032$ and was re-scaled to the value above (Guy 2012, private communication).}
\end{deluxetable*}

The population model parameters listed in Table \ref{table:pop_model} were determined by comparing data and Monte-Carlo simulations in K13, and from examining the extreme tails of the nearby sample. The global \salt parameters $\alpha$, $\beta$, $M_0$ and $\sigma_M$, however, are derived from a Hubble diagram fit, with fits to different data sets resulting in somewhat different values of the parameters (Table \ref{table:salt2_globalvars}). Such discrepancies arise either because surveys are biased, or because the true, underlying parameters are redshift-dependent, and different surveys probe different redshift ranges. Table \ref{table:salt2_globalvars} only lists analyses which used the G10 version of the \salt model or the SiFTO model, where SiFTO parameters were re-scaled to the corresponding \salt values (G10).

For the population model, we rely on the results of the most recent SNLS3 analyses, $\alpha = 0.13$, $\beta = 3.2$ and $M_0=-19.095$ \citep[G10;][]{sullivan_11_snls3, conley_11_snls_3year}. These SNLS3 results, however, are in tension with the most recent analysis of SDSS data \citep{campbell_13_sdss} who find an unusually high value of $\alpha = 0.22 \pm 0.02$ in their photometric sample, $4.5 \sigma$ off the SNLS3 value. However, they suspect that the discrepancy may be due to non-Ia contamination, and find a lower value of $\alpha=0.16 \pm 0.02$ when only fitting spectroscopically confirmed events.

\begin{figure}
\centering
\includegraphics[trim = 82mm 17mm 90mm 3mm, clip, scale=0.55]{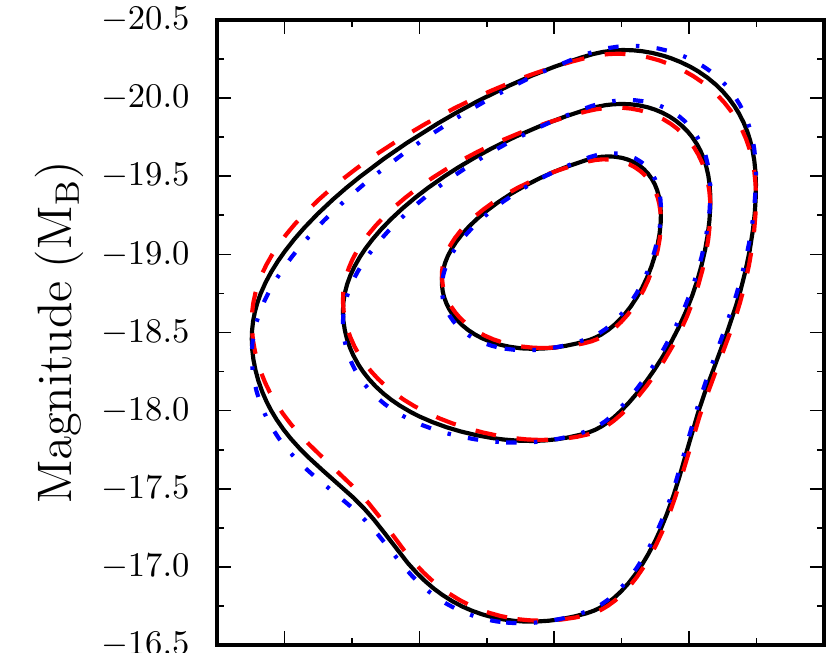}
\includegraphics[trim = 190mm 17mm 0mm 3mm, clip, scale=0.55]{\figdir/PopModel_VariedAlpha.\figext}
\includegraphics[trim = 82mm 4mm 90mm 3mm, clip, scale=0.55]{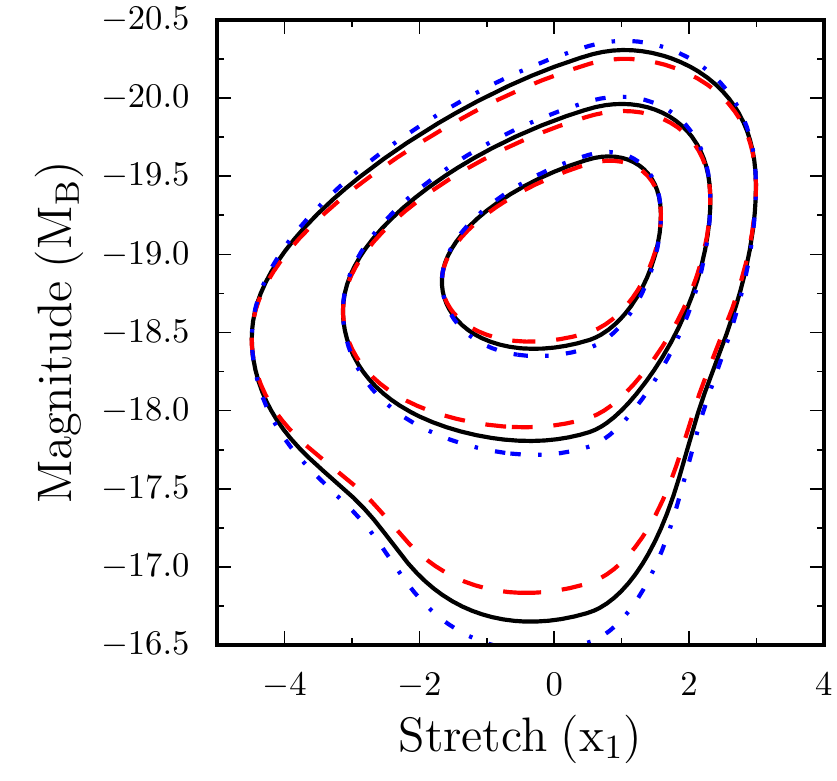}
\includegraphics[trim = 190mm 4mm 0mm 3mm, clip, scale=0.55]{\figdir/PopModel_VariedBeta.\figext}
\caption{Sensitivity of the population model to changes in $\alpha$ and $\beta$. Each row shows stretch--magnitude and color--magnitude contours for the fiducial population model ($\alpha=0.13$, $\beta=3.2$), as well as contours with $\alpha \pm 2 \sigma_{\alpha}$ (top panels) and $\beta \pm 2 \sigma_{\beta}$ (bottom panels), using the uncertainties from G10. The population model is remarkably stable against those changes in $\alpha$ and $\beta$.}
\label{fig:pop_model_varied}
\end{figure}

In order to quantify the impact of the uncertainties on $\alpha$ and $\beta$, we plot confidence contours of the population model for values which differ from the fiducial values by $2 \sigma$, using the uncertainties from G10. The top two panels of Figure \ref{fig:pop_model_varied} show stretch--magnitude and color--magnitude contours for our fiducial population model ($\alpha=0.13$ and $\beta=3.2$), as well as contours for $\alpha=0.106$ and $\alpha=0.154$. The contours are barely distinguishable from the fiducial model. In the bottom panels of Figure \ref{fig:pop_model_varied} we show the same plots, with $\beta = 2.94$ and $\beta=3.46$. The differences to the contours of the fiducial model are marginally larger than when varying $\alpha$, but still small, particularly for the 68\% and 95\% contours. In conclusion, we find that our values for $\alpha$ and $\beta$ reflect the most recent observational data, and that our population model is barely sensitive to their uncertainties. 

While the choice of $\alpha$ and $\beta$ does not have a significant impact on our population model, the choice of $M_0$ clearly does. Its impact would manifest itself in Figure \ref{fig:pop_model_varied} by shifting the contours up and down in magnitude, while the explosion model $\Mb$ would stay the same. As the scatter in \sna magnitudes is only $\sigma_{\rm M} = 0.13$, fractions of a magnitude matter. We note that, in the context of the \salt model, $M_0$ is the uncorrected ($x_1=c=0$) absolute $B$-band magnitude of \sne, {\it not} their average magnitude. While the magnitude distribution of \sne has been quantified \citep{yasuda_10, li_11_loss2}, $M_0$ must be derived from a Hubble diagram fit to \salt fitted \sne. Unfortunately, $M_0$ is often ignored in cosmological analyses since it is degenerate with $H_0$ and can be scaled out without influencing cosmological results.

As with $\alpha$ and $\beta$, we choose the value of $M_0$ found in the most recent SNLS 3-year analysis of G10. It is crucial to note that this choice of $M_0$ corresponds to their assumed Hubble constant, $H_0 = 70 \kmsmpc$ (G10). The exact value of $H_0$ has recently been subject to debate; while the final $WMAP$9 analysis of \citet{hinshaw_12_wmap9} found $H_0 = 69.3 \pm 0.8$, the newer Planck data favor a lower value of $67.80 \pm 0.77$ \citep{planck_13_cosmo}. In general, the values inferred from cosmic microwave background analyses are in tension with local universe measurements \citep[e.g., $73.8 \pm 2.4$ from Cepheid variables,][]{riess_11_hubble}. In light of such large systematic differences, we conclude that our value of $H_0 = 70 \kmsmpc$ represents a reasonable compromise between the current measurements. We do, however, caution that significant shifts in the best estimate of $H_0$ could change $M_0$ noticeably, namely by 0.03 magnitudes for each $1 \kmsmpc$ change in $H_0$. For example, if the Planck value of $67.8$ turned out to be correct, $M_0$ would shift by $0.07$ magnitudes. This implicit uncertainty in $M_0$ should be seen as part of our standard deviation in magnitude, $\sigma_{\rm M}$.

Table \ref{table:salt2_globalvars} shows that different analyses arrive at different conclusions regarding the intrinsic scatter in magnitude, $\sigma_{\rm M}$, which is an important input parameter to our population model since it directly determines the model's tolerance for magnitudes offset from the mean magnitude--stretch and magnitude--color relations. Our value for the intrinsic scatter in magnitude, $\sigma_{\rm M} = 0.13$, represents the coherent scatter model of K13. They found that this model (a wavelength-independent scatter of 0.13 magnitudes) reproduces the scatter observed in the Hubble diagram fit to SDSS and SNLS data. 

\subsubsection{Host Galaxy Extinction}
\label{sec:disc:popmodel:extinction}

Host galaxy extinction could affect our population model in two separate ways. First, the color distribution derived in K13 includes the intrinsic population convolved with host galaxy extinction, meaning that the true, underlying distribution might contain more blue \sne than our population model implies. Secondly, the model assumes that the color-magnitude relation in \sne is intrinsic, and should thus be obeyed by explosion models. Let us discuss these aspects in turn.

Constraining the amount of dust extinction in \sne has been a long-standing challenge \citep{riess_96_mlcs, riess_96_colorlaw}, partially because the extinction law parameter, $R_{\rm V}$, and thus the amount of color variation, is difficult to constrain \citep{jha_07_mlcs2k2}. Nevertheless, we attempt a simple estimate of the impact of extinction on the color distribution. We use the extinction distribution of \citet{woodvasey_07_essence} which is based on theoretical work by \citet{hatano_98_extinction} and \citet{riello_05_extinction}, as well as a fiducial value of $R_{\rm V} = 3.1$, and compute the shift in the underlying color distribution which reproduces the observed distribution when convolved with the extinction distribution. We find that the center of the distribution remains at $0$, but that virtually all \sne with $c > 0.2$ are accounted for by extinction, not intrinsic color variations. This result implies that either (1) there are no intrinsically red \sne or (2) the assumed distribution overestimates the amount of extinction significantly. 

Given these fundamental uncertainties, we choose to remain agnostic and ignore host galaxy extinction in our population model. We did, however, compute all figures of merit based on the extinction-corrected population model described above. The average $\fpavav$ of all models is lowered only slightly, from $0.701$ to $0.689$. The KRW09 models are barely affected since they lie towards the blue end of the color distribution. Those models with red colors, however, are strongly affected by the correction. For example, the average $\fpav$ of the few-bubble pure deflagration models drops from $0.57$ to $0$.

Besides the uncertainty in the underlying color distribution, extinction changes the interpretation of the nature of color variations in \sne. When using the population model to evaluate simulated \sne light curves, we assume that the color--magnitude relation is intrinsic, and should thus be obeyed by explosion models. If, however, this relation was caused by processes other than the \sna explosion itself, such as dust extinction, explosion models should show discrepancies, such as never producing highly reddened events. The assumption of an intrinsic color--magnitude relation can only render our model too lenient toward explosion models. Consider the extreme case of a color--magnitude relation which is entirely caused by dust reddening; in that case, explosion models should occupy one particular locus in color--magnitude space which would still be allowed by the population model (the locus of unreddened \sne). However, the model would also allow a range of other values of color and magnitude which correspond to the relation imposed by dust reddening. This extreme scenario has been ruled out, for example by \citet{maeda_11_asym_color} who investigated a sample of \sne which were deemed to be essentially free of host reddening, and found strong evidence for an intrinsic color--magnitude relation. Nevertheless, one should keep this caveat in mind when constraining explosion models using the color--magnitude relation.

\subsubsection{Dependence on Redshift and Host Galaxy Type}
\label{sec:disc:popmodel:other}

We note various systematic trends with redshift and host galaxy type which might affect the accuracy of our population model. For example, \citet{conley_11_snls_3year} found a mild redshift dependence in the best-fit values for $\alpha$ and $\beta$. Similarly, it is unclear how exactly the stretch and color populations of \sne depend on host galaxy, and thus redshift \citep{sullivan_06_hostgal, lampeitl_10_hostgal, sullivan_10_hostgal, kelly_10_hostgal, smith_11_hostgal}. The observed evolution of $x_1$ and $c$ with redshift can mainly be ascribed to Malmquist bias, but even after accounting for this type of bias the population model parameters from SDSS and SNLS are somewhat inconsistent, with SNLS producing more bright, blue explosions (K13). This disagreement is qualitatively explained by the observation that young, star-forming galaxies tend to produce brighter and bluer SNe \citep{sullivan_06_hostgal}. Nevertheless, the best-fit parent populations in color and stretch are slightly different for SDSS and SNLS. 

\subsubsection{Comparison with the Observed Magnitude Distribution}
\label{sec:disc:popmodel:lumdist}

The distribution of $\Mb$ in the population model is derived from the color and stretch distributions, the \salt prescription for the correlations between color, stretch, and magnitude (Equation (\ref{equ:salt2_MBfit2})), as well as the chosen values of $\alpha$, $\beta$ and $M_0$. Thus, comparing the $\Mb$ distribution with observed samples is an independent test of the accuracy of the population model.

We compared the distribution of peak magnitudes from our population model with the observed distribution of \citet{li_11_loss2}, a sample of 74 very nearby \sne from the Lick Observatory Supernova Search \citep[LOSS,][]{leaman_11_loss1}. The distributions generally agree, but the LOSS sample shows an excess at dim magnitudes. The sample was taken at $z \approx 0$ (at a maximum distance of 80 $\rm Mpc$), whereas the population model is trained on SDSS and SNLS data sets whose mean redshift is larger. Thus, the discrepancy could be a redshift effect, since more high-stretch, bright \sne are found at high redshift \citep{smith_11_hostgal}. Furthermore, LOSS was not a blind survey, meaning that host galaxy selection could affect the luminosity distribution. Regardless of these potential explanations, the LOSS sample is too small to conclude whether the dimmer magnitudes are a statistical fluctuation or not. 

The limited available data sets also pose the most important restriction on the accuracy of the color and stretch distributions, as the amount of data used for the analysis of K13 only allowed for a rough fit of the simulated distributions of $x_1$ and $c$ to those observed in SDSS and SNLS. With larger future data sets, a population model should be constructed from fitting functions with large numbers of degrees of freedom, including a dependence on redshift and host galaxy properties. Such an investigation is beyond the scope of this paper, but is a promising avenue for future research, particularly once DES data will be available.

\subsection{The Perils of Visual Light Curve Comparisons}
\label{sec:disc:directcomp}

\begin{figure}
\centering
\includegraphics[trim = 2mm 15mm 3mm 5mm, clip, scale=0.7]{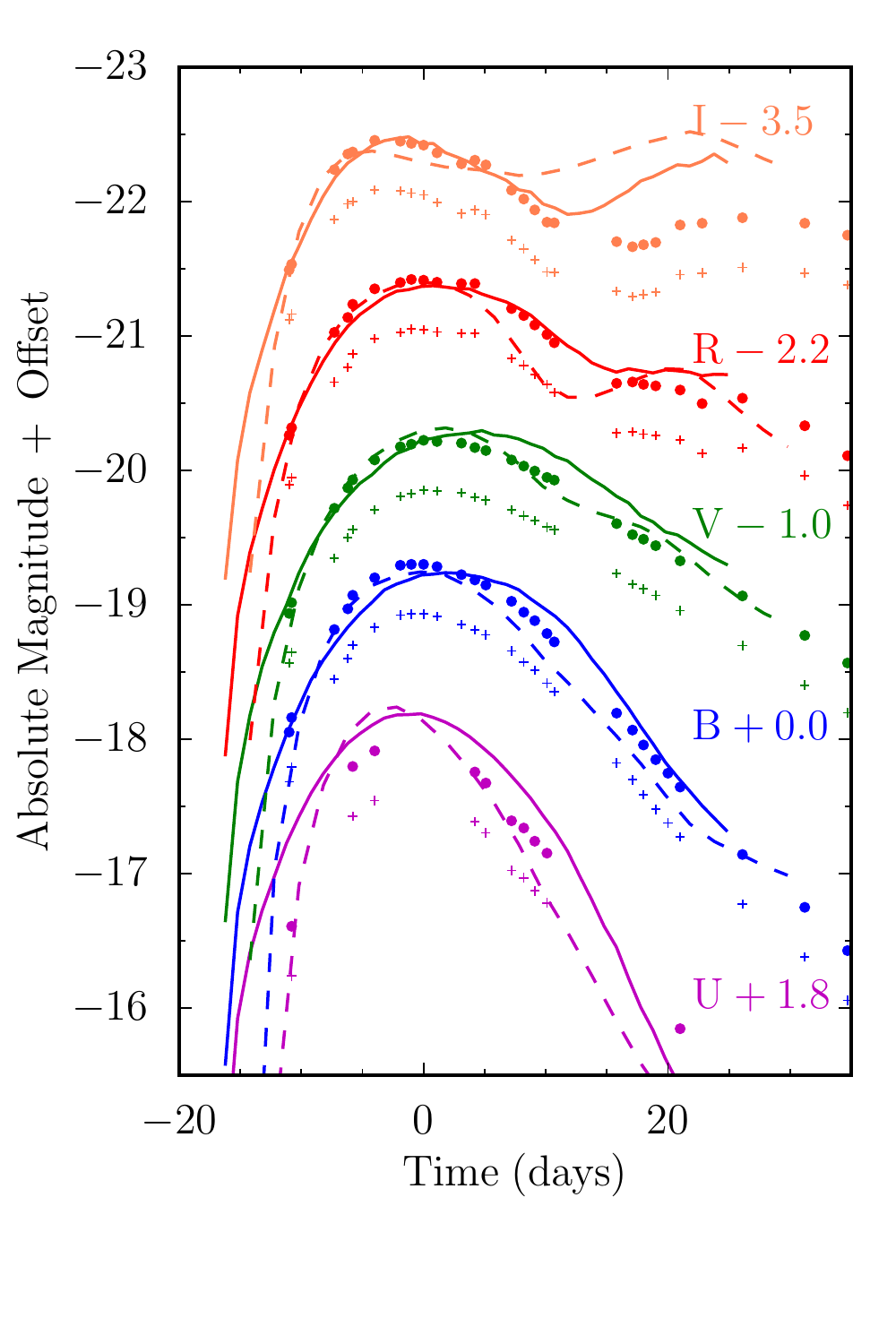}
\caption{Visual light curve comparison. The dots and crosses show the light curves of 2003du \citep{stanishev_07_2003du} for the two different distance moduli suggested in their paper, 32.79 (dots) and 32.42 (crosses). The solid lines show light curves of the 88\degree\ viewing angle of the KRW09\_iso\_6\_dc\_2 model. The dashed lines show light curves of the W7 model.}
\label{fig:directcomp_2003du}
\end{figure}

\begin{figure}
\centering
\includegraphics[trim = 85mm 4mm 90mm 3mm, clip, scale=0.56]{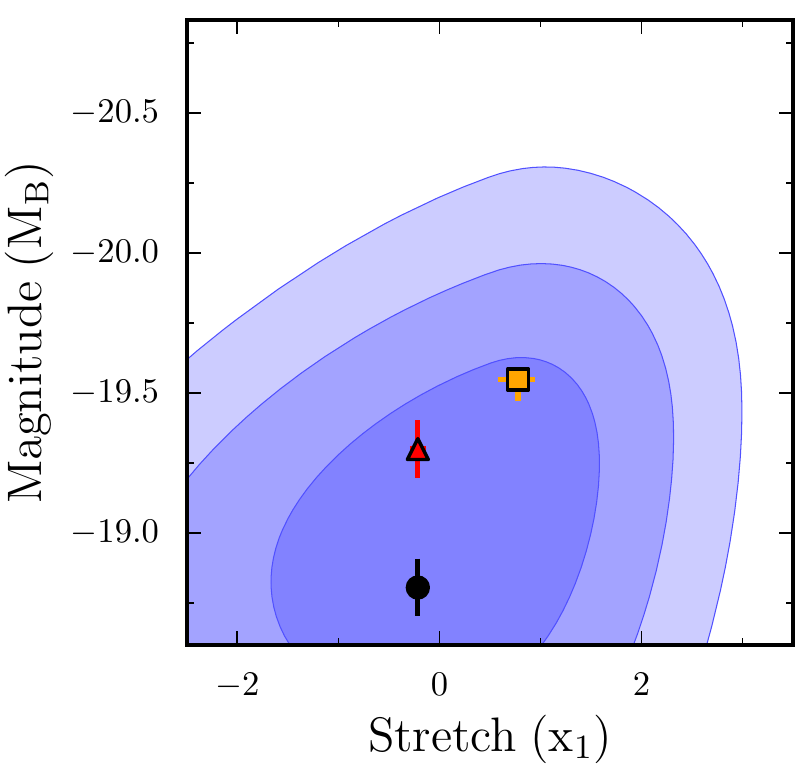}
\includegraphics[trim = 190mm 4mm 3mm 3mm, clip, scale=0.56]{\figdir/PopModel_2003du.\figext}
\caption{Fit results for the KRW09\_iso\_6\_dc\_2 delayed-detonation model (orange square) and the observed SN 2003du (CfA3 data, black circle). If we assume the absolute magnitude resulting from the data of \citet{stanishev_07_2003du} and their suggested distance modulus of $\mu=32.79$, the point shifts to $\Mb = -19.3$ (red triangle). While the fitted parameters of both KRW09\_iso\_6\_dc\_2 and 2003du are in good agreement with observations, they occupy different loci on the parameter planes, depending on which distance modulus is assumed. Furthermore, the assumed distance modulus determines whether 2003du is a ``normal'' SN in the sense that it lies close to the center of the color-magnitude distribution. See Section \ref{sec:disc:directcomp} for a detailed discussion.}
\label{fig:pop_model_2003du}
\end{figure}

We have previously alluded to issues with visual comparisons between simulated and observed light curves. In some cases, only a rough comparison with light curves is desired, for example if an explosion model is not intended to represent normal \sne but a sub-group such as sub-luminous events. In such cases, visual comparisons are perfectly appropriate \citep[e.g.][]{kromer_13_failed}. Often, however, the purpose of visual comparisons is to determine whether an explosion model reproduces the light curves of normal \sne \citep[see, e.g.,][]{woosley_07, kromer_10_doubledet, jack_11, pakmor_12_dd}. Here we investigate a particular example of this technique. Both KRW09 and \citet{blondin_11_comparison} overplot light curves of the KRW09\_iso\_6\_dc\_2 model (88\degree\ viewing angle) and the observed \sna 2003du, and note that there is good visual agreement. In general, our analysis agrees, since the particular viewing angle receives a figure of merit for goodness-of-fit of $\fxiav = 0.83$. However, on closer inspection of the light curve fit (third column of Figure \ref{fig:fits_krw}), we note that the fit is $\approx 1.6 \sigma$ off the best-fit \salt model on average, and shows somewhat poor agreement at early times. In this section, we investigate the reasons why those discrepancies are not apparent in the visual comparison of light curves. We emphasize that we pick the example of KRW09\_iso\_6\_dc\_2 because the explosion model light curves have already been analyzed in this paper, not because it represents a particularly good or bad choice of data to compare to. 

First, and most importantly, the distance modulus to 2003du, and thus its magnitude, are highly uncertain. \citet{stanishev_07_2003du} quote two different distance moduli of $\mu=32.42$ and $\mu=32.79$, leading to a 0.37 magnitude difference. The reason for this uncertainty is the uncertain peculiar velocity of the host galaxy of 2003du which could be a few hundred $\kms$, corresponding to a redshift uncertainty $\sigma_z \approx 0.001$. Given that the redshift of 2003du is $z=0.007$, this uncertainty leads to a 0.3 magnitude uncertainty. Clearly, using very nearby SNe for absolute magnitude comparisons is dangerous. Using $\mu=32.79$, the value adopted in \citet{kasen_09_models}, 2003du has $\Mb = -19.30$ \citep{stanishev_07_2003du}, a value which coincides well with the KRW09\_iso\_6\_dc\_2 model. Figure \ref{fig:directcomp_2003du} shows a visual comparison of the $UBVRI$ light curves of 2003du and the 88\degree\ viewing angle of the KRW09\_iso\_6\_dc\_2 model. For comparison, the 2003du light curves are also plotted using the alternative distance modulus of $\mu=32.42$ which leads to a catastrophic fit.

Secondly, the light curve fit to the $88\degree$ viewing angle of KRW09\_iso\_6\_dc\_2 is not as good as the visual comparison suggests, particularly at early epochs (Figure \ref{fig:fits_krw}). Returning to the visual comparison shown in Figure \ref{fig:directcomp_2003du}, we notice that the rising slope of KRW09\_iso\_6\_dc\_2 is indeed quite shallow compared to the W7 model (dashed lines). The shallow rising slope of some of the KRW09 models was pointed out in Figure 8 of \citet{blondin_11_comparison}, but the comparison to 2003du does not highlight this issue, since there are few data points at early times. Our fitting method captures the different rising slopes by assigning the KRW09\_iso\_6\_dc\_2 model a larger stretch than 2003du (Figure \ref{fig:pop_model_2003du}).

Finally, we need to address the question whether 2003du is a truly ``photometrically normal'' \sna as claimed by \citet{stanishev_07_2003du}. Figure \ref{fig:pop_model_2003du} shows the fitted parameters for 2003du, using CfA3 observations. Given the redshift assumed in the survey data, the fitted absolute magnitude of 2003du is $-18.81$. With this magnitude, 2003du lies at the center of the stretch--magnitude distribution, but somewhat off the main color--magnitude relation and has a figure of merit of $\fxi=0.60$ (black circle). If we use the peak magnitude derived from the larger distance modulus, $\Mb=-19.3$, 2003du moves toward the center of the color--magnitude relation and receives $\fxi=0.79$ (red triangle). Either way 2003du is located at the blue end of the bulk distribution, and while it may qualify as a ``photometrically normal'' \sna, it is not an ideal candidate for light curve comparisons. 

In summary, there are good reasons to be careful with visual light curve comparisons, most importantly that features such as a shallow rising slope can easily be missed. 2003du is a particularly unfortunate case where a large uncertainty in the distance modulus makes any inference about explosion models difficult.


\section{Conclusion} 
\label{sec:conclusion}

We have presented a new method to evaluate explosion model light curves using observed data. We rely on a data-driven model to represent the heterogeneous family of \sna light curves. For this purpose, we choose the most recently trained model, \salt. For each explosion model, we perform multiple fits to the \salt model using a wide range of reasonable choices for filter bands and epoch range, since those choices affect the fit results. For each fit, we derive fit results in the \salt parameter space of stretch, color and magnitude. We take the weighted mean of those parameters to be the best estimate of the fit results for a given set of explosion model light curves. We compare the fitted parameters to a parametric population model in stretch--color--magnitude space which we construct from the color and stretch distributions of K13, and values for $\alpha$, $\beta$, and $M_0$ from the literature. We extend this population model with small outlier populations in color and magnitude to account for observed outlier events in a nearby data set. The final figure of merit is composed of two individual figures of merit for the goodness-of-fit of a light curve, and for the likelihood of observing an event with the fitted stretch, color and magnitude in nature. 

We have used this method to evaluate a variety of explosion models, and found their figures of merit to range from $0.124$ to $0.909$, given an allowed range of 0 (worst) to 1 (best). We found the delayed-detonation models of KRW09 to agree well with observed light curves, particularly the isotropic models. We confirmed that the W7 model also matches observed light curves well. We found a wide range of figures of merit for the pure-deflagration models analyzed, clearly distinguishing between the few-bubble and many-bubble variations of those models. Finally, we found that a suite of off-centered detonation models does not reproduce the stretch--magnitude and color--magnitude relations.

We discussed uncertainties on our population model, and concluded that our model is remarkably stable within the current uncertainties on the \salt parameters $\alpha$ and $\beta$. However, due to the uncertainty in $H_0$ it is still somewhat difficult to define the absolute magnitudes of \sne to the desired accuracy. We conclude that our population model is valid within the latest cosmological limits on $H_0$, but that future measurements might shift the overall magnitude distribution. Furthermore, we investigated some seemingly contradictory inferences from visual light curve comparisons and our method. We concluded that visual light curve comparisons can be misleading, particularly if the absolute magnitude of the observed \sna is not well determined.

We have left various promising avenues of research for future investigations, and briefly discuss a few of them. First, we did not consider the IR regions of the explosion model light curves since they are notoriously difficult to model accurately in radiative transfer calculations. Our method could easily be extended by using a data-driven model which covers the IR, such as SNooPy. Using a different data-driven model, however, would mean having to develop a new population model in the new parameter space. 

Secondly, our analysis neglected the time of explosion which is known for explosion models. While the time between explosion and peak brightness is unknown for individual observed \sne, recent observations have shown the slope of the rising light curves to closely follow a $t^2$ law. This behavior was demonstrated for observations of a single well-observed \sna \citep{nugent_11_2011fe}, the statistical average of a large sample \citep{hayden_10_sdss_lcs, ganeshalingam_11_rise} and through analytical modeling \citep[][and references therein]{piro_12_rise}. The $t^2$ dependence could be used to extrapolate the \salt model to very early times, and thus include the time of explosion as a data point in the light curve fit. The constraint of zero flux at explosion would place tight constraints on the rising light curves produced by explosion models, which have been found to be too shallow in various explosion models.

Third, there are uncertainties in our population model which are difficult to address with current data, such as the best functional form to parameterize the distributions of stretch and color, the tails of these distributions at extreme, rare values, as well as the dependence of the population model on redshift. Given the larger data sets of the future, we envision a global fit of the data to a function with many free parameters. As with the analysis of K13, such strategies still have to rely on an accurate model of survey biases. 

Finally, data-driven models do not parameterize peculiar \sne. Thus, our method assigns explosion models with peculiar light curves a relatively low figure of merit, even though they might match peculiar \sne observed in nature. For explosion models which might reproduce peculiar \sne such as sub-luminous Type Iax, we could use the photometric classification algorithm of \citet{sako_11}. In this analysis, core-collapse SN templates are replaced with templates for peculiar \sne, and explosion models are identified as either a normal \sna or any of the known peculiar-Ia. Such a method could also be applied to the light curves of double-degenerate explosion models to investigate whether they resemble normal Type Ia, or peculiar objects.


\section*{Acknowledgments}

We thank Dan Kasen for providing the SEDs of the KRW09 explosion models. We are grateful to Sean Couch and Emille Ishida for a careful reading of the draft and their comments. We are indebted to Tom Loredo for fruitful discussions about the statistical aspects of our method, and to Mark Sullivan for his comments on extinction. R.K. is grateful for the support of National Science Foundation grant 1009457, and of the Kavli Institute for Cosmological Physics. This work was supported in part by the University of Chicago and the Department of Energy under section H44 of Department of Energy Contract No. DE-AC02-07CH11359 awarded to Fermi Research Alliance LLC, and by the National Science Foundation under grant AST-0909132, as well as NASA award NNX09AK60G.


\bibliographystyle{apj}
\bibliography{../../_LatexInclude/sn.bib}


\appendix
 
\section{Survey Data}
\label{sec:app:data}

Here we briefly review the survey data which were used to check the consistency of our population model  in Section \ref{sec:method:popmodel}, and to extend it to represent outliers in color and magnitude. We describe the SDSS and CfA3 surveys, the selection cuts applied to these samples, and our procedure to reconstruct the rest-frame peak magnitude. We note that the population model of K13 itself was based on both the SDSS and SNLS (G10) surveys, and refer the reader to those papers for detailed information on the data sets.

\subsection{The SDSS Survey}
\label{sec:app:data:sdss}

The SDSS-II Supernova Survey \citep{york_00_sdss, frieman_08_sdss} used the SDSS camera and telescope \citep{gunn_98_sdsscam, gunn_06_sdsstelescope} to search for SNe in three seasons from 2005 to 2007 \citep{sako_08_sdss}. The survey scanned a 300 deg$^2$ region called Stripe 82 with a typical cadence of four nights, obtaining images in the $ugriz$ filter bands \citep{fukugita_96_sloanfilters}. The final \sna photometry was derived using the Scene Modeling Photometry technique \citep{holtzman_08_smp}, and used in this paper as well as K13. The SDSS survey discovered about 500 spectroscopically confirmed \sne.

\subsection{The CfA3 Survey}
\label{sec:app:data:cfa3}

The CfA3 survey \citep{hicken_09_cfa3} was conducted between 2001 and 2008 at the F. L. Whipple Observatory of the Harvard-Smithsonian Center for Astrophysics (CfA). The survey recorded a total of 11,500 observations of 185 \sne below $z=0.08$. Over its lifetime, CfA3 used two cameras, 4Shooter and Keplercam. The Johnson $U$, $B$ and $V$ filters were used on all three cameras, but the IR filters were changed during the survey. Unlike SDSS, the CfA3 survey did not discover \sne itself, but followed up discoveries by other groups and amateur astronomers, about half of them from the LOSS \citep{filippenko_01_loss}. \citet{hicken_09_cfa3} emphasize that the CfA3 sample is not representative of the underlying \sna population since the observing strategy favored young and extreme events. 

\subsection{Selection Cuts}
\label{sec:app:data:cuts}

The SDSS and CfA3 samples used in Section \ref{sec:method:popmodel} were fit with \salt using \snana. In order to ensure reliable light curve fits, we applied the following selection cuts:
\begin{itemize}
\item At least three filters need to have at least one observation each with signal-to-noise ratio $> 5$. 
\item There must be at least one observation before $B$-band peak, and at least one later than 10 rest-frame days after peak. This criterion ensures that $x_1$ is somewhat constrained.
\item There must be a total of at least 12 observations in all filters combined. 
\item The fit probability (as derived from $\chi^2/N_{\rm dof}$) is at least 1\%.
\item The SN must be a confirmed Type Ia event.
\end{itemize}
These cuts are deliberately not too strict, since we do not want to exclude interesting populations which might deviate from our population model.

\subsection{Reconstruction of Rest-frame Magnitudes}
\label{sec:app:data:restframe}

A \salt fit to observed \sne returns values for $\mb$, the peak magnitude in a {\it redshifted} $B$-band, for $x_1 = c = 0$. For the purpose of comparing explosion models to data, however, we are interested in the {\it rest-frame} characteristics of an explosion, namely the best-fit, rest-frame, $B$-band peak magnitude, $\Mb^{\rm rest}$. As we generate our mock light curves of the explosion models at a negligible redshift, $\Mb^{\rm rest} \simeq \mb - \mu$. For observed \sne at higher redshift, however, the filter transmission function for $B$-band is redshifted between the rest and observer frames, meaning that K-corrections are necessary to translate between $\mb$ and $\Mb^{\rm rest}$. For observed \sne at higher redshift, appropriate K-corrections are used to compute $\Mb^{\rm rest}$.

\section{Population Model}
\label{sec:app:popmodel}

In Section \ref{sec:method:popmodel} we stated the probability distribution for our population model. Here, we derive it more rigorously, and compute marginalized probability distributions in two variables. We assume that the variables $\mathbf{x} = (x_1, c, \delM)$ are independent and Gaussian. For now, we ignore their asymmetric $\sigma$ values and their potentially non-zero means. We can write the probability density as a trivariate Gaussian, $dP(x_1, c, \delM) = N e^{\frac{1}{2}\mathbf{x}^T C^{-1} \mathbf{x}} dx_1\, dc\, d\delM$ with $C^{-1} = \rm{diag}(\sigma_{x1}^{-2}, \sigma_{c}^{-2}, \sigma_{M}^{-2})$. The transformation to the desired parameter space $\mathbf{y} =(x_1, c, \Mb)$ is given by $\mathbf{y} = T \mathbf{x}$, where $T$ and the new inverse covariance matrix $D^{-1}$ are 
\begin{equation}
T = \left( \begin{array}{ccc}
1 & 0 & 0 \\
0 & 1 & 0 \\
-\alpha & \beta & 1 \end{array} \right)\,\Longrightarrow\,D^{-1} = (T^{-1})^T C^{-1} (T^{-1}) = \left( \begin{array}{ccc}
\sigma_{x1}^{-2} + \alpha^2 \sigma_{M}^{-2} & -\alpha \beta \sigma_{M}^{-2} & \alpha \sigma_{M}^{-2} \\
-\alpha \beta \sigma_{M}^{-2} & \sigma_{c}^{-2} + \beta^2 \sigma_{M}^{-2} & -\beta^2 \sigma_{M}^{-2}  \\
\alpha \sigma_{M}^{-2}  & -\beta^2 \sigma_{M}^{-2}  & \sigma_{M}^{-2}  \end{array} \right)\,.
\end{equation}
The probability density is now
\begin{equation}
dP(x_1, c, \Mb) = N \exp \left[\frac{1}{2}\mathbf{y}^T D^{-1} \mathbf{y} \right] dx_1\, dc\, d\Mb = N \exp \left[\frac{x_1^2}{\sigma_{x_1}^2} + \frac{c^2}{\sigma_{c}^2} + \frac{(\Mb + \alpha x_1 - \beta c)^2}{\sigma_{M}^2} \right] dx_1\, dc\, d\Mb \,.
\end{equation}
Inserting the mean values $\mu_{x1}$, $\mu_c$ and $\mu_M$ recovers Equation (\ref{equ:popmodel_basic}). As explained in Section \ref{sec:method:popmodel}, one Gaussian probability function per parameter does not capture outliers in color and magnitude. Thus, we represent each parameter by a sum of Gaussians with independent means, lower and upper variances such that
\begin{equation}
\frac{dP(\theta)}{d \theta}  = \sum_{i} \sqrt{\frac{2}{\pi}} \frac{w_i}{(\sigma_{\theta+,i} + \sigma_{\theta-,i})} \left\{
  \begin{array}{l l l}
    \mathrm{exp}\left(\frac{(\theta - \mu_{\theta,i})^2}{2 {\sigma_{\theta+,i}}^2} \right) & \forall & \theta > \mu_{\theta,i} \\
    \mathrm{exp}\left(\frac{(\theta - \mu_{\theta,i})^2}{2 {\sigma_{\theta-,i}}^2} \right) & \forall & \theta < \mu_{\theta,i} \\
  \end{array} \right.
\end{equation}
where the weights $w_i$ sum to 1. Due to the linearity of the system, all the above equations still hold for the sums of Gaussians. The remaining challenge is to marginalize the distribution over each one of the variables to obtain probability contours in planes of two variables for plotting. For any combination of $x_1$, $c$ and the corrected magnitude, $\Mb + \alpha x_1 - \beta c$, their probability distributions are linear sums of independent Gaussians (Figure \ref{fig:pop_model2}). In the cases of the $x_1-\Mb$ and $c-\Mb$ planes, however, we need to integrate over the asymmetric Gaussian distributions of both the third variable and $\Mb$ (since it is not independent of the third variable). For the case of integrating over $i$ Gaussians in $x_1$, their $\pm$ asymmetry, and $j$ (symmetric) Gaussians in $\Mb$, we obtain
\begin{eqnarray}
\label{equ:dPcM}
\frac{dP(c, \Mb)}{dc\, d\Mb} &=& \frac{dP(c)}{dc} \sum_i \sum_j \sum_{\pm} \sqrt{\frac{2}{\pi}} \frac{w_{x1,i}}{(\sigma_{x1,+,i} + \sigma_{x1,-,i})}  \frac{w_{M,j}}{2 \sigma_{M,j}} \left(\frac{1}{\sigma_{x1,i\pm}^2} +\frac{\alpha^2}{\sigma_{M,j}^2} \right)^{-\frac{1}{2}} \nonumber \\
&& \times\, \mathrm{exp} \left[\frac{(\Mb - \mu_{M,j} + \alpha \mu_{x1,i} - \beta c)^2}{(\sigma_{M,j}^2 + \alpha^2 \sigma_{x1,i\pm}^2)} \right] \times \left[ 1 \pm \mathrm{erf} \left( \frac{\alpha \sigma_{x1,i\pm}}{\sqrt{2} \sigma_{M,j}} \frac{(\Mb - \mu_{M,j} + \alpha \mu_{x1,i} - \beta c)}{\sqrt{\sigma_{M,j}^2 + \alpha^2 \sigma_{x1,i\pm}^2}}  \right)  \right]
\end{eqnarray}
and similarly
\begin{eqnarray}
\label{equ:dPxM}
\frac{dP(x_1, \Mb)}{dx_1\, d\Mb} &=& \frac{dP(x_1)}{dx_1} \sum_i \sum_j \sum_{\pm} \sqrt{\frac{2}{\pi}} \frac{w_{c,i}}{(\sigma_{c,+,i} + \sigma_{c,-,i})}  \frac{w_{M,j}}{2 \sigma_{M,j}} \left(\frac{1}{\sigma_{c,i\pm}^2} +\frac{\beta^2}{\sigma_{M,j}^2} \right)^{-\frac{1}{2}} \nonumber \\ 
&& \times\, \mathrm{exp} \left[\frac{(\Mb - \mu_{M,j} + \alpha x_1 - \beta \mu_{c,i})^2}{(\sigma_{M,j}^2 + \beta^2 \sigma_{c,i\pm}^2)} \right] \times \left[ 1 \mp \mathrm{erf} \left( \frac{\beta \sigma_{c,i\pm}}{\sqrt{2} \sigma_{M,j}} \frac{(\Mb - \mu_{M,j} + \alpha x_1 - \beta \mu_{c,i})}{\sqrt{\sigma_{M,j}^2 + \beta^2 \sigma_{c,i\pm}^2}}  \right)  \right]\,.
\end{eqnarray}
The confidence contours for these probability distributions were computed numerically, and are shown in Figure \ref{fig:pop_model}(c).

\end{document}